\documentclass[11pt,preprint]{aastex}

\newcommand{\hcccn}{HC$_3$N}
\newcommand{\cch}{C$_2$H}
\newcommand{\kms}{km s$^{-1}$}
\newcommand{\kmspc}{km s$^{-1}$ pc$^{-1}$}

\shorttitle{Kinematics and Magnetic Field Studies of Molecular
Clouds}
 \shortauthors{Li et al.}

\begin{document}

\title{Large-Scale Kinematics, Astrochemistry and Magnetic Field Studies of Massive
Star-forming Regions through \hcccn, HNC and \cch\ Mappings}

\author{Juan Li\altaffilmark{1, 2}, Junzhi Wang\altaffilmark{1, 2}, Qiusheng Gu\altaffilmark{1, 2}, Zhi-yu Zhang\altaffilmark{3, 4, 5}, Xingwu Zheng\altaffilmark{1, 2}}

\altaffiltext{1}{School of Astronomy \& Space Science, Nanjing
University, 22 Hankou RD, Nanjing 210093, China; ljuan@nju.edu.cn}
\altaffiltext{2}{Key Laboratory of Modern Astronomy and Astrophysics
(Nanjing University), Ministry of Education, Nanjing 210093, China}
\altaffiltext{3}{Purple Mountain Observatory, CAS, 2 West Beijing
Road, Nanjing, 210008, China} \altaffiltext{4}{Max-Planck-Institut
f\"{u}r Radioastronomie, Auf dem H\"{u}gel 69, D- 53121 Bonn,
Germany} \altaffiltext{5}{Graduate School of the Chinese Academy of
Sciences, 19A Yuquan Road, P.O. Box 3908, Beijing 100039, China}

\begin{abstract}

 We have mapped 27 massive star-forming regions associated with water masers
 using three dense gas tracers: \hcccn\ 10-9, HNC 1-0 and \cch\ 1-0.
The FWHM sizes of HNC clumps and \cch\ clumps are about 1.5
and 1.6 times higher than those of \hcccn, respectively,
which can be explained by the fact that \hcccn\ traces more
dense gas than HNC and \cch. We found evidence for increase in
optical depth of \cch\ with `radius' from center to outer regions in
some targets, supporting the chemical model of \cch. The \cch\
optical depth is found to decline as molecular clouds evolve to
later stage, suggesting that \cch\ might be used as ``chemical
clock" for molecular clouds. Large-scale kinematic structure of
clouds was investigated with three molecular lines. All these
sources show significant velocity gradients. The magnitudes of
gradient are found to increase towards the inner region, indicating
differential rotation of clouds. Both the ratio of rotational to
gravitational energy and specific angular momentum seem to decrease
toward the inner region, implying obvious angular momentum transfer,
which might be caused by magnetic braking. The average magnetic
field strength and number density of molecular clouds is derived
using the uniformly magnetic sphere model. The derived magnetic
field strengths range from 3 to 88 $\mu$G, with a median value of 13
$\mu$G. The mass-to-flux ratio of molecular cloud is calculated to
be much higher than critical value with derived parameters, which
agrees well with numerical simulations.

\end{abstract}

\keywords{ ISM: clouds - ISM: kinematics and dynamics - ISM:
molecules - stars: formation }

\section{Introduction}
\qquad

%Large-Scale Rotation and Magnetic field Studies of Molecular Clouds
%through \hcccn, HNC and \cch\ Mapping Observations

Massive stars play an important role in the evolution of the
Universe as the principal source of heavy elements and UV radiation
and profoundly affect the physical, chemical and morphological
structure of galaxies (e.g., Zinnecker et al. 2007). Understanding
the formation of massive stars is also crucial to an improved
understanding of galaxy formation. However, limited by distance,
complexity, and the rapid evolution of young stars, our knowledge of
the massive star formation process is still crude, in both
observational and theoretical aspects. It is generally accepted that
giant molecular clouds (GMC) are birthplace of massive stars, thus
studies of GMCs are vital to our understanding of the physical and
chemical conditions where massive star-formation occurs. Molecular
line observations could provide plenty of information about
temperature, density, mass, kinematics, magnetic field and so on.
Molecular lines are not only important diagnostic of the chemistry
and physics of molecular clouds, studies of molecular lines
themselves are also of astro-chemical importance since the time
dependence of the chemical composition of GMC provides potential
means to infer the evolutionary timescales of those  sources (e.g.,
van Dishoeck 1998; Gwenlan et al. 2000 ).

Interstellar cyanoacetylene (\hcccn), which was discovered by Turner
(1971), is an excellent dense gas tracer (Morris et al. 1976, 1977;
Chung et al. 1991; Bergin et al. 1996). This molecule belongs to an
important group of interstellar molecules - the cyanopolyynes,
HC$_{2n+1}$N. It has high electric dipole moments ($\mu$ = 3.72
Debye), thus it traces high density regions. Though the abundance of
\hcccn\ is high enough that this molecule is easily detected in many
sources, the lines above J = 5-4 are usually optically thin, which
enables us to study the inner detailed structure of molecular clouds
(Chung et al. 1991).
%Given that the rotational constant of \hcccn\ (B$_0$ =
%4549.058 MHz) is small, and the spectrum is compressed in frequency
%relative to lighter molecules, \hcccn\ has many transitions at cm
%and mm wavelengths than other gas density probing molecules, and use of
%many lines enables more detailed analysis on physical condition
%(e.g., Wyrowski et al. 2003; Pardo et al. 2004).
Observations of the
Orion A GMC complex show that the [\hcccn]/[CN] abundance ratio
varies from values of $\sim$ 10$^{-3}$ for the ionization fronts
surrounding the H II regions, to 100 for the hot core in Orion,
indicating that \hcccn\ might be easily destroyed by ultraviolet
(UV) photons from central ionizing sources, thus the [\hcccn]/[CN]
abundance ratio is proposed to be an excellent tracer of
photon-dominated regions (PDRs) and hot cores within regions of
massive star formation (Rodriguez-Franco et al. 1998). In addition,
the fractional abundances of large massive dense cores seem to be
smaller than that of low mass dense core, supporting the destruction
of \hcccn\ by UV photons (Chung et al. 1991). However, observations
of a large size of star-forming regions are still needed to investigate the effect of
ultraviolet (UV) photons on \hcccn.

The ethynyl radical (\cch) was first detected in interstellar
clouds by Tucker et al. (1974). It is an important molecule for
probing chemical evolution of molecular clouds, since it is a
crucial intermediate molecular in the interstellar chemistry leading
to long chain carbon compounds (Padovani et al. 2009). \cch\ is of
special interest because the rotational transitions are splited into
hyper-fine structure components. At 3 mm, six components with line
strengths varying by over an order of magnitude could be observed,
and their relative intensities could be compared extremely
precisely, which allows for a precise determination of optical depth
(Padovani et al. 2009). Observations indicate that \cch\ is almost
omnipresent toward evolutionary stages from infrared dark clouds
(IRDC) via high-mass protostellar objects (HMPO) to ultracompact HII
regions (UCHII) (Huggins et al. 1984; Beuther et al. 2008; Padovani
et al. 2009; Walsh et al. 2010). Beuther et al. (2008) proposed that
\cch\ is abundant from the onset of massive star formation, but
later it is rapidly transformed to other molecules in the core
center. In the outer regions the abundance of \cch\ remains
high due to constant replenishment of elemental carbon from CO being
dissociated by the interstellar UV photons. However, only a few
mapping observations have been carried out and more observations are
required to study the large-scale spatial distribution and chemical
evolution of \cch\ in massive star-forming regions.

The extensive velocity information contained in spectral line maps
provides good opportunity to analyze kinematics of molecular clouds,
and the presence of rotation has been found in many molecular clouds
in this way (e.g. Zheng et al. 1985; Arquilla et al. 1986; Goodman
et al. 1993; Philips 1999; Liu et al. 2010; Pirogov et al. (2003);
Higuchi et al. (2010); Zhang et al. (2011)). Previous studies have
demonstrated that angular momentum of rotating molecular cloud is
many orders of magnitude higher than that of the stars that
eventually form in this cloud (e.g., Spitzer 1978). Star-formation
models must account for this and several mechanisms are proposed,
e.g., magnetic braking or magnetocentrifugally driven outflows
(e.g., Weise et al. 2010). However, it is still unclear whether
there is one dominant process for dispersing angular momentum during
the entire star-formation process. To solve this problem it is
necessary to measure kinematic structure of a large sample of
molecular clouds in both small and large scales. Previous studies
have come primarily from observations of a single species.
Systematic studies with different molecular tracers are required to
characterize the large scale dynamical structure of molecular clouds
and to investigate the angular momentum problem of molecular clouds.
HNC, one of the most commonly used dense gas tracer (n$_{crit} >
10^4$ cm$^{-3}$), could be observed simultaneously with \hcccn\ 10-9
within 1 GHz band. Thus the HNC 1-0 transition was also observed to
investigate the kinematic structure of molecular clouds by combining
with observations of \hcccn\ and \cch.

In this paper we present mapping observations of \hcccn, HNC and
\cch\ of 27 massive star-forming regions with the Purple Mountain
Observatory 13.7m telescope. The influence of HII region
on \hcccn, optical depth of \cch\,
kinematic structure and magnetic field of GMC are investigated. We first
introduce the observations and data reductions in \S\
\ref{observation}. In \S\ \ref{result}, we present the observational
results. The analysis and discussion of the results are presented in
\S\ \ref{discussion}, followed by a summary in \S\ \ref{summary}.

\section{OBSERVATIONS AND DATA REDUCTION }
\label{observation}

\subsection{PMO 14m Observations}

We performed mapping observations of \hcccn\ 10-9 (90978.99 MHz),
HNC 1-0 (90663.59 MHz) and \cch\ 1-0 (87317.05 MHz) with the Purple
Mountain Observatory 13.7m telescope (PMO 14m) located in Delingha,
China in May and June, 2010. Position-switch mode was used for all
the mapping observations. The main beam size is about 55\arcsec,
corresponding to spatial resolution between 0.015 and 0.24 pc for all
sources. The pointing accuracy is estimated to be better than 9\arcsec. A
cooled SIS receiver working in the 80-115 GHz band was employed. A
fast Fourier transform spectrometer (FFTS) of 16,384 channels with
bandwidth of 1 GHz was used, supplying a velocity resolution of
about 0.21 \kms. \hcccn\ 10-9 and HNC 1-0 were observed
simultaneously. Typical system temperature was around 150-350 K,
depending on the weather conditions. Typical on source time for each
position is about 3 minutes, resulting in rms noise level of about
0.1$\sim$0.2 K (T$_A^*$) per channel (0.21 \kms). The mapping step
is 30\arcsec\ for all observations. The mapping size
was extended to about one fifth of the peak strength.

Most objects in our sample come from Shirley et al. (2003) of
massive star-forming cores associated with water masers. We first
searched for \hcccn\ 10-9 emission toward the whole sample of
Shirley et al. (2003). \hcccn\ emission was detected above 0.5 K in
25 sources ($\sim$41\%). Then we conducted mapping observations of
\hcccn, HNC and \cch\ toward these sources. The observing center was
the water maser position from the catalog of Cesaroni et al. (1988).
These sources have been mapped with CS and HCN transitions, as well
as dust emission (Shirley et al. 2003; Mueller et al. 2002; Wu et
al. 2010). All of the objects mapped are listed in Table 1, where we
give the source name in column (1), the ($\alpha$, $\delta$)
coordinates in column (2) and (3), the distance $D$ in column (4),
the reference of distance in column (5), and the galactocentric
distance $D_g$ in column (6). To enlarge the sample, we also
included three more massive star-forming regions (sources below the
blank line in Table 1). The distances were determined from an
extensive literature search. Trigonometric parallax distances were
used if available. The galactocentric distance, $D_g$, is derived by
adopting a distance of 8.5 kpc to the solar circle.

%Two sources have been mapped in \hcccn\ 10-9 transition by Chung et
%al. (1991); however, we still observe them for systematical
%investigation and comparison.

The data processing was conducted using Gildas\footnote{\tt
http://www.iram.fr/IRAMFR/GILDAS.}. Linear baseline subtractions
were used for most spectra, while sin baseline subtractions were
used for spectra showing obvious standing wave components. The line
parameters (central velocity, width, and peak intensity) are
obtained by Gaussian fitting. The data are presented in the unit of
antenna temperature
(T$_A^*$).

\subsection{VLA Archive Data}

To characterize the evolution stage of the molecular clouds,
continuum data at X or C band, have been gathered from NRAO DATA
Archive System\footnote{\tt
http://archive.nrao.edu/archive/e2earchivex.jsp.}. Table
\ref{tab:vlaobs} summarizes the detailed information of VLA archive
data, including observing date in column (2), observing
configuration in column (3), observing frequency in column (4), the
synthesized beam in column (5), and project code in column (6). The
data processing was conducted using AIPS\footnote{\tt
http://www.aips.nrao.edu.}. Depending on source intensities, the
integration times of observations range from several minutes to
several hours. The resulted rms noise levels range from 0.01 to 50
mJy beam$^{-1}$.

\section{RESULTS}
\label{result}

\subsection{Molecular Line Results}

Figure 1 shows spectra of \hcccn\ 10-9, HNC 1-0 and \cch\ 1-0 (J =
3/2 $\rightarrow$ 1/2, F = 2 $\rightarrow$ 1) in the observing
center for four sources. The line profiles of \hcccn\ and \cch\
could be well fitted with a single Gaussian. The line profiles of
HNC could be fitted with a single Gaussian in about half sources.
Blue asymmetric structure, named `blue profiles', a
combination of a double peak with a brighter blue peak or a skewed
single blue peak in optically thick lines (Mardones et al. 1997),
has been found in HNC lines in five sources: Cep A, W33cont, W75OH,
G121.30+0.66 and W44. Three of these sources have been identified as
infalling candidates by Sun et al. (2009) except for W33cont and
G121.30+0.66. `Red profiles' are found in source G14.33-0.64,
G35.20-0.74, S87 and ON2S. The origin of `red profile' remains
unclear, which might be caused by outward motion or rotation. Figure
\ref{intensity1} - \ref{intensity9} show contour maps of \hcccn\
10-9, HNC 1-0 and \cch\ 1-0 (J = 3/2 $\rightarrow$ 1/2, F = 2
$\rightarrow$ 1) imposed on continuum images in gray-scale for 27
clumps. The contours are between 30\% and 90\% of the peak, in steps
of 20\% of the peak intensity. The heavy lines in contour maps
indicate the half-peak contours. The clumps
show various morphologies. Most of them have single peaks, % Higuchi et al. 2010, ApJ, 719, 1813
similar to previous CS and HCN observations (Shirley et al. 2005; Wu
et al. 2010). The peak antenna temperature (T$_A^*$), integrated
line intensity ($\int$T$_A^*$d$\nu $), LSR velocity, and full-width
half-maximum (FWHM) linewidth of \hcccn\ 10-9, HNC 1-0 and \cch\ 1-0
(J = 3/2 $\rightarrow$ 1/2, F = 2 $\rightarrow$ 1) at the central
(0, 0) position of the massive dense cores are present in Table
\ref{tab:dlhresult}. All the results are obtained by Gaussian
fitting. It should be noted that a single-gaussian fit is
insufficient for the optical thick molecular lines of HNC in some
sources. Sources that can not be fitted well with a single gaussian
are marked with `$*$' in Table \ref{tab:dlhresult}. \hcccn\ 10-9
maps of DR21S and S140 have also been observed by Chung et al.
(1991) with Nobeyama 45m telescope. Despite the different beam sizes
of the Nobeyama 45m and PMO 14m telescope, the morphology and size
from two observations are quite consistent. \cch\ 2-1 emission of
DR21S have been observed by Tucker \& Kutner (1978) with NRAO 11m
telescope, while \cch\ 2-1 and \hcccn\ 10-9 maps of Cep A have been
obtained by Bergin et al. (1996, 1997) with 14m FACRO telescope. The
morphology and size from these observations are also consistent with
our observations.

Figure \ref{intencomp} shows comparison of integrated intensities
averaged over the entire regions for three species. A tight
correlation between \hcccn\ and HNC (r=0.92) was found. The result given
by least-squares fitting is:

I(HNC) = (3.8 $\pm$ 0.4) $\times$ I(\hcccn).

A very high correlation coefficiency (r=0.88) was also found for
average integrated intensities of \cch\ and \hcccn\ (Figure
\ref{intencomp}). The result given by least-squares fitting is:

I(\cch) = (1.3 $\pm$ 0.2) $\times$ I(\hcccn).

The tight correlation of intensities between \hcccn, HNC and \cch\
imply correlation of the abundance of three species in the early
evolutionary stage of massive star-formation.

%Under assumptions of optically thin and Local Thermodynamic
%Equilibrium (LTE), the total column density of molecular gas could
%be calculated using (Scoville et al. 1986):
%\begin{equation}
%    N = \frac{3k}{8 \pi ^{3}B \mu ^{2}} \frac{e^{hBJ_l (J_l +1)/kT_{ex}} }{J_l + 1 } \frac{T_{ex} + hB/3k }{ 1 - e^{-h \nu /kT_{ex}}} \int \tau_{\nu} d v,
%\end{equation}
%where B is the rotational constant of the molecule, $\mu$ is
%permanent dipole moment, and $J_l$ is the rotational quantum number
%of the lower state in the observed transitions. The abundance ratio
%of HNC to \hcccn\ is about 20:1 while adopting a typical $T_{ex}$ of
%20 K, HNC optical depth of 2, and \hcccn\ optical depth of 0.2.

The size of clouds is characterized using beam deconvolved angular
diameter and linear radius of a circle with the same area as the half
peak intensity:
\begin{equation}
\theta_{transition} = 2(\frac{A_{1/2}}{\pi} -
\frac{\theta_{beam}^2}{4})^{1/2},
\end{equation}
\begin{equation}
R_{transition} = D(\frac{A_{1/2}}{\pi} -
\frac{\theta_{beam}^2}{4})^{1/2},
\end{equation}
where $A_{1/2}$ is the area within the contour of half peak
intensity, $\theta_{beam}$ is the FWHM beam size, and $D$ is the
distance of the source. The centroid of velocity-integrated
intensity maps, the linear FWHM size and angular FWHM size of clouds
for three species are tabulated in Table \ref{tab:size}.
%Figure \ref{historadius} shows
%distribution of FWHM radius.
The radius of \hcccn\ ranges from 0.05 to 1.33 pc, with a median
value of 0.24 pc. Nearly half of cloud sizes derived from \hcccn\
are smaller than the telescope beam size. The radius of HNC ranges
from 0.21 to 2.08 pc, with a median value of 0.44 pc. The FWHM sizes
of HNC are well above the beam size. The radius of \cch\ clouds
ranges from 0.11 to 2.17 pc, with a median value of 0.49 pc. Nearly
all the FWHM sizes of \cch\ are well above the beam size. Figure
\ref{sizecomp} shows comparison of FWHM size of three species.
Tight correlation was found between FWHM sizes of HNC and \hcccn\
(r=0.93), as well as between the sizes of \cch\ and \hcccn\
(r=0.85). Results given by least-squares fitting are:

R(HNC) = (1.68 $\pm$ 0.17) $\times$ R(\hcccn),

R(\cch) = (1.90 $\pm$ 0.18) $\times$ R(\hcccn).

The FWHM sizes shrink as excitation requirements increase,
indicating that the clumps are centrally condensed, with either a
smooth decrease in density with increasing radius, or an increased
fraction of the volume filled by dense ``raisins" (Mueller et al.
2002; Wu et al. 2010).

The FWHM linewidths of \hcccn, HNC and \cch\ at (0, 0) range from
1.82 to 9.84 \kms, 2.07 to 9.96 \kms, and 1.07 to 8.68 \kms, with
median values of 3.18, 4.07 and 3.75 \kms, respectively. The largest
value of linewidths, 9.96 \kms, comes from HNC observation of W51M,
while the smallest value, 1.07 \kms, comes from \cch\ observations
of AFGL 490. The linewidths of HNC lines are about 1.2 times larger
than those of \hcccn, which should be caused by large optical depth
of HNC.
%The linewidths of \cch\ lines are about 1.05 times larger
%than those of \hcccn.
The linewidth of \hcccn\ is larger than that of \cch\ or HNC
in sources such as G8.67-0.36, G10.6-0.4 and W51M, indicating more
violent turbulence in \hcccn\ emission region than the outward.
%For a Gaussian line shapes, the
%broadening of optical depth could be calculated using:
%\begin{equation}
%    \frac{\triangle v}{\triangle v_0} = \frac{1}{\sqrt{ln 2}} \sqrt{ln\{\tau /[ln \frac{2}{1 + exp (-\tau)}] \} },
%\end{equation}
%where $\triangle v_0$ is the optically thin linewidth (Philips et
%al. 1979). The HNC to \hcccn\ linewidth ratio corresponding to an
%average optical depth of 2.05 for HNC.

\subsection{Radio Continuum Emission}

In Table \ref{tab:vlaresult}, we list the peak intensity $S_p$ in
column (2), flux density $S_{\nu}$ in column (3), the deconvolved
size in column (4), position angle in column (5), linear size in
column (6), and classification of HII regions in column (7). All the
parameters are obtained by fitting one single Gaussian component to
the strongest emission peak with AIPS task IMFIT, and the linear
size of HII region is calculated using the fitting results of major
axis. No prominent continuum emission was detected in G14.33-0.64
($\sigma \sim$ 0.04 mJy), G59.78+0.06 ($\sigma \sim$ 0.008 mJy) and
W75OH ($\sigma \sim$ 0.4 mJy),
%According to literature search, no
%obvious radio continuum emission has been reported for these
%sources, which is consistent with our results.
suggesting that these sources are in early evolutionary phase.

Based on the 2 cm size measurements and densities, stellar
ionized regions regions are usually classified as hyper-compact
(HCHII) ($\leq$ 0.01 pc), ultra-compact (UCHII) ($\leq$0.1 pc),
compact (CHII) ($\leq$0.5 pc) and extended HII regions ($>$0.5 pc or
clearly associated with a classical HII regions) (Wood \& Churchwell
1989; Kurtz 2000, 2002; Churchwell 2002), which probably represents
an evolutionary sequence. It should be noted that the morphology of
HII regions is strongly dependent upon the UV-coverage (e.g., long
baselines are not sensitive to the extended emission), which makes
the classification of HII regions ambiguous.
%For those in
%which multiple continuum sources including UCHII and HCHII regions
%are detected (e.g., W51M, S140), we regard them as UCHII for
%simplicity.
Previous classification is always adopted if available. Most of our
sample are classified as UCHII regions, and three sources (S231,
S235 and S76E) are classified as extended HII regions.

%Both S231 and S235 have low $\tau_{aver}$
%(0.10 and 0.45), the only exception is S76E, which has a high
%$\tau_{aver}$ of 2.01.

\subsection{Individual Targets}

Comparison with previous CS, HCN and 350 $\mu$m dust continuum
emission (Mueller et al. 2002; Shirley et al. 2003; Wu et al. 2010)
indicates that molecular lines observed here exhibit similar
morphologies to CS and HCN for most sources. The molecular
emission morphology has a very centrally condensed structure for
about half of the sample, with emission peak near to the HII region
and the maxima of dust continuum emission. Such a simple structure
suggests simple density gradient increasing from the edge of the
dense core to the center. Sources with distinctive distributions are
discussed below.

%G8.67-0.36: Molecular lines observed exhibit similar morphologies to
%CS and HCN emission. There are single cores elongated from southwest
%to northeast in spectral line maps, with emissions peak toward the
%UCHII region and the maxima of dust continuum emission.

%G10.6-0.4: The molecular emission morphology in G10.6-0.4 has a very
%centrally condensed structure. Molecular line emissions are intense
%toward the CHII region and the maxima of dust continuum emission.
%Such a simple structure suggests simple density gradient increasing
%from the edge of the dense core to the center.

%G12.42+0.50: Centrally condensed structure was also observed in this
%source. The molecular line emissions peak toward the UCHII region
%and the maxima of dust continuum emission.

%W33cont: Molecular line emissions show centrally condensed structure
%in this source. The molecular line emission peaks near the maxima of
%radio continuum emission.

%W33A: W33A HII region is classified as an HCHII region because of
%small size and rising spectra up to 43 GHz (Rengarajan \& Ho 1996;
%van der Tak \& Menten 2005). Molecular lines observed show similar
%morphology, with a single peak near the HCHII region and the maxima
%of dust continuum emission.

%G14.33-0.64: \hcccn\, \cch\ and CS clouds show centrally condensed
%structure, with single cores peak toward the maxima of dust
%continuum emission. HNC shows a marked difference from \hcccn\ and     not special
%\cch, with a single core elongated from north to southwest.

W42: The radio continuum map consists of a UCHII and a CHII region.
All the molecular line emissions except \cch\ peak towards the UCHII
region (Shirley et al. 2003; Wu et al. 2010). Both the HNC and \cch\
exhibit a single core elongating from southeast to northwest in
velocity-integrated intensity maps, while the \hcccn\ emission is
seen to avoid the CHII region. The different spatial distribution of
three species around HII regions can be explained by the
destruction of \hcccn\ by UV photons.

W44: The radio continuum map consists of a UCHII region and a
supernova remnant. Molecular line emissions peak toward the UCHII
region. HNC, \cch, HCN and CS maps show weak tails toward supernova
remnant (Wu et al. 2010).

S76E: An extended HII region is shown in VLA map of S76E. There is a
single core in spectral line maps. The molecular line emissions have
an offset from peak of the radio continuum emission, which might be
caused by relative motion between young stellar objects and
molecular clouds. It is also possible that the HII region is not
associated with molecular cloud.

G35.20-0.74: Molecular line emissions exhibit similar centrally
condensed structure. \hcccn\ and HNC emissions peak toward CHII
region, while \cch\ and CS emissions peak to the southeast of CHII
region (Shirley et al. 2003; Wu et al. 2010).

W51M: High-resolution centimeter observations reveal several compact
continuum components in W51M (e.g., Gaume et al. 1993; Shi et al.
2010). HNC, high J transitions of CS and HCN and dust emission
reveal two cores, with the southeast core associated with water
masers (Shirley et al. 2003; Wu et al. 2010; Mueller et al. 2002).
\hcccn, \cch\ and CS emissions peak towards the southeast core,
while HNC and HCN peak toward the northwest core. The \hcccn\
linewidth is widest among all our sample, indicating rather violent
turbulence in this source.

G59.78+0.06: No radio continuum emission was detected in this source.
Both \hcccn\ and \cch\ emissions are relatively weak in
G59.78+0.06. HNC shows similar morphology to HCN and CS emission (Wu et al. 2010),
with a single peak away from water masers.

S87: VLA maps of S87 HII region reveal a radio continuum source
consisting of a compact core and a fan-shaped tail extending to the
southeast (Bally \& Predmore 1983; Barsony 1989). There are single
core elongated from northeast to southwest in spectral line maps.
The peak molecular line emissions obviously offset from both the
maxima of radio continuum emission and dust continuum emission
(Mueller et al. 2002).

%ON1: ON 1 is one of the smallest UCHII regions in the Galaxy and
%exhibits various star formation signposts (Zheng et al. 1985; Su et
%al. 2009). Molecular line emissions exhibit simple centrally
%condensed structure, with a single peak near the maxima of radio
%continuum emission and dust continuum emission.

ON2S: The map consists of two molecular clouds. The northern
molecular cloud (ON2N) contains the G75.78+0.34 UCHII region excited
by an early B star, and the southern cloud (ON2S) contains the
G75.77+0.35 HII region excited by an O star (e.g., Matthews \&
Spoelstra 1983; Dent et al. 1988). Molecular line emissions show
similar morphology, with emission peaks slightly offset from UCHII
region and the maxima of dust continuum emission (Mueller et al.
2002). It should be noted that the peculiar morphology of \cch\ map
is mainly caused by the low signal-to-noise ratio (SNR).

%W75N: Molecular line emissions show simple morphology, with a single
%peak near the UCHII region.

DR 21S: The DR 21S massive star-forming region contains two cometary
HII regions, aligned nearly perpendicular to each other on the sky.
The molecular line emissions peak to the south of HII regions.
\hcccn\ has the most compact emission morphology, while the \cch\
has the most extended emission.

W75OH: We did not detect prominent radio continuum emission
associated with this source. \hcccn, HNC, CS and HCN emissions show
similar morphology, with filaments structure consisting of two cores
extending from south to north in spectral line maps (Shirley et al.
2003; Wu et al. 2010). \cch\ emission differs significantly from
other species, with integrated intensities varying slowly with
position, which agrees well with observations of Tucker \& Kutner
(1978). They attributed it to that \cch\ emission is produced almost
entirely in the extended molecular clouds containing the rich
molecular cores, with little or no emission arising from the dense
cores themselves.

%Cep A: Cep A is a complex molecular cloud condensation which
%contains no fewer than 14 HII regions (e.g., Hughes \& Wouterloot
%1984; Hughes et al. 1995). There are single cores elongated from
%northeast to southwest in spectral line maps, with emission peak
%near HII region and dust continuum emission.

W3(OH): Molecular line emissions show similar morphology. Obvious
offset between \hcccn\ emission peak emission and HII regions was
found, while other molecular line emissions peak near to HII
regions.

S231: The high resolution VLA observation seems to resolve the
extended emission, so we adopt it as an extended HII region,
following the classification of Israel \& Felli (1978). There are
single cores in spectral line maps, with emission peak near to HII
region and maxima of dust continuum emission (Mueller et al. 2002).

S235: This is an extended HII region. The spectral line maps offset
significantly from peak radio continuum emission, which might be
caused by the relative motion between young stellar embedded in HII
regions and their molecular clouds. The relative motions have been
proposed while inspecting the velocity difference between HII
regions and molecular clouds, which are found to range from 0.5 to
10 \kms\ (Zuckerman 1973; Churchwell et al. 2010). The radio
recombination line (RRL) of S235 is observed to be -23.07 \kms\
(Quireza et al. 2006), thus the radial velocity difference between
HII region and \hcccn\ cloud is 6.28 \kms, implying possibility of
relative motion. Separation between \hcccn\ centroid and peak
emission of HII regions is 28\arcsec. At the assumed distances of
1.6 kpc, the separation correspond to 0.22 pc. Adopting a proper
motion velocity of 1 \kms, we could obtain a timescale of
2$\times$10$^5$ yr, which is consistent with the evolution timescale
of UCHII regions (e.g., Mellema et al. 2006). The molecular line
emissions peak near to the maxima of dust continuum emission
(Mueller et al. 2002).

S255: There are single cores elongated from northwest to southeast
in spectral line maps. The \hcccn\ and HNC emissions peak toward
UCHII region, which is 60\arcsec\ north of water masers. CS and HCN
peak close to water masers, while the maxima of \cch\ emission lies
in between water masers and UCHII region.

S140: There are single cores elongated from northeast to southwest
in molecular line maps. \hcccn\ and HNC emission peak toward radio
continuum emission and water masers, but slightly offset from the
maxima of dust continuum emission (Wu et al. 2010). \cch\ emission
peak offset from HII region and water masers by about half beam.

Previous observations suggest that \hcccn\ is easily destroyed by UV
photons (e.g., Rodriguez-Franco et al. 1998; Gwenlan et al. 2000).
Though angular resolution seems too coarse to investigate this
issue, observations present here could give us some hints, e.g.,
separations larger than $\theta_{mb}/2$ are found between the maxima
of radio continuum emission and \hcccn\ emission in S87 and S235.
Different spatial distribution between \hcccn\ and other molecular
lines in W42 also support that point.

\section{DISCUSSIONS}
\label{discussion}

\subsection{Central Depletion of \cch}

We can use the hyperfine components to obtain the \cch\ optical
depth. In Figure \ref{specc2h} we show an example of a \cch\ spectra
for Cep A. The spectra are averaged over the entire emitting region.
As is shown in the figure, \cch\ 1-0 has six hyperfine components (J
= 3/2 $\rightarrow$ 1/2, F = 1 $\rightarrow$ 1; J = 3/2
$\rightarrow$ 1/2, F = 2 $\rightarrow$ 1; J = 3/2 $\rightarrow$ 1/2,
F = 1 $\rightarrow$ 0; J = 1/2 $\rightarrow$ 1/2, F = 1
$\rightarrow$ 1; J = 1/2 $\rightarrow$ 1/2, F = 0 $\rightarrow$ 1; J
= 1/2 $\rightarrow$ 1/2, F = 1 $\rightarrow$ 0). If the \cch\ 1-0
line is optically thin and the hyperfine levels are populated
according to LTE, the line ratios between the six hyperfine
components approximate 1:10:5:5:2:1. The intrinsic relative
intensities of the hyperfine components are taken from Tucker et al.
(1974). We estimated the optical depth of \cch\ 1-0 (J = 3/2
$\rightarrow$ 1/2, F = 2 $\rightarrow$ 1) by fitting the lines using
METHOD HFS of the CLASS program, which is part of the GILDAS
package. This method assumes that all the hyperfine components have
the same excitation temperature and broadening with fixed separation
according to the laboratory value. Note that CLASS HFS METHOD could
only provide correct fitting results for optical depth larger than
0.1.

In Table \ref{tab:opdepth}, we list optical depth of spectral line
averaging over the whole emitting region $\tau_{aver}$ in column
(2), optical depth of spectral line at (0, 0) $\tau_{0,0}$ in column
(3), maximum optical depth $\tau_{max}$ in column (4), and offset of
spectral line with $\tau_{max}$ in column (5). Note that results
present here are optical depth of the main hyperfine line.
$\tau_{aver}$ ranges from 0.10 to 5.37, with a median value of 1.16.
$\tau_{0,0}$ is smaller than $\tau_{aver}$ for source W33cont, ON1,
DR21S, S140 and NGC2264. For most sources the offset of $\tau_{max}$
is different from \cch\ centroid. Such trend is also reflected in
the opacity maps. Figure \ref{opdepth1} - \ref{opdepth3} show
velocity integrated contour maps of \cch\ 1-0 (F = 3/2, 2
$\rightarrow$ 1/2, 1) superimposed on opacity maps of \cch\ 1-0 (J =
3/2 $\rightarrow$ 1/2, F = 2 $\rightarrow$ 1) in gray-scale. The
optical depth has obvious dip toward the emission peak of \cch\ in
eight sources, such as DR21S, Cep A, W51M, ON1 and NGC2264. Since
optical depth is a direct measurement of column density, thus the
\cch\ column density should decrease toward the centroid for these
sources, which is well consistent with chemical model of \cch\
(Beuther et al. 2008; Padovani et al. 2009; Walsh et al. 2010).
However, some sources donot show signature of \cch\ depletion toward
hot cores or HII regions, such as G14.33-0.64, G8.67-0.36, W33A.
Limited by low resolution (FWHP$\sim$55\arcsec) and high rms noise
level ($\sim$0.2 K with velocity resolution of $\sim$0.21 km
s$^{-1}$) of our observations, it is hard to tell whether this is
intrinsic to these sources or not.

Figure \ref{opdistribution} shows $\tau_{(0, 0)}$ versus
$\tau_{aver}$ for observing sources. Sources in which no radio
continuum emission has been detected and those associated with HCHII
regions are plotted as circles. Sources associated with UCHII region
are shown in squares, while sources associated with CHII or HII
region are plotted as star. One could see clearly from the figure
that those ``young" molecular clouds that associate with HCHII or
not associated with detectable HII regions have larger optical depth
than ``elder" molecular clouds that associated with UCHII, CHII or
extended HII regions. The optical depth of our sample is much
smaller than two prestellar cores observed by Padovani et al.
(2009), in which the total optical depth ($\tau_{total} =
\frac{\tau_{main}}{0.4167}$) ranges from 13.5 to 29.4. These results
suggest that \cch\ abundance decreases as molecular clouds evolve
from hot cores to extended HII regions. Thus \cch\ might be used as
``chemical clock" for molecular clouds.

%Both S231 and S235 have low $\tau_{aver}$
%(0.10 and 0.45), the only exception is S76E, which has a high
%$\tau_{aver}$ of 2.01.

\subsection{Large-Scale Kinematic Structure of Molecular Clouds} \label{kinematics}

\subsubsection{Velocity Gradient}

Molecular line observations of different density tracers provide
good opportunity to analyze the kinematic structure of GMC. Figure
\ref{velspec} shows spectra of \hcccn\ (10-9) at the offset position
of $\triangle Dec = 0$ for W51M and Cep A. The dashed lines, solid
lines and dotted lines are used to mark centroid velocities for
spectral lines at different offset in right ascension. The centroid
velocity shifts are about 9.3 \kms\ and 1.5 \kms\ for W51M and Cep
A, respectively, implying presence of large-scale rotation in
molecular clouds (Mardones et al. 1997). We obtain velocity
gradients with the gradient calculation method formulated in Goodman
et al. (1993) by fitting the function:
\begin{equation}
v_{LSR} = v_0 + a \vartriangle \alpha + b \vartriangle \beta,
\end{equation}
where $v_{LSR}$ represents an intensity-weighted average velocity
along the line of sight through the cloud, $v_{0}$ is the systematic
velocity of the cloud, with respect to the local standard of rest,
$\vartriangle \alpha$ and $\vartriangle \beta$ represent offsets in
right ascension and declination in radians, and a and b are the
projections of the gradient per radian on the $\alpha$ and $\beta$
axis. The magnitude of the velocity gradient, in a cloud at distance
$D$, is
\begin{equation}
Grad = \frac{(a^2 + b^2)^{1/2}}{D},
\end{equation}
and its direction (the direction of increasing velocity, measured
east of north) is given by
\begin{equation}
\theta = tan^{-1} \frac{a}{b}.
\end{equation}

Only points with integrated intensities above 3$\sigma$ levels are
used in the calculation. As is stated in Goodman et al. (1993), maps
with fewer than nine detections could not be fitted reliably, so we
excluded results of any fits to fewer than nine points. Result of
gradient fitting for three species, including magnitude of velocity
gradient ($Grad$), direction of velocity gradient ($\theta$), ratio
of rotation kinetic energy to gravitation energy ($\beta$), and
specific angular momentum ($J/M$) are tabulated in Table
\ref{tab:velgrad}. Significant gradients ($Grad \geq 3
\sigma_{Grad}$) were detected in 17, 26 and 18 sources for \hcccn,
HNC and \cch, respectively. Note that only significant gradient
($Grad \geq 3 \sigma_{Grad}$) are used in the following discussions.
Fewer detections in \hcccn\ are mainly caused by small size of
\hcccn\ maps, some of which donot have enough data points. This is
also the case for HNC map of ON2S and ON2N. The high detection rate
with HNC ($\sim$100\%) suggests that rotation should be quite common
and present in nearly all molecular clouds.

The derived gradients of \hcccn\ range from 0.24 to 6.03 \kmspc,
with a median value of 1.43 \kmspc. The detected gradients of HNC
range from 0.13 to 3.58 \kmspc, with a median value of 0.76 \kmspc.
The detected gradients of \cch\ range from 0.25 to 3.70 \kmspc, with
a median value of 0.85 \kmspc. Velocity gradients in some of our
targets have already been detected in previous studies, such as
S235, S140, AFGL490 (Higuchi et al. 2010), G10.6-0.4 (e.g., Keto et
al. 1987; Liu et al. 2010) and W75OH (Schneider et al. 2010). The
velocity gradients derived using Nobeyama 45m observations of
H$^{13}$CO$^{+}$ range from 0.5 \kmspc\ to 4.3
\kmspc\ (Higuchi et al. 2010), which are in agreement with results present here.
%The molecular outflows in our targets have already been
%detected in previous studies. Higuchi et al. (2010) pointed out that
%there are no correlations between the orientation of the outflows
%and those of the velocity gradients.

Figure \ref{graddirec} shows distribution of direction of detected
gradients. The gradient directions show no signature of alignment
with the overall direction of Galactic rotation, which is consistent
with observations of early-type stars and field Ap stars, in which
random orientation of rotational axes are observed (eg., Fleck \&
Clark 1981; Imara \& Blitz 2011). The result supports the idea that
the initial cloud angular momentum is unlikely to come from Galactic
rotation.

Figure \ref{graddirecomp} shows comparison of gradient direction
between HNC and \hcccn, as well as \cch\ and \hcccn. Results of
least-squares fitting and calculated correlation coefficients ($r$)
are as follows:

$\theta_{HNC}$ = (1.01$\pm$0.12)$\times \theta_{HC_3N}$, $r$ = 0.91,

$\theta_{C_2H}$ = (1.10$\pm$0.23)$\times \theta_{HC_3N}$, $r$ =
0.80.

Since \hcccn\ traces inner denser region, results above indicate
than the gradients orientation are preserved over a range of density
and these tracers are indeed mapping out a real physical entity.
This is consistent with results of Goodman et al. (1993).

Figure \ref{gradcomp} shows comparison of gradient magnitude of HNC,
\hcccn\ and \cch. Tight correlation between three species were found
and the least-squares fitting results and correlation coefficients
are as follows:

$Grad_{HNC}$ = (0.63$\pm$0.08)$\times Grad_{HC_3N}$, $r$ = 0.89,

$Grad_{C_2H}$ = (0.68$\pm$0.10)$\times Grad_{HC_3N}$, $r$ = 0.91.

It is evident that the magnitude of gradient increase toward the
inner region, $\omega \propto \frac{1}{R}$, thus most of those
molecular clouds should experience differential rotation, rather
than rigid-body rotation ($\omega$ constant). In addition, the
rotation velocity $v$ remains constant. These results are opposite
of what is expected based on angular momentum conservation. It seems
to be well consistent with MHD simulation of Mellon \& Li (2009).
They studied the effects of ambipolar diffusion on magnetic braking
and disk formation and evolution during the main accretion phase of
star formation. Their results show that rotation velocity $v$ is
nearly constant in the region $6 \times 10^{15}$ cm $\leq r \leq 1
\times 10^{17}$ cm (see Figure 3 of Mellon \& Li 2009) because of
angular momentum loss caused by magnetic braking. The specific
angular momentum was computed in Section 4.2.3 to better illustrate
this point.

\subsubsection{Ratio of Rotation Kinetic Energy to Gravitational Energy}

The parameter $\beta$, which is defined as the ratio of rotational
energy to the gravitational potential energy, is always used to
quantify the dynamical role of rotation in a cloud. It could be
written as (Goodman et al. 1993):
\begin{equation}
\beta= \frac{(1/2)I\omega^2}{qGM^2/R} =
\frac{1}{2}\frac{p}{q}\frac{\omega^2R^3}{GM},
\end{equation}
where I is the moment of inertia, which is given by $I=pMR^2$ for
the particular shape and density distribution of the cloud, and
$qGM^2/R$ represents the gravitational potential energy of mass $M$
within radius $R$. For a uniform density sphere, $q=\frac{3}{5}$ and
$p=\frac{2}{5}$, thus $p/q=\frac{2}{3}$. For a sphere with an
density profile of $r^{-2}$, $p/q=0.22$, which reduces $\beta$ for
fixed $M$, $\omega$ and $R$, by a factor of 3 comparing with the
uniform density case.

For a sphere with constant density $\rho_0$, it could be written as
(Goodman et al. 1993):
\begin{equation}
\beta= \frac{1}{4\pi G\rho_0}\omega^{2} =  \frac{1}{4\pi
G\rho_0}\frac{Grad^2}{sin^2 i},
\end{equation}
where sin i is assumed to be 1 in our calculation. The density of
clouds are calculated using virial mass (e.g., Shirley et al. 2005).
The virial mass of the clouds with power-law density distribution is
given by

\begin{equation}
M_{Vir}(R) = \frac{5R\triangle v^2}{8a_1a_2Gln2} = 209\frac{(R/1
pc)(\triangle v/1 km s^{-1})}{a_1a_2} M_{\odot};
\end{equation}

\begin{equation}
a_1 = \frac{1-p/3}{1-2p/5}, p < 2.5,
\end{equation}
where $a_1$ is the correction for power-law density distribution and
$a_2$ is the correction for a nonspherical shape (Bertoldi \& McKee
1992). For aspect ratios less than 2, $a_2 \sim$ 1 and can be
ignored for our sample. An average $p$ value of 1.77 from Mueller et
al. (200) is adopted to calculate $a_1$ for all clouds. The FWHM
linewidth of clouds, $\triangle v$, are determined from a Gaussian
fitting to the average spectra over regions of half power contour.

The calculated $\beta$ for three species are listed in Table
\ref{tab:velgrad}. $\beta_{HC_3N}$ range from 0.0006 to 0.028, with
a median value of 0.011, $\beta_{HNC}$ range from 0.0001 to 0.067,
with a median value of 0.018, and $\beta_{C_2H}$ range from 0.0012
to 0.032, with a median value of 0.012. Results presented here seem
to be consistent with previous observations of NH$_3$, in which
molecular clouds were observed to have $\sim$2\% rotational energy
against gravitational energy (Goodman et al. 1993). $\beta_{C_2H}$
seem to be smaller than $\beta_{HNC}$, implies that \cch\ should
arise from more extended and diffuse components, which is consistent
with observations of Tucker \& Kutner (1978). It is noted that
Goodman et al. (1993) have pointed out that gradients determined
using low-density tracers can be contaminated by gas at the size
scale most populated by that tracer, and thus can give inappropriate
estimate of $\beta$, a result also suggested by the analysis in
Arquilla \& Goldsmith (1986).

Goodman et al. (1993) found that $\beta$ is roughly independent of
size in their analysis of NH$_3$ observations, but they caution that
it is necessary to measure $\beta$ using several density tracers in
individual cores in order to avoid density selection effects. In
Figure \ref{betacomp} we show comparison of $\beta$ among three
molecular lines. Correlation coefficient between $\beta_{HNC}$ and
$\beta_{HC_3N}$ is 0.62. We also carried out least-squares fitting
to data of HNC and \hcccn:

$\beta_{HNC}$ = (1.48$\pm$0.52)$\times$ $\beta_{HC_3N}$.

Correlation coefficient between $\beta_{C_2H}$ and $\beta_{HC_3N}$
is only 0.52, so we did not carry out least-square fitting. A dotted
line with a slope of 1 is plotted for reference (Figure
\ref{betacomp} (right)). We can see from the figure that
$\beta_{C_2H}$ is larger than $\beta_{HC_3N}$ for 10 sources (67\%).
Thus we conclude that the ratio of rotational kinetic energy to
gravitational energy does depend on size and decreases toward the
center. These results indicate that the dynamical importance of
rotation decreases toward the central core.

\subsubsection{Specific Angular Momentum}

The specific angular momentum $J/M$ is calculated using $\frac{J}{M}
= p\omega$R$^2$, $p=\frac{2}{5}$ for a uniform density sphere. The
decrease of $J/m$ from large to small scales has been found in
protostellar sources (Goodman et al. 1993; Ohashi et al. 1997; Liu
et al. 2010). However, most of these studies are derived using
single tracers and careful examination of a large sample of clouds
are required to characterize the evolution of $J/M$ within single
cores.

The calculated $J/M$ for three species are listed in Table
\ref{tab:velgrad}. $J/M$ of \hcccn\ range from 0.0072 to 1.68
\kmspc, with a median value of 0.16 \kmspc, corresponding to
5.1$\times$ 10$^{22}$ cm$^2~$s$^{-1}$; $J/M$ of HNC range from
0.0024 to 2.50 \kmspc, with a median value of 0.28 \kmspc,
corresponding to 8.5$\times$ 10$^{22}$ cm$^2$~s$^{-1}$; and $J/M$ of
\cch\ range from 0.014 to 1.8 \kmspc, with a median value of 0.24
\kmspc, corresponding to 7.5$\times$ 10$^{22}$ cm$^2$~s$^{-1}$.
Results presented here are an order of magnitude larger than
previous results (Goodman et al. 1993; Caselli et al. 2002). Since
most of sources in Goodman et al. (1993) are low-mass dense cores,
massive star-forming regions seem to have larger specific angular
momentum. Seven objects of our sample, including G121.3+0.66,
W3(OH), AFGL 490, S231, NGC 2264, W75OH and S140, have been observed
in N$_2$H$^+$ 1-0 transition by Pirogov et al. (2003). Velocity
gradients, ratio of rotation to gravitational energy and specific
angular momentum of these objects were derived with
N$_2$H$^+$ observations. Their results are not exactly the
same as results present here since N$_2$H$^+$ and \hcccn, HNC and
\cch\ trace different gas, and the beam size of their observations
are also different from our observations. However, they all have
same trends.

Figure \ref{jmcomp} shows comparison of J/M between three species.
Correlation coefficient between $J/M$ of HNC and \hcccn\ is 0.92.
Least-Squares fitting result is as follows:

$J/M_{HNC}$ = (1.54$\pm$0.17)$\times$ $J/M_{HC_3N}$.

Correlation coefficient between $J/M$ of \cch\ and \hcccn\ is 0.92.
Least-Squares fitting result is as follows:

$J/M_{C_2H}$ = (1.10$\pm$0.13)$\times$ $J/M_{HC_3N}$.

These results imply obvious decline of $J/M$ toward central regions
from $\sim$2 pc to $\sim$ 0.1 pc. Thus the angular momentum transfer
not only occurs at small scale ($<$ 0.1 pc) (Liu et al. 2010), but
also occurs at large scale. Magnetic braking and outflow are thought
to be possible mechanism for angular momentum transfer (e.g.,
Mouschovias \& Paleologou 1979; Fleck \& Clark 1981; Mestel \& Paris
1984; Rosolowsky et al. 2003). Since outflow is unlikely to be
responsible to angular momentum transfer in large scale, magnetic
braking, in which magnetic field lines anchoring molecular cloud to
the ambient interstellar matter provide the tension necessary to
slow down rotation, seems to be the most possible explanation for
the decrease of specific angular momentum. Results present here
allow us to put some constraints on the magnetic field strength.
Numerical simulation indicates that whether the evolution of a
collapsing core is regulated by the centrifugal forces or by
magnetic forces depends on the value of the rotation velocity and
magnetic field strength $\omega/B$ (Machida et al. 2005; Girart et
al. 2009). If the measured $\omega/B$ is larger than
$\frac{\omega_{st}}{B}_{crit} =
1.69\times10^{-7}\times(\frac{c_s}{0.19km/s})^{-1}$ year$^{-1} \mu$
G$^{-1}$, where $c_s$ is the sound speed and $\omega$ is the angular
velocity, the centrifugal forces dominate the dynamics. Otherwise
the magnetic forces regulate the dynamics. The median velocity
gradient derived using \hcccn\ is 1.43 \kmspc, corresponding to
$1.45\times10^{-6}$ year$^{-1}$. For $c_s$=0.19 km s$^{-1}$, we
found a critical magnetic field strength of 9$\mu$G. Thus, the
magnetic field should be larger than several $\mu$G to dominate the
dynamics of the collapse.

\subsection{Implication for Magnetic Field of Molecular Clouds}

Magnetic field is believed to play an important role in the
evolution of molecular clouds and star formation. Magnetic fields
appear to provide the only viable mechanism for transporting excess
angular momentum from collapsing cores, thus allowing continuous
accretion, and they may play a significant role in the physics of
bipolar outflows and jets that accompany protostar formation (e.g.,
Falgarone et al. 2008; Girart et al. 2009). The magnetic field of a
cloud is probably its most difficult property to obtain. For many
years observational studies of magnetic fields in molecular clouds
are restricted to probe the line-of-sight component via the Zeeman
effect of molecular lines (eg., HI, OH, CN), or the plane of sky
component via polarization measurements of the thermal radiation
emitted by magnetically aligned aspherical dust grains. All these
measurements require observations with both high stability and high
signal to noise ratio, and are therefore quite difficult and time
consuming (e.g., Bourke \& Goodman 2004; Bergin \& Tafalla 2007).

The idea that interstellar magnetic fields may be coupled to gas
motions was first discussed by Alfv$\acute{e}$n (1943) and Fermi
(1949). When the interstellar medium contains a well-coupled
magnetic field, disturbances should propagate at the
Alfv$\acute{e}$n velocity, $V_A = B/(4 \pi mn)^{1/2}$, where $B$
represents the mean magnetic field strength, $m$ represents the mean
molecular mass (2.33 amu), and $n$ represents the number density.
Myers \& Goodman (1988) develops a uniformly magnetic sphere model
to derive magnetic field strength and number density of the dense
cores from the FWHM linewidth. In this model, the supersonic
linewidth is attributed to magnetic motions, and the upper limit of
the magnetic field could be estimated via comparing the nonthermal
contribution of a cloud's velocity dispersion, $\sigma_{NT}$, to its
Alfv$\acute{e}$n velocity. The model cloud is a uniform
self-gravitating sphere of mass $M$, radius $R$, number density $n$,
temperature $T$, and mean magnetic field strength $B$. It is
surrounded by a medium of negligible kinetic and magnetic pressure.
It is in virial equilibrium between self-gravity, magnetic energy
and its internal random motions,
\begin{equation}
\label{eq:virequi1}
 3nkT + \frac{B^2}{4\pi} -
\frac{3GMmn}{5R} = 0,
\end{equation}
where $m$ is the mean molecular weight. Equation above
could also be written as:
\begin{equation}
\label{eq:virequi2} \frac{GM}{5R} = \sigma _{T}^2 + \sigma_{NT}^2.
\end{equation}
This relation models the virial equilibrium trend reported by Larson
(1981). The nonthermal velocity dispersion is taken to be the
difference between the observed velocity dispersion and the thermal
component:
\begin{equation}
\label{eq:calsig1} \sigma_{NT}^2  = \sigma_{obs}^2 - \sigma_{T}^2 =
\frac{\triangle V_{obs}^2}{8 ln2} - \frac{kT}{\mu_{obs}},
\end{equation}
where $\triangle V_{obs}$ is the FWHM linewidth of the observed
molecule, $\mu_{obs}$ is the molecule's mass, and $T$ is the kinetic
temperature of the gas. The nonthermal kinetic energy density is
assumed to be equal to the magnetic energy density, as given by
Spitzer (1978):
\begin{equation}
\sigma_{NT}^2  = \frac{2}{3}\frac{B^2}{8\pi mn} = \frac{v_A^2}{3}.
\end{equation}
Finally, magnetic field strength and number density could be
obtained by solving equations above:
\begin{equation}
B  = \frac{3 \sigma_{NT}}{R}[
\frac{5}{G}(\sigma_T^2+\sigma_{NT}^2)]^{1/2}.
\end{equation}
\begin{equation}
n = \frac{15}{4}\frac{1}{\pi GR^2m}(\sigma_T^2 + \sigma_{NT}^2).
\end{equation}
Zheng \& Huang (1996) derived the magnetic field strength of the
dense cores in the Orion B molecular cloud from the FWHM linewidth
based on the model of a uniformly magnetic sphere. Average magnetic
field strength of 110 $\mu$G is obtained, which is agreed closely to
those derived from observations of Zeeman splitting of HI and OH.
They also obtained an average number density of 8$\times 10^4$
cm$^{-3}$, which agreed with those derived using observations of
NH$_3$. They suggested that this method could be applicable to the cores of
R$>$0.2 pc.

Optically thick lines should be avoided to obtain the true velocity
dispersion from the linewidth, thus only \hcccn\ observations are
suitable to derive the magnetic field strength and number density.
Rotation could also contribute to the observed linewidth for
rotating clouds. For a cloud at a distance of $D$, a 55\arcsec\ beam
would give $\sigma_{rot} = D \times \frac{55}{60\times60}\times
\frac{\pi}{180}\times Grad$.
%Figure \ref{rotlinebroad} shows
%histogram of line broadening due to rotation calculated using
%velocity gradients of \hcccn.
The rotation line broadening is calculated to range from 0.14 to
3.43 \kms, with a median value of 0.76 \kms. The rotational line
broadening and rotational energy should be taken into account for
clouds in which velocity gradients are detected. Thus Equation
(\ref{eq:virequi1}), (\ref{eq:virequi2}) and (\ref{eq:calsig1})
should be written as:
\begin{equation}
\label{eq:virequi3}
 3nkT + \frac{B^2}{4\pi} -
\frac{3GMmn}{5R} + 2 \times \frac{\frac{1}{2}I \omega^2}{V} = 3nkT +
\frac{B^2}{4\pi} - \frac{3GMmn}{5R} + \frac{2mnR^2 \omega^2}{5} = 0,
\end{equation}
\begin{equation}
\label{eq:virequi4} \frac{GM}{5R}- \frac{2R^2\omega^2}{15} = \sigma
_{T}^2 + \sigma_{Turb}^2,
\end{equation}
\begin{equation}
\label{eq:calsig2} \sigma_{Turb}^2  = \sigma_{obs}^2 - \sigma_{T}^2
- \sigma_{rot}^2 = \frac{\triangle V_{obs}^2}{8 ln2} -
\frac{kT}{\mu_{obs}} - \sigma_{rot}^2,
\end{equation}
where $\sigma_{Turb}$ represents turbulent line broadening.

Magnetic field strength and number density could be obtained by
solving Equation (\ref{eq:virequi3}), (\ref{eq:virequi4}) and
(\ref{eq:calsig2}):
\begin{equation}
B  = \frac{3 \sigma_{Turb}}{R}[
\frac{5}{G}(\sigma_T^2+\sigma_{Turb}^2 + \frac{2}{15}R^2\omega^2
)]^{1/2}.
\end{equation}
\begin{equation}
n = \frac{15}{4}\frac{1}{\pi GR^2m}(\sigma_T^2 + \sigma_{Turb}^2 +
\frac{2}{15}R^2\omega^2).
\end{equation}

Single-dish ammonia observations of water maser sources reveal
molecular gas temperature of 10-20 K (Codella et al. 1997). Zheng \&
Huang (1996) found that both the magnetic field strength and number
density are insensitive to temperature. For a dense core with R =
0.25 pc, $\delta$ v = 1.5 \kms, the difference is only 8\% for
magnetic field strength derived using 20K and 10K, while the
difference is only 5\% for number densities derived using 20K and
10K. Thus a temperature of 20 K was adopted here. Gradients derived
using \hcccn\ are adopted if available, otherwise Gradients derived
using HNC will be used. Table \ref{tab:magnetic} presents the
rotation line broadening $\sigma_{rot}$ in column (2), turbulence
broadening $\sigma_{Turb}$ in column (3), derived magnetic field
strength in column (4), and number density in column (5). The
magnetic field strength ranges from 3 $\mu$G to 88 $\mu$G, with a
median value of 13 $\mu$G. Falgarone et al. (2008) carried out CN
Zeeman measurements toward 14 star-forming regions and obtained a
median value of 560 $\mu$G. It is no wonder that results derived
here are much smaller, since it is averaged over the FWHM contours.
The major uncertainty comes from the volume filling factor, which
makes it difficult to unambiguously determine source radius R. Thus
B and n are probably under estimate of actual values. The strongest
magnetic field comes from W51M, which has been indicated in previous
polarization observations of water maser and thermal dust emission
of this region (Lepp$\ddot{a}$nen et al. 1998; Lai et al. 2001).
This source would be an excellent candidate for future Zeeman
measurements of CN or \cch. The derived number densities range from
2.4$\times$10$^4$ to 1.2$\times$10$^6$ cm$^{-3}$, with a median
value of 1$\times$10$^5$ cm$^{-3}$, which agrees closely with
critical density of \hcccn\ 10-9 (Chung et al. 1991).

The mass-to-flux ratio $M/\phi_B$ is a quantitative measure
of whether the molecular cloud is magnetically supercritical (i.e.,
the magnetic field cannot resist the gravitational collapse) or
subcritical (i.e., the magnetic field could resist the gravitational
collapse) (e.g., Crutcher et al. 1996, 1999). The critical value of
ratio, $M/\phi_B = c_{\phi}/\sqrt{G}$, with $c_{\phi}=0.126$ for
disks with only thermal support along field lines, was derived by
Mouschovias \& Spitzer (1976). Tomisaka et al. (1988) found a
$c_{\phi}$ of 0.12 from extensive numerical calculations. The
mass-to-flux ratio $M/\phi_B$, in units of the critical value of
$M/\phi_B$, is $10^{-20}\times N(H_2)/|B|$ cm$^2 \mu$ G, where
$N(H_2)$ is the average column density and could be written as
$(4/3)Rn$. Due to difficulties in magnetic field strength
measurements, $M/\phi_{B}$ is still uncertain. Many cores have only
lower limits. The uniformly magnetic sphere model appears to provide
a reasonable estimate of $M/\phi_{B}$ as it gives the average
magnetic field strength and number density. $M/\phi_B$ is calculated
to be 72 times critical with the median magnetic field strength of
13 $\mu$G, median number density of 1$\times$10$^5$ cm$^{-3}$ and
median radius of 0.24 pc. This means that the molecular clouds are
strongly supercritical and magnetic fields alone are insufficient to
support clouds against gravity. V$\acute{a}$zquez-Semadeni et al.
(2011)  have recently found that clouds formed by supercritical
inflow streams proceed directly to collapse, while clouds formed by
subcritical streams first contract and then re-expand, oscillating
on the scale of tens of Myr. Their simulations with initial magnetic
field strength of several $\mu$G also show that only supercritical lead
to reasonable star forming rates. This result is not altered by the
inclusion of ambipolar diffusion. Thus we could see that both
magnetic field strength and number density derived using the
uniformly magnetic sphere model are not in contradiction with
observations and numerical simulations, indicating possible magnetic
field origin of turbulence, as is suggested by Zeeman measurements
of magnetic field strengths (Crutcher et al. 1999; Falgarone et al.
2008). Turbulence is another important support against gravity
besides magnetic field, understanding of the origin of turbulence is
critical to our understanding of massive star-formation. More
observations are needed to further investigate the relationship
between magnetic field and the origin of turbulence.

%\subsubsection{Relationship between $J/M$ and $\sigma_{Turb}$}

%The origin of rotation of molecular clouds remains uncertain, and the
%random distribution of angular momentum vectors for stars and
%molecular clouds deny the galactic rotation origin of angular
%momentum (e.g., Philips 1999; Imara \& Blitz 2011). Fleck \& Clark
%(1981) have suggested that rotation in clouds may derive from the
%vorticity associated with interstellar turbulence. In this case
%close correlation between specific angular momentum $J/M$ and
%$\sigma_{Turb}$ would be expected. Figure \ref{jmv} shows $J/M$
%versus $\sigma_{Turb}$. Correlation coefficients between specific
%angular momentum and $\sigma_{Turb}$ are 0.56, 0.39 and 0.46 for
%\hcccn, HNC and \cch, respectively. Our results imply possible but
%weak correlation between origin of angular momentum and turbulence.

\subsection{Galactic Trends}

The FWHM size, linewidth, velocity gradient, $\beta$ and $J/M$ are
plotted versus Galactic radius in Figure \ref{galactrend}. There is
little evidence for any trend in velocity gradient (r=0.13, 0.07 and
0.26 for \hcccn, HNC and \cch), $\beta$ (r=-0.13, -0.24 and 0.09 for
\hcccn, HNC and \cch). There may be weak anti-correlations of $J/M$
(r=-0.30, -0.26 and -0.29 for \hcccn, HNC and \cch). Possible
anti-correlations are found for FWHM size (r=-0.48, -0.51 and -0.57
for \hcccn, HNC and \cch), FWHM linewidth of \hcccn\ (r=-0.50), as
well as $\sigma_{Turb}$ (r=-0.53). We just plot FWHM
linewidth and $\sigma_{Turb}$ for \hcccn\ since only \hcccn\
emission is optically thin among three molecular line emissions,
thus only line profile of \hcccn\ avoids being affected by
self-absorption. The large spike in FWHM linewidth near $D_g=6.4
kpc$ is due to observations of W51M. These results suggest decline
of turbulence and dense gas abundance toward outer Galaxy. Our
results seem to be inconsistent with observations of CS, in which
few trends with galactocentric distance was found (e.g., Zinchenko
1995, 1998; Shirley et al. 2003). Observations of larger samples are
needed to support these results.

\section{SUMMARY}
\label{summary}

We conducted mapping observations toward 27 massive star-forming
regions associated with water masers with the PMO 14m telescope in
the \hcccn\ 10-9, HNC 1-0 and \cch\ 1-0 lines. Maps of \hcccn\
10-9, HNC 1-0 and \cch\ 1-0 main lines are presented in this paper. The
radio continuum emission is searched to characterize the
evolutionary stage of molecular clouds. The optical depth of \cch\ 1-0 main line
is derived using its hyperfine structure. The opacity maps of \cch\ 1-0 main line
are also presented. The large-scale kinematic structure of molecular
clouds is investigated with three species. Velocity gradients,
ratios of gravitational energy to gravitational energy, specific
angular momentums are calculated. The uniform magnetic field sphere
model is inspected and average magnetic field strength of molecular
clouds is derived. Our main results are summarized as follows.

The integrated intensities of \hcccn, HNC and \cch\ correlate well
with each other, implying abundance correlation of three species in
massive star-forming regions. Comparison of deconvolved FWHM sizes
indicate that \hcccn\ traces dense gas, while \cch\ trace extended gas.

The derived optical depth of \cch\ main line range from 0.1 to 6.5.
Obvious central depletion is observed in \cch\ opacity maps, which
is consistent with chemical model of \cch. Abundance decline of
\cch\ as molecular clouds evolve is noticed, implying significant
destruction of \cch\ by UV photons. These results suggest that \cch\
is a potential ``chemical clock" for molecular clouds.

Velocity gradients are detected in nearly all the sources, implying
that rotation is a common feature of molecular clouds. The direction
of velocity gradients are nearly the same for three species, and the
magnitude of velocity gradients increase toward the center, implying
differential rotation rather than rigid-body rotation.

Both the ratio of rotational energy and specific angular momentum
are found to decrease toward the center for most clouds, implying effective
angular momentum transfer in large scale. The angular momentum transfer
is possiblly caused by magnetic braking. Magnetic field
strength of several $\mu$G was obtained to regulate rotation and dominate the dynamics of
molecular clouds, which is consistent with Zeeman measurements of
magnetic field in molecular clouds.

The magnetic field strength and density of molecular clouds are
derived from the observed radius and FWHM linewidth based on the
uniformly magnetic field sphere model. An average magnetic field strength
of 13 $\mu$G and average density of 1 $\times$10$^5$ cm$^{-3}$ are
obtained. The dense cores are calculated to be supercritical
with derived parameters, which is consistent with numerical simulations.
These results suggest a magnetic field origin of turbulence.

No trends in velocity gradient, or ratio of rotational kinetic
energy to gravitational energy with galactocentric radius are
apparent. Weak decrease in FWHM sizes, linewidth, and specific
angular momentum with galactocentric distance are observed. These
results suggest decrease of turbulence and dense gas abundance toward
outer Galaxy.

\acknowledgments We thank the anonymous referee for constructive comments that greatly improved the manuscript. This work was supported by the Natural Science Foundation of
China under grants of 10803002, 10833006, 10878010 and 11103006. We would like to thank Key Laboratory of Radio Astronomy, Chinese Academy of Sciences. We are very grateful to the staff of Qinghai
Station of Purple Mountain Observatory for their assistance with the
observations and data reductions. JL thank Jingjing Li, Bing Jiang,
Yuanwei Wu for help with data reduction, and thank Shaobo Zhang for
providing vfit program.

%\appendix
%
%\section{Radio Continuum Emission}

\begin{table}
\scriptsize
    \begin{center}
%      \begin{minipage}{105mm}
      \caption{Source List}\label{tab:source}
      \begin{tabular}{lcccccc}
      \\
    \hline
    \hline
Source Name      & RA(J2000)     & DEC(J2000)       & D   & Ref. & D$_g$    \\
                 &               &                  & (kpc) &      & (kpc)    \\
            \hline
 G8.67-0.36     &   18:06:18.87 &-21:37:37.9       &  4.5 &  1  & 4.1   \\
 G10.6-0.4      &   18:10:28.70 &-19:55:48.7       &  6.5 &  2  & 2.4  \\
 G12.42+0.50    &   18:10:51.80 &-17:55:55.9       &  2.1 &  3   & 6.5  \\
 W33cont        &   18:14:13.67 &-17:55:25.2       &  4.1 &  2  & 4.6  \\
 W33A           &   18:14:39.30 &-17:52:11.3       &  4.5 &  4  & 4.2  \\
 G14.33-0.64    &   18:18:54.71 &-16:47:49.7       &  2.6 &  1  & 6.0 \\
 W42            &   18:36:12.46 &-07:12:10.1       &  9.1 &  5   & 3.8   \\
 W44            &   18:53:18.50 &  01:14:56.7      &  3.7 &  2   & 5.8   \\
 S76E           &   18:56:10.43 &  07:53:14.1      &  2.1 &  6    &  7.0   \\
 G35.20-0.74    &   18:58:12.73 &  01:40:36.5      &  1.99 & 7   & 6.9  \\
 W51M           &   19:23:43.86 &  14:30:29.4      &  5.41 & 8    & 6.3 \\
 G59.78+0.06    &   19:43:11.55 &  23:43:54.0      &  2.2  & 6      &  7.6  \\
 S87            &   19:46:20.45 &  24:35:34.4      &  1.9  & 9    &  7.6  \\
 ON1            &   20:10:09.14 &  31:31:37.4      &  2.57 & 10   & 7.9  \\
 ON2S           &   20:21:41.02 &  37:25:29.5      &  5.5  & 6     & 8.9  \\
 W75N           &   20:38:36.93 &  42:37:37.5      &  3.0  & 11     & 8.6  \\
 DR21S          &   20:39:00.80 &  42:19:29.8      &  3.0  & 11    & 8.6  \\
 W75OH          &   20:39:01.01 &  42:22:49.9      &  3.0  & 11     & 8.6   \\
 Cep A           &   22:56:18.14 &  62:01:46.3      &  0.55 & 7     &  8.7  \\
 G121.30+0.66   &  00:36:47.51  & 63:29:02.1       &   1.2 &  6      &  9.2  \\
 W3(OH)         &  02:27:04.69  & 61:52:25.5       &   3.42 & 7    & 11.1  \\
 S231           &  05:39:12.91  & 35:45:54.1       &   2.3  & 12     & 10.8  \\
 S235           &  05:40:53.32  & 35:41:48.8       &   1.6  & 12      & 10.1  \\
 S255           &  06:12:53.72  & 17:59:22.0       &   1.59 & 10   & 10.1  \\
\hline
 S140       &  22:19:19.140 & 63:18:50.30   &   0.76  & 13    &  8.7    \\
 AFGL490        &  03:27:38.80  & 58:47:00.0      &  1& 14     &  9.3 \\
 NGC2264   &  06:41:09.80  &   09:29:32.0     &  0.950 & 15    &  9.4  \\
             \hline
      \end{tabular}
%    \end{minipage}
  \end{center}
Notes: -Columns are (1) Source Name, (2) Right ascension, (3) declination, (4) distance D,
(5) references for D, (6) galactocentric distance D$_g$. \\
References: (1) Val'tts et al. 2000; (2) Solomon et al. 1987; (3)
Zinchenko et al. 1994; (4) Braz \& Epchtein 1983; (5) Downes et al.
1980; (6) Plume et al. 1992; (7) Reid et al. 2009; (8) Sato et al. 2010; (9) Brand \&
Blitz 1993; (10) Rygl et al. 2010; (11) Genzel \& Downes 1977; (12)
Blitz et al. 1982; (13) Hirota et al. 2008; (14) Snell et al. 1984;
(15) Baxter et al. (2009).
\end{table}

\begin{table}
    \begin{center}
\scriptsize
      \caption{Observing Information for VLA Archive Data}\label{tab:vlaobs}
      \begin{tabular}{lccccccccc}
      \\
    \hline
    \hline
Source    & Observing Date  &  Configuration &  Frequency &  Synthesized Beam          &  Project       \\
          &                 &                &  (GHz)     &  (\arcsec $\times$ \arcsec, $^{\circ}$)     &       \\
            \hline
G8.67-0.36    &  Apr 27, 1986     &  VLA:A:1      &  4.8   &  0.7$\times$0.6, -14       &  AW158         \\
 G10.6-0.4    & Jan 2, 2005       & VLA:A:1       &  8.4   &  0.8$\times$0.7, -6.7      &   AS824          \\
 G12.42+0.50  &  May 22, 2007      &  VLA:D:1      &  8.4   &  2.3$\times$2.0, 5.6      &    TLS30        \\
 W33cont      &  Aug 13, 1992     &  VLA:D:1      &  8.4   &  14$\times$8, 20            &   AB642        \\
 W33A         & Apr 18, 1990      &  VLA:A:1      &  8.4   &  0.2$\times$0.1, -26         &  AH391         \\
 G14.33-0.64  &   July 30, 1989  &  VLA:BC:1     &  4.8   &  8.6$\times$4.2, -20       &   AH361         \\
 W42           &  Jun 23, 1989     &   VLA:BC:1   &   4.8  &  5.2$\times$3.8, -8        &   AB544        \\
 W44            &  Dec 16, 1997    &  VLA:D:1     &   8.4  &  12$\times$9, -23          &   AD406       \\
 S76E           &   Nov 21, 1997   &   VLA:D:1    &  8.4   &  14.4$\times$8.7, 50       &   AR390       \\
 G35.20-0.74   &  Mar 31, 1991     &   VLA:D:1    &  8.4   &  9.7$\times$7.9, -0.3      &   AB601         \\
 W51M           &  Mar 16, 1993    &  VLA:B:1     &  8.4   &  1.2$\times$1, 70          &   AM374        \\
 G59.78+0.06   &  Mar 6, 2005     & VLA:B:1     &  8.4   &   0.9$\times$0.7, 72        &   AS831                    \\
 S87           &  Dec 13, 1998    &   VLA:C:1     &  8.4   & 2.3$\times$2.1, 6          &   AK477        \\
 ON1           & Mar 25, 2005     &  VLA:B:1      &  8.4   &   0.7$\times$0.6, -20       &   AS830         \\
 ON2N         & Jan 12, 1993       &  VLA:A:1     &   8.4  &   0.6$\times$0.4, -84       &   AR283         \\
 W75N         & Nov 24, 1992       &  VLA:A:1     &  8.4   &  0.3$\times$0.3, -88         &   AT141        \\
 DR21S         &  Aug 20, 1996     &   VLA:D:1    &  8.4   &  8.8$\times$7.2, 89        &   AW443        \\
W75OH        &   Sep 16, 2004      &  VLA:A:1     &  8.4  &   0.5$\times$0.5, 80        &    AP480        \\
 Cep A         &  May 4, 1991      &  VLA:D:1     &  8.4   &    8.8$\times$6.6, -22     &  AH429          \\
 G121.30+0.66   &   Oct 31, 1993   &   VLA:D:1    &   8.4  &  10.4$\times $8.2, 62.5     &   AH497      \\
 W3(OH)         &  Nov 15, 1996     &  VLA:A:1    &  8.4   &  0.5$\times$0.4, -57         &   AR363       \\
S231           &   Jun 21, 2003   &   VLA:A:1    &   8.4  &   0.4$\times$0.3, 76         &    AR513                        \\
 S235           &  Feb 27, 2004     &  VLA:BC:1   &  4.8   &  4.7$\times$4.2, -8.5        &   AM786        \\
 S255           &  Jun 15, 2003     &  VLA:A:1    &  8.4   &  1.6$\times$1.3, 71          &    AH819      \\
 S140           &  Nov 24, 1992     &  VLA:A:1    &  8.4   &  0.6$\times$0.4, 87         &  AT141            \\
 AFGL490         &  Jun 3, 1985     &  VLA:B:1    & 15    &  0.65$\times$0.5, 90        &  AC110              \\
 NGC2264     & Mar 4, 2002     &   VLA:A:1   &  8.4  & 0.30$\times$0.29, -58      &  AR465                    \\
             \hline
      \end{tabular}
%    \end{minipage}
  \end{center}
Notes: -Columns are (1) Source name, (2) observing date, (3) observing configuration, (4) observing frequency, (5) synthesized beam, (6) project code.
\end{table}

\begin{deluxetable}{lrrrrrrrrrrrr}
\tabletypesize{\scriptsize} \rotate
\tablecaption{\label{tab:dlhresult}Observational results of \hcccn
10-9, HNC 1-0 and \cch\ 1-0 (J = 3/2$\rightarrow$1/2, F =
2$\rightarrow$1) Transitions at (0, 0).} \tablewidth{0pt}
\tablehead{ \multicolumn{1}{l}{Source} &\multicolumn{4}{c}{\hcccn\
10-9}&\multicolumn{4}{c}
{HNC 1-0}&\multicolumn{4}{c}{\cch\ 1-0}\\
\colhead{}             & \colhead{T$_{A}^{*}$}  &
\colhead{$\int$T$_{A}^{*}$d$\nu$}  & \colhead{V$_{LSR}$}    &
\colhead{FWHM}         & \colhead{T$_{A}^{*}$}  &
\colhead{$\int$T$_{A}^{*}$d$\nu$}  & \colhead{V$_{LSR}$}    &
\colhead{FWHM}         & \colhead{T$_{A}^{*}$}  &
\colhead{$\int$T$_{A}^{*}$d$\nu$}  & \colhead{V$_{LSR}$}    &
\colhead{FWHM}         \\
\colhead{}             & \colhead{(K)}          &
\colhead{(Kkms$^{-1}$)}  & \colhead{(kms$^{-1}$)}   &
\colhead{(kms$^{-1}$)}   & \colhead{(K)}          &
\colhead{(Kkms$^{-1}$)}  & \colhead{(kms$^{-1}$)}   &
\colhead{(kms$^{-1}$)}   & \colhead{(K)}          &
\colhead{(Kkms$^{-1}$)}  & \colhead{(kms$^{-1}$)}   &
\colhead{(kms$^{-1}$)} } \startdata
  G8.67-0.36     &  1.2(.1)&   5.7 (.2)     &    34.97(.06) &   4.47(.16) &   2.0(.2)      &     8.2(.3)    &      33.69(.06) &    3.93(.18)  &   0.7(.2) &   3.6(.3)&   35.01(.18) &   4.94(.56)        \\
 G10.6-0.4      &  1.3(.2)&   8.9 (.3)     &   -2.93(.11) &   6.23(.27) &   4.0(.2)      &    29.3(.3)    &      -3.11(.04) &    6.94(.10)  &     1.1(.2) &   6.5(.3)&   -2.74(.14) &   5.52(.36)    \\
 G12.42+0.50    &  0.7(.1)&   2.1 (.1)     &    17.76(.08) &   2.81(.18) &   2.3(.1)      &     8.8(.1)    &      17.66(.03) &    3.58(.07)  &   0.8(.2)  &  2.2(.2) &  18.06(.15) &   2.67(.32)       \\
 W33cont        &  2.4(.1)&   9.8 (.2)     &    34.99(.03) &   3.81(.08) &   3.7(.1)      &    23.0(.2)    &      34.86(.03) &    5.81(.08)  &    2.1(.3) &   9.5(.4)&   34.91(.08) &   4.35(.22)       \\
 W33A           &  0.9(.1)&   3.5 (.2)     &    37.07(.08) &   3.67(.20) &   1.2(.1)      &     7.6(.3)    &      36.35(.08) &    5.85(.28)  &    0.6(.2) &   3.9(.3)&   36.65(.19) &   5.72(.57)     \\
 G14.33-0.64    &  1.5(.1)&   3.9 (.1)     &    22.11(.03) &   2.53(.08) &   1.8(.1)      &     6.0(.2)    &      23.82(.04) &    3.13(.11)  &    0.6(.2) &   2.5(.2)&   22.36(.18) &   3.93(.44)     \\
 W42            &  0.8(.1)&   3.7 (.1)     &    110.2(.07) &   4.12(.20) &   1.4(.1)      &     7.2(.2)    &      111.1(.05) &    4.76(.16)  &    0.6(.1) &   2.7(.2)&   110.7(.15) &   4.04(.42)       \\
 W44            &  1.5(.1)&   8.3 (.1)     &    58.42(.04) &   5.20(.10) &   4.4(.1)      &    16.7(.2)    &      56.53(.02) &    3.58(.04)  &    1.5(.2) &   6.1(.2)&   57.53(.07) &   3.84(.19)      \\
 S76E           &  1.5(.1)&   4.5 (.1)     &    32.65(.03) &   2.88(.07) &   3.0(.1)      &     9.5(.1)    &      33.12(.02) &    2.97(.04)  &    1.3(.2) &   3.4(.2)&   33.01(.06) &   2.53(.15)     \\
 G35.20-0.74    &  1.1(.1)&   4.6 (.2)     &    33.92(.07) &   3.87(.15) &   2.0(.1)      &    12.3(.2)    &      33.97(.05) &    5.85(.10)  &    0.8(.2) &   3.5(.2)&   33.91(.13) &   4.02(.30)     \\
 W51M           &  1.1(.1)&  11.7 (.2)     &    57.22(.07) &   9.84(.15) &   3.2(.1)      &    34.0(.2)    &      56.90(.03) &    9.96(.06)  &    1.4(.3) &   12.(.5)&   56.88(.17) &   8.68(.36)      \\
 G59.78+0.06    &  0.5(.1)&   1.2 (.1)     &    22.43(.09) &   2.08(.22) &   2.0(.1)      &     5.6(.1)    &      22.64(.03) &    2.58(.07)  &   0.4(.2)  & 1.8(.4) &   22.40(.3) &  3.76(1.27)     \\
 S87            &  0.7(.1)&   1.7 (.1)     &    23.39(.09) &   2.32(.26) &   2.2(.1)      &     9.4(.1)    &      22.82(.03) &    4.03(.07)  &    1.1(.2) &   4.2(.2)&   22.64(.10) &   3.6(.22)       \\
 ON1            &  0.5(.1)&   2.0 (.1)     &    11.46(.11) &   3.65(.27) &   1.5(.1)      &     8.4(.2)    &      12.22(.05) &    5.36(.13)  &    0.5(.2) &   3.0(.2)&   11.23(.21) &   5.33(.52)       \\
 ON2S           &  0.6(.1)&   2.0 (.2)     &   -1.50(.14) &   3.43(.31) &   1.2(.1)      &     6.5(.2)    &      -1.30(.08) &    5.03(.16)  &    0.5(.2) &  2.4(.2) &  1.62 (.20) &  4.11(.52)    \\
 W75N           &  1.0(.1)&   3.4 (.1)     &    9.46(.05) &   3.18(.13) &   2.6(.1)      &    12.5(.2)    &       9.53(.03) &    4.56(.07)  &     0.9(.2) &   3.5(.2)&    9.30(.11) &   3.45(.30)      \\
 DR21S          &  2.1(.1)&   5.7 (.1)     &   -2.12(.03) &   2.55(.07) &   3.1(.1)      &    13.9(.2)    &      -2.32(.03) &    4.19(.07)  &     2.0(.2) &   6.0(.3)&   -2.28(.59) &   2.90(.15)      \\
 W75OH          &  1.6(.1)&   7.0 (.1)     &   -3.23(.04) &   4.16(.10) &   2.5(.1)      &    16.5(.2)    &      -3.20(.03) &    6.32(.07)  &     1.6(.2) &   8.0(.3)&   -3.14(.10) &   4.68(.23)      \\
 Cep A           &  0.9(.1)&   3.1 (.1)     &   -10.55(.07) &   3.35(.15) &   2.0(.1)      &    10.5(.2)    &     -10.84(.04) &    4.84(.06)  &    1.0(.2) &   3.6(.2)&  -10.62(.10) &   3.56(.22)        \\
 G121.30+0.66   &  0.7(.1)&   1.7 (.1)     &   -17.58(.07) &   2.29(.16) &   2.5(.1)      &     7.3(.1)    &     -17.87(.03) &    2.76(.07)  &  0.7(.1) & 2.1(.1)&  -17.48(.09) &  2.65(.21)       \\
 W3(OH)         &  0.6(.1)&   2.3 (.1)     &   -47.28(.08) &   3.62(.19) &   2.0(.1)      &     8.8(.1)    &     -47.58(.03) &    4.17(.07)  &    1.0(.2) &   4.1(.2)&  -47.36(.11) &   3.88(.28)       \\
 S231           &  0.7(.1)&   2.0 (.1)     &   -16.56(.07) &   2.72(.16) &   2.0(.1)      &     8.5(.1)    &     -16.58(.03) &    4.07(.06)  &    0.7(.1) &   3.0(.2)&  -16.56(.12) &   3.75(.25)      \\
 S235           &  0.6(.1)&   1.4 (.1)     &   -16.79(.06) &   2.21(.13) &   2.3(.1)      &     6.2(.1)    &     -16.91(.02) &    2.53(.05)  &    1.1(.2) &   3.3(.2)&  -16.81(.07) &   2.80(.17)     \\
 S255           &  1.0(.1)&   2.3 (.1)     &    7.15(.03) &   2.16(.08) &   2.6(.1)      &     7.6(.1)    &       7.10(.01) &    2.77(.03)  &     1.7(.2) &   4.2(.2)&    7.25(.05) &   2.33(.12)      \\
 S140      &  2.0(.1)&   4.8 (.1)     &   -6.83(.01) &   2.32(.03) &   5.1(.1)      &    15.1(.1)    &      -6.77(.06) &    2.79(.01)  &     1.9(.1) &   5.4(.1)&   -6.90(.01) &   2.69(.03)        \\
 AFGL490        &  0.4(.1)&   0.8 (.1)     &   -13.32(.12) &   1.82(.28) &   1.4(.1)      &     3.0(.1)    &     -13.20(.04) &    2.07(.09)  &     0.8(.2) & 1.0(.1)&   -13.23(.08) &  1.07(.20)      \\
 NGC2264    & 1.6(.1) &  3.5 (.1)      &   8.143(.03)  &  2.11(.07)  &  4.0(.1)        &   14.5(.1)     &     7.90(.01)  &    3.39(.03)  &   1.6(.2)  &  4.7(.3) &   8.10(.07)  &  2.86(.18)    \\

      \enddata \\
Note: -Columns are (1) Source name, (2) intensities of \hcccn, (3)
integrated intensities of \hcccn, (4) centroid
velocities of \hcccn, (5) FWHM linewidth of \hcccn, (6) intensities of HNC, (7) integrated
intensities of HNC, (8) centroid velocities of HNC, (9) FWHM linewidth of HNC, (10)
intensities of \cch, (11) integrated intensities of
\cch, (12) centroid velocities of \cch, (13)
FWHM linewidth of \cch.
\end{deluxetable}

\begin{table}
\scriptsize
    \begin{center}
%      \begin{minipage}{105mm}
      \caption{The Centroid and FWHM Sizes of Clouds for \hcccn\ 10-9, HNC 1-0 and \cch\ 1-0 (J = 3/2$\rightarrow$1/2, F =
2$\rightarrow$1) Transitions.}\label{tab:size}
      \begin{tabular}{lccccccccc}
             \hline
    \hline
Source      & Centroid  & R$_{(HC_3N)}$ & $\theta_{(HC_3N)}$ & Centroid  &  R$_{(HNC)}$ & $\theta_{(HNC)}$  & Centroid  & R$_{(C_2H)}$ & $\theta_{(C_2H)}$   \\
                 &   (\arcsec)      &   (pc)   &     (\arcsec)   &  (\arcsec)  &    (pc)    &    (\arcsec)  &  (\arcsec)   &  (pc)   &  (\arcsec)      \\
            \hline
 G8.67-0.36     & (0,0) &  0.59      &    54.4     & (+30,0)    &  $\geq$1.00 &  $\geq$91.8 & (0,0) &    1.05    &   96.1     \\
 G10.6-0.4      & (0,0) &  0.43       &    27.5    & (0,0)    &   0.56      &   35.5       & (0,0) &    1.18      &   75.0      \\
 G12.42+0.50    & (0,0) &  0.18       &    35.0    & (0,0)    &    0.24     &  47.3        & (0,0) & $\geq$0.33    & $\geq$65.7        \\
 W33cont        & (0,0) &  0.62      &    62.7     & (+30,0)    & $\geq$0.99  & $\geq$99.2 & (0,0) &    1.07    &   107.4     \\
 W33A           & (0,0) &  0.55     &    50.2      & (0,0)    &  $\geq$1.00 & $\geq$92.0   & (0,0) &    0.93   &   85.4      \\
 G14.33-0.64    & (0,0) &  0.24      &    38.4     & (0,0)    & $\geq$0.52  & $\geq$82.3   & (0,-30) &    0.48    &   76.8  \\
 W42            & (0,0) &  0.79      &    35.9     & (0,0)    &   1.99      &   90.1       & (-30,+30) &    2.54    &   114.9      \\
 W44            & (0,0) &  0.54     &    59.8      & (0,0)    &   0.71      &   79.1       & (0,0) &    1.20    &   133.3      \\
 S76E           & (0,0) &  0.22      &    43.7     & (0,-30)    &   0.38      &   74.9       & (0,0) &    0.50    &   98.5        \\
 G35.20-0.74    & (0,0) &  0.22     &    45.3      & (0,0)    &   0.42      &   87.7       & (0,-30) &    0.45   &   92.1      \\
 W51M           & (0,0) &  1.33       &    101.4   & (0,0)    &   2.08      &   158.5      & (0,0) &    2.17     &   166.1    \\
 G59.78+0.06    & (0,0) &  0.15     &    27.5      & (-30,0)    &  $\geq$0.34 & $\geq$64.5   & (0,+30) & $\geq$0.37    & $\geq$70.3    \\
 S87            & (+30,+30) &  0.26      &    56.6     & (+30,+30)    &   0.39      &   83.6  & (+30,+30) &    0.49    &   105.9        \\
 ON1            & (0,0) &  0.37      &    58.9     & (0,0)    &   0.48      &   76.5       & (0,0) &    0.17    &   27.5     \\
 ON2S           & (-30,-30) &  0.65      &    89.8  & (-30,-30)    &   0.68      &  93.0   & (0,-30) & $\geq$0.67  & $\geq$92.3     \\
 W75N           & (0,0) &  0.43      &     59.3    & (0,0)    &    0.55     &    75.7      & (0,0) &     0.60  &    82.0  \\
 DR21S          & (0,0) &  0.20    &     27.5      & (0,0)    &    0.44     &    60.6      & (0,-30) &     1.02   &    140.1     \\
 W75OH          & (0,-30) &  0.49    &     67.7    & (0,0)    &    0.80     &    110.6     & (0,0) &     1.08    &    148.8       \\
 Cep A          & (+30,0) &  0.11     &     84.9     & (0,0)    &    0.21     &    156.1   & (0,-30) &     0.16   &    120.7     \\
 G121.30+0.66   & (0,0) &  0.12     &    41.5      & (0,0)    &    0.22     &    75.6      & (0,0) &  $\geq$0.18   &  $\geq$61.4     \\
 W3(OH)         & (0,-30) &  0.39    &     47.2     & (0,-30)    &    0.81     &    97.7   & (0,0) &     0.95  &    115       \\
 S231           & (0,0) &   0.23     &     41.5    & (0,0)    &    0.40     &    71.2      & (0,0) &     0.27   &    48.1  \\
 S235           & (0,0) &   0.12    &     29.8     & (0,0)    &  $\geq$0.24 &  $\geq$62.1  & (0,0) &     0.22  &    55.7        \\
 S255           & (0,+60) &   0.22     &     57.5    & (0,+60)    &    0.36     &    94.0  & (0,30) &     0.42   &    109.4     \\
 S140       & (0,0) &   0.16    &     88.7     & (0,0)    &  0.21       & 116.2        & (30,0) &     0.20   &    106.3       \\
 AFGL490        & (0,0) &   0.05   &     20.3      & (0,0)    & $\geq$0.21  & $\geq$87.1   & (0,0) &   $\geq$0.11   &  $\geq$46.5       \\
 NGC2264   & (0,0) &    0.17     &  75.3      & (0,0)    &  0.29       &  130.4       & (+30,0) &  0.21    &   95.3           \\

             \hline
      \end{tabular}
%    \end{minipage}
  \end{center}
  Note: -Columns are (1) Source name, (2) centroids of \hcccn\ clouds, (3)
FWHM linear radius of \hcccn\ clouds, (4) FWHM angular radius of
\hcccn\ clouds, (5) centroids of HNC clouds, (6)
FWHM linear radius of HNC clouds, (7) FWHM
angular radius of HNC clouds, (8) centroids of \cch\
clouds, (9) FWHM linear radius of \cch\ clouds,
(10) FWHM angular radius of \cch\ clouds.
\end{table}

\begin{table}
    \begin{center}
\scriptsize
      \caption{Results of Radio Continuum Emission.}\label{tab:vlaresult}
      \begin{tabular}{lccccccccc}
      \\
    \hline
    \hline
Source        &  S$_p$     &  S$_{\nu}$     & Deconvolved Size              &  P.A.             &  Linear Size   & Classfication      \\
              &  (mJyB$^{-1}$)    &  (mJy)    &  (\arcsec $\times$ \arcsec)                 &    ($^{\circ}$)   &  (pc)          &     \\
            \hline
G8.67-0.36    &  34.6(.5)    & 426(6)     &     1.82(.02)$\times$1.60(.02)    &  22(4)            &  0.0397(.0004)   &  UCHII     \\
G10.6-0.4     &  190(2)     &  1645(16)    &    2.59(.02)$\times$1.56(.02)       &  134(1)           &  0.0816(.0006)   &  CHII   \\
G12.42+0.50  & 12.0(.1)     &  15.8(.2)    &  1.73(.03)$\times$0.59(.06)      &   42(1)    &   0.0176(.0003) &   UCHII      \\
W33cont       &  8689(48)    & 18598(142)    &  12.2(.1)$\times$8.3(.13)      &   110(1)  &  0.243(.002)   &  UCHII   \\
W33A          &  0.59(.03)   &  0.97(0.08)  &  0.38(.03)$\times$0.21(.02)      & 154(7)            &  0.0083(.0007)   &   HCHII$^{a, b}$ \\
G14.33-0.64 &  $<$0.2       &     -         &       -                        &          -          &      -           &    -        \\
W42           &   33(1)      &  39(2)       & 3.4(.2)$\times$0.53         &  165(4)  &  0.150(.009)   &   UCHII     \\
W44           &    2051(3)   & 2512(5)      &  6.0(.1)$\times$2.9(.1)    & 96(1) &  0.108(.002) & CHII   \\
S76E          &   133(1)     &  2437(28)    &    49.9(.5)$\times$43.4(.4)     &  49(2)            &  0.5080(.0051)   &   HII     \\
G35.20-0.74   &   6.77(.08)  &  13.4(0.2)   &    15.8(.2)$\times$2.1(.3)      &   1(1)  &  0.153(.002)   &  CHII    \\
W51M          &     208(4)   &  9206(174)   &   7.9(.1)$\times$6.7(.1)    &   99(3) &  0.207(.003)   &  CHII   \\
G59.78+0.06  &   $<$0.04      &     -         &           -                  &    -     &      -             &      -       \\
S87           &   43.7(.5)$^e$   &  54(1)       &    1.49(.04)$\times$0.57(.07)    & 13(2)&  0.0137(.0004)  &  UCHII \\
ON1           &  99.6(.2)    &  148.8(.5)   &     0.529(.002)$\times$0.364(.003)  &    103(1)         &  0.0066(.0001)   &  UCHII$^c$    \\
ON2N        &  1.60(.04)$^e$  &   4.6(.1)     &    0.89(.02)$\times$0.51(.02)    & 86(2.)   &  0.0129(.0003)   &  CHII \\
W75N          &   5.6(.2)    &   6.7(.3)    &    0.17(.02)$\times$0.06     &  160(9) & 0.0025(.0003)   &   UCHII$^d$    \\
DR21S         &   5781(6)    &  13932(19)   &   11.1(.1)$\times$7.7(.1)     &  133(1) & 0.161(.002) & UCHII     \\
W75OH        &   $<$2      &    -          &         -                    &    -     &        -       & -     \\
Cep A         &   4.81(.05)  &   10.3(.1)   &    11.0(.1)$\times$5.2(.2)     &  121(1)& 0.0293(.0003) &  UCHII      \\
G121.30+0.66  &   0.63(.03)   &  0.83(.07)  &     8.4(.9)$\times$2.6(1.0)      &   57(7)           &  0.049(.005)   &  UCHII     \\
W3(OH)        &   11.54(.06)  &  122.6(.7)   &    1.71(.01)$\times$1.12(.01)    &   69(1)           &  0.0282(.0002)  & UCHII     \\
S231         &   0.49(.06)    &  0.48(.10)   &   0.45(.05)$\times$0.24(.03)    &    72(8)     &   0.005    &   HII$^d$   \\
S235          &  4.01(.03)    &  36.0(.3)    &     13.1(.1)$\times$12.0(.1)     &   171(3)          &  0.1016(.0008)  &  HII     \\
S255          &    10.6(.2)  & 27.9(.6)     &    2.81(.04)$\times$1.00(.03)     & 75(1) &  0.0217(.0003)   &  UCHII    \\
S140      &     5.5(.2)  &  8.3(.4)     &      0.54(.02)$\times$0.09     &  44(3)  &  0.0020(.0001)   &  UCHII    \\
AFGL490       &    0.59(.09) & 1.9(.2)      &    1.0(.2)$\times$2.0(.3)     &   105(10)  &  0.010(.002) &  UCHII  \\
NGC2264       &   0.360(0.009) & 0.62(.02)  &   0.40(.01)$\times$0.14(0.02)  &  117(3)   &  0.0018(.0001)    &  UCHII      \\
\hline
      \end{tabular}
%    \end{minipage}
  \end{center}
Notes: -Columns are (1) Source name, (2) peak intensities S$_p$, (3) flux densities S$_{\nu}$, (4) deconvolved size, (5) position angle, (6) linear size, (7) classification of HII regions. \\
  a: van der Tak \& Menten et al. 2005; b: Rengarajan \& Ho 1996; c: Zheng et al. 1985; d: Hunter et al. 1994; d:
Israel \& Felli 1978; e: no flux scaling
\end{table}

\begin{table}
\scriptsize
    \begin{center}
%      \begin{minipage}{105mm}
      \caption{Optical Depth of \cch\ 1-0 (J = 3/2$\rightarrow$1/2, F = 2$\rightarrow$1).}\label{tab:opdepth}
      \begin{tabular}{lcccccc}
      \\
    \hline
    \hline
Source      & $\tau_{aver}$ & $\tau_{(0, 0)}$    &  $\tau_{max}$  &  Offset$_{\tau_{max}}$ \\
            \hline
 G8.67-0.36      &  0.25(0.003)  &   2.52(1.03)   &   2.52(1.03)   &  (0, 0)   \\
 G10.6-0.4      &  0.36(0.52)  &   2.40(0.32)    &   2.40(0.32)   &    (0, 0)      \\
 G12.42+0.50    &  0.10(20.92)  &  0.29(2.29)    &  2.28(2.98)     &   (+30, 0)          \\
 W33cont        &  2.07(0.26)   &  0.36(0.47)    &  11.63(3.74)     &  (-60, +60)       \\
 W33A           &  2.14(0.59)   &  2.93(0.99)    &  3.72(1.45)     &    (+30, 0)         \\
 G14.33-0.64    &  5.37(0.76)   &  6.46(1.36)    &  11.33(7.30)     &   (-60, 0)     \\
 W42            &  1.50(0.39)   &  1.94(1.14)    &  3.93(1.24)     &   (-30, +30)         \\
 W44            &  1.16(0.21)   &  1.02(0.58)    &  3.78(1.25)     &   (+60, 0)          \\
 S76E           &  2.01(0.31)   &  2.06(0.75)    &  13.61(7.88)       &    (+60, -60)          \\
 G35.20-0.74    &  1.43(0.36)   &  2.81(0.98)    &  4.78(1.64)     &     (-30, 0)        \\
 W51M           &  0.57(0.03)   &  0.82(0.36)    &  26.93(24.41)     &   (-150, -30)        \\
 G59.78+0.06    &  3.42(1.42)   &  2.46(3.19)    &  2.46(3.19)     &    (0, 0)    \\
 S87            &  0.10(0.53)   &  0.38(0.69)    &  2.04(1.68)     &     (0, +90)         \\
 ON1            &  1.92(0.76)   &  0.10(0.95)    &  6.08(3.76)     &   (+30, 0)       \\
 ON2S           &  0.76(0.47)   &  1.80(1.76)    &   -    &  -    \\
 W75N           &  0.89(0.31)   &  1.53(0.80)    &   4.00(1.38)    &  (-30, +30)        \\
 DR21S          &  1.23(0.18)   &  0.10(0.02)    &  4.83(2.61)     &    (-90, +60)     \\
 W75OH          &  1.99(0.15)   &  4.33(0.24)    &   4.69(0.92)    &   (-30, +120)           \\
 Cep A          &  1.31(0.22)   &  2.75(0.99)    &  6.33(2.10)     &    (-60, -60)          \\
 G121.30+0.66   &  2.33(0.79)   &  2.12(1.14)    &  8.68(1.05)     &    (0, -30)     \\
 W3(OH)         &  0.42(0.27)   &  0.55(0.75)    &  3.46(1.78)     &    (-30, -60)        \\
 S231           &  0.10(0.59)   &  0.18(0.79)    &  2.65(2.18)     &   (-30, 0)      \\
 S235           &  0.45(0.31)   &  0.74(0.73)    &  8.54(6.06)     &    (0, -90)           \\
 S255           &  0.14(0.01)   &  0.10(18.81)   &  4.88(3.27)     &   (-60, +60)           \\
 S140       &  0.68(0.11)   &  0.10(0.001)   &  2.58(0.53)     &     (+90, 0)           \\
 AFGL490        &   0.11(6.65)   &  0.68(1.82)   &   -     &    -         \\
 NGC2264    &  1.35(0.21)   &  0.19(0.21)   &  3.02(0.77)     &    (+60, 0)            \\
             \hline
      \end{tabular}
%    \end{minipage}
  \end{center}
Notes: -Columns are (1) Source name, (2) optical depth of spectral line averaging over the whole emitting
region $\tau_{aver}$, (3) optical depth of spectral line at (0, 0), (4)
maximum optical depth $\tau_{max}$, (5) offset of spectral line with $\tau_{max}$.
\end{table}

\begin{deluxetable}{lrrrrrrrrrrrrrrrrr}
\tabletypesize{\scriptsize} \rotate
\tablecaption{\label{tab:velgrad}Results of Gradient Fitting for
\hcccn\ 10-9, HNC 1-0 and \cch\ 1-0 main lines. } \tablewidth{0pt}
\tablehead{ \multicolumn{1}{l}{Source} &\multicolumn{4}{c}{\hcccn\
10$\rightarrow$ 9}&\multicolumn{4}{c}
{HNC 1$\rightarrow$ 0}&\multicolumn{4}{c}{\cch\ 1 $\rightarrow$ 0}\\
\colhead{}             & \colhead{$G$}  & \colhead{$\theta$} &
\colhead{$\beta$} & \colhead{$J/M$}    & \colhead{$G$}  &
\colhead{$\theta$} & \colhead{$\beta$} & \colhead{$J/M$}   &
\colhead{$G$}  & \colhead{$\theta$} &
\colhead{$\beta$} & \colhead{$J/M$}    \\
\colhead{}             & \colhead{(\kmspc)}          &
\colhead{(deg)} & \colhead{ }   &  \colhead{(\kms\ pc)} &
\colhead{(\kmspc)}          & \colhead{(deg)} & \colhead{ }   &
\colhead{(\kms\ pc)} & \colhead{(\kmspc)}          & \colhead{(deg)}
& \colhead{ }   &  \colhead{(\kms\ pc)} } \startdata
 G8.67-0.36  &   1.50(0.12)   &   58(4)  &  0.013 & 0.21   &  2.06(0.10) &  100(2) &  0.067    &  0.83 &  1.07(0.04) &  60(5)          &  0.020 &  0.47   \\
 G10.6-0.4   &   0.26(0.22)$^*$&  296(36)&       - &      -  &  0.24(0.05)  &  1(7)   &  0.0003    &  0.030 &   0.29(0.07)  &  10(9)        &  0.0018  &  0.16   \\
 G12.42+0.50 &   -            &        - &       - &      -  &  1.40(0.20)  &  28(6)  &  0.0060    &  0.032 &   1.08(0.61)$^*$   &   9(25)  &  -        &       -    \\
 W33cont     &  0.57(0.05)   &  193(3)   &  0.0037 & 0.089  &   0.77(0.02) & 156(1)  &  0.017    &  0.30 &  0.45(0.03)   &   124(3)      &  0.0068&  0.21    \\
 W33A        &  -            &       -   &       - &      -  &  0.46(0.11)  & 278(10) &  0.0065    &  0.19 &  0.02(0.17)$^*$    &   87(321)&  -        &       -    \\
 G14.33-0.64 &  0.22(0.18)$^*$& 157(35)  &       - &      -  &  1.37(0.11)  & 158(4)  &  0.042    &  0.15 &  0.29(0.23)$^*$   &   128(33) &  -        &       -    \\
 W42         &  0.11(0.09)$^*$&   205(34)&       - &      -  &  0.25(0.04)  & 326(6)  &  0.0071    &  0.40 &  0.11(0.06)$^*$   &   289(21) &  -        &       -    \\
 W44         &  0.31(0.05)   &  350(6)   &  0.0006 & 0.036  &  0.31(0.02)  &  64(2)  &  0.0010    &  0.062 &  0.32(0.03)   &  321(4)       &  0.0031&  0.18     \\
 S76E        &0.40(0.17)$^*$ &  139(17)  &       - &      -  &  0.37(0.05)  &  62(5)  &  0.0012    &  0.022 &  0.28(0.09)  &   52(12)       &  0.0012 &  0.028     \\
 G35.20-0.74 & 1.43(0.25)   &  99(7)     &  0.0033 & 0.028  &  1.68(0.08)   &  107(2)&  0.017    &  0.12 & 1.69(0.20) &   112(4)         &  0.019 &  0.13     \\
 W51M        & 2.38(0.03)  &  273(1)     &  0.036 & 1.7  & 1.45(0.01)  &  298(1)  &  0.033    &  2.5 & 0.95(0.03) &    331(1)        &  0.016&  1.80      \\
 G59.78+0.06 &     -        &        -   &       - &      -  &  0.97(0.10)   & 166(4) &  0.012    &  0.046 &  2.21(0.78)$^*$ &  176(14)    &  -        &       -     \\
 S87         & 1.95(0.18)   & 15(4)      &  0.028 & 0.053  &  1.79(0.06)  &  49(1)  &  0.052    &  0.11 &   1.11(0.08)  &   39(3)       &  0.032 &  0.11     \\
 ON1         &  1.70(0.34)  &  239(8)    &  0.016 & 0.092  &  1.18(0.04)  &  231(2) &  0.013    &  0.11 &  1.40(0.27) &  245(8)         &  0.0023 &  0.016     \\
 ON2S        &     -        &    -       &       - &      -  &     -      &     -     &  -         &       - &    -      &    -              &  -        &       -     \\
 W75N        &  0.85(0.09)  &  264(4)    &  0.0065 & 0.063  &  0.71(0.03)  &  288(2) &  0.0074    &  0.086 &  0.44(0.06) &   313(6)        &  0.0033 &  0.063       \\
 DR21S       &  0.50(0.06)  &  178(6)    &  0.0009 & 0.0080  &  0.42(0.03)  & 189(3)  &  0.0029    &  0.033 &  0.25(0.03) &  181(4)         &  0.0055 &  0.10      \\
 W75OH       &  0.24(0.05)  &  218(9)    &  0.0006 & 0.023  & 0.46(0.02)  & 245(1)   &  0.0061    &  0.12 &  0.32(0.03) &  242(3)         &  0.0053 &  0.15        \\
 Cep A       &  6.03(0.44)  &  285(2)    &  0.024 & 0.03`  & 3.58(0.18)  & 285(2)   &  0.028    &  0.062 &  3.70(0.24) &  284(3)         &  0.018 &  0.038      \\
 G121.30+0.66&  1.90(0.56)  &  107(12)   &  0.0045 & 0.011  & 0.80(0.08)  & 122(4)   &  0.0027    &  0.016 &  2.18(0.52) &  328(10)        &  0.013  &  0.028       \\
 W3(OH)      & 1.10(0.14)   &  235(5)    &  0.0080 & 0.067  & 0.74(0.03)  & 245(2)   &  0.016    &  0.19 & 0.61(0.06)  &  224(4)         &  0.015 &  0.22      \\
 S231        & 1.26(0.17)   &  58(5)     &  0.0045 & 0.027  & 0.49(0.05)  & 30(4)    &  0.0020    &  0.031 &   0.40(0.17)$^*$  &  350(19)  &  -        &       -      \\
 S235        & -           &   -         &       - &      -  & 0.51(0.09)  & 153(7)   &  0.0016    &  0.012 &  0.75(0.18)  &  116(9)        &  0.0027  &  0.014      \\
 S255        & 1.98(0.15)  &   348(4)    &  0.011 & 0.039  & 1.57(0.04)  & 342(1)   &  0.018    &  0.083 &  1.33(0.09)  &   342(3)       &  0.017 &  0.095      \\
 S140    & 0.67(0.11) &   278(7)     &  0.0013 & 0.0072  &  0.13(0.04) & 144(13)  &  0.0001    &  0.0024 &  0.14(0.13)$^*$    &   98(40) &  -        &       -       \\
 AFGL490     &    -       &       -      &       - &      -  & 0.50(0.16) &  306(12)  &  0.0008    &  0.0089 &  -             &  -           &  -        &       -      \\
 NGC2264  & 2.11(0.30)  &  200(6)    &  0.012 & 0.023   &  1.65(0.07) &  243(2) &  0.023    &  0.055  &  1.30(0.15) &   246(5)       & 0.0076 &  0.023        \\
        \enddata \\
Notes: -$^*$: Gradients with G/$\sigma_G <$ 3. \\
Columns are (1) Source name, (2) velocity gradients $G$ of \hcccn\ clouds,
(3) directions of velocity gradient $\theta$ of \hcccn\
clouds, (4) ratio of rotation kinetic energy to
gravitation energy $\beta$ of \hcccn\ clouds, (5)
specific angular momentum $J/M$ of \hcccn\ clouds, (6)
$G$ of HNC clouds, (7) $\theta$ of HNC clouds,
(8) $\beta$ of HNC clouds, (9) $J/M$ of HNC
clouds, (10) $G$ of \cch\ clouds, (11)
$\theta$ of \cch\ clouds, (12) $\beta$ of \cch\ clouds, (13) $J/M$ of \cch\ clouds.
\end{deluxetable}

\begin{table}
\scriptsize
    \begin{center}
%      \begin{minipage}{105mm}
      \caption{Results of Uniformly Magnetic Sphere Model Fitting}\label{tab:magnetic}
      \begin{tabular}{lcccccc}
      \\
    \hline
    \hline
Source  &  $\sigma_{rot}$ &  $\sigma_{Turb}$  & Magnetic Field  &  Number Density                         \\
              &   (\kms)        &  (\kms)           &     ($\mu$G)   &      (cm$^{-3}$)                         \\
            \hline
G8.67-0.36     &   1.80(.14)        &     0.59(0.64)     &    10        &              5.0$\times$10$^4$                       \\
 G10.6-0.4      &   0.66(.14)$^*$    &    2.61(0.13)     &    61         &             1.8$\times$10$^5$                        \\
 G12.42+0.50    &   0.78(.11)$^*$    &    0.90(0.20)     &    10          &                2.1$\times$10$^5$                        \\
 W33cont        &   0.62(.05)        &    1.49(0.06)     &    21         &                3.3$\times$10$^4$                        \\
 W33A           &   0.55(.13)$^*$    &    1.46(0.14)     &    20         &                3.9$\times$10$^4$                        \\
 G14.33-0.64    &   0.95(.08)$^*$    &    0.50(0.24)     &    5           &               9.6$\times$10$^4$                       \\
 W42            &   0.61(.10)$^*$    &    1.64(0.13)     &    25         &                2.4$\times$10$^4$                         \\
 W44            &   0.31(.05)        &    2.18(0.05)     &    43         &            8.1$\times$10$^4$                         \\
 S76E           &   0.21(.03)$^*$    &    1.20(0.04)     &    13         &                  1.5$\times$10$^5$                          \\
 G35.20-0.74    &   0.76(.13)        &    1.45(0.15)     &    21         &             2.7$\times$10$^5$                          \\
 W51M           &   3.43(.04)        &    2.38(0.18)     &    88         &             4.8$\times$10$^4$                       \\
 G59.78+0.06    &   0.57(.06)$^*$    &    0.67(0.17)     &     5          &            1.7$\times$10$^5$                              \\
 S87            &   0.99(.09)        &    -             &    -          &             6.9$\times$10$^4$                   \\
 ON1            &   1.16(.23)        &    1.02(0.43)     &    14         &             8.4$\times$10$^4$                    \\
 ON2S           &     -              &    1.45(0.13)     &    19         &             2.4$\times$10$^4$                   \\
 W75N           &   0.68(.07)        &    1.16(0.10)     &    14         &             4.8$\times$10$^4$                           \\
 DR21S          &   0.40(.05)        &    1.00(0.05)     &    10          &             1.4$\times$10$^5$                   \\
 W75OH          &   0.19(.04)        &    1.75(0.05)     &    27         &             6.3$\times$10$^4$                    \\
 Cep A           &  0.88(.06)        &    1.11(0.13)     &    14         &             8.1$\times$10$^5$                     \\
 G121.30+0.66   &   0.61(.18)        &    0.75(0.23)     &     6          &            3.2$\times$10$^5$                    \\
 W3(OH)         &   1.00(.13)        &    1.16(0.22)     &    16         &             7.5$\times$10$^4$                    \\
 S231           &   0.77(.10)        &    0.86(0.18)     &     9          &            1.2$\times$10$^5$                         \\
 S235           &   0.22(.04)$^*$    &    0.91(0.07)     &     8          &            2.9$\times$10$^5$               \\
 S255           &   0.84(.06)        &    0.36(0.23)     &     3          &            8.4$\times$10$^4$                     \\
 S140           &   0.14(.02)        &    0.98(0.02)     &     9          &            1.8$\times$10$^5$                      \\
 AFGL490        &   0.13(.04)$^*$    &    0.76(0.13)     &     5          &            1.2$\times$10$^6$                      \\
 NGC2264        &   0.51(.07)        &    0.73(0.08)     &      6         &        1.3$\times$10$^5$                         \\
              \hline
      \end{tabular}
%    \end{minipage}
  \end{center}
Notes: -Columns are (1) Source name, (2) the rotation line broadening $\sigma_{rot}$, (3) turbulence broadening $\sigma_{Turb}$, (4) derive magnetic field strength, (5) derived number densities. \\
  $^*$: calculated using gradients of HNC.
\end{table}

\begin{figure}[]
\begin{center}
\includegraphics[width=2.7in]{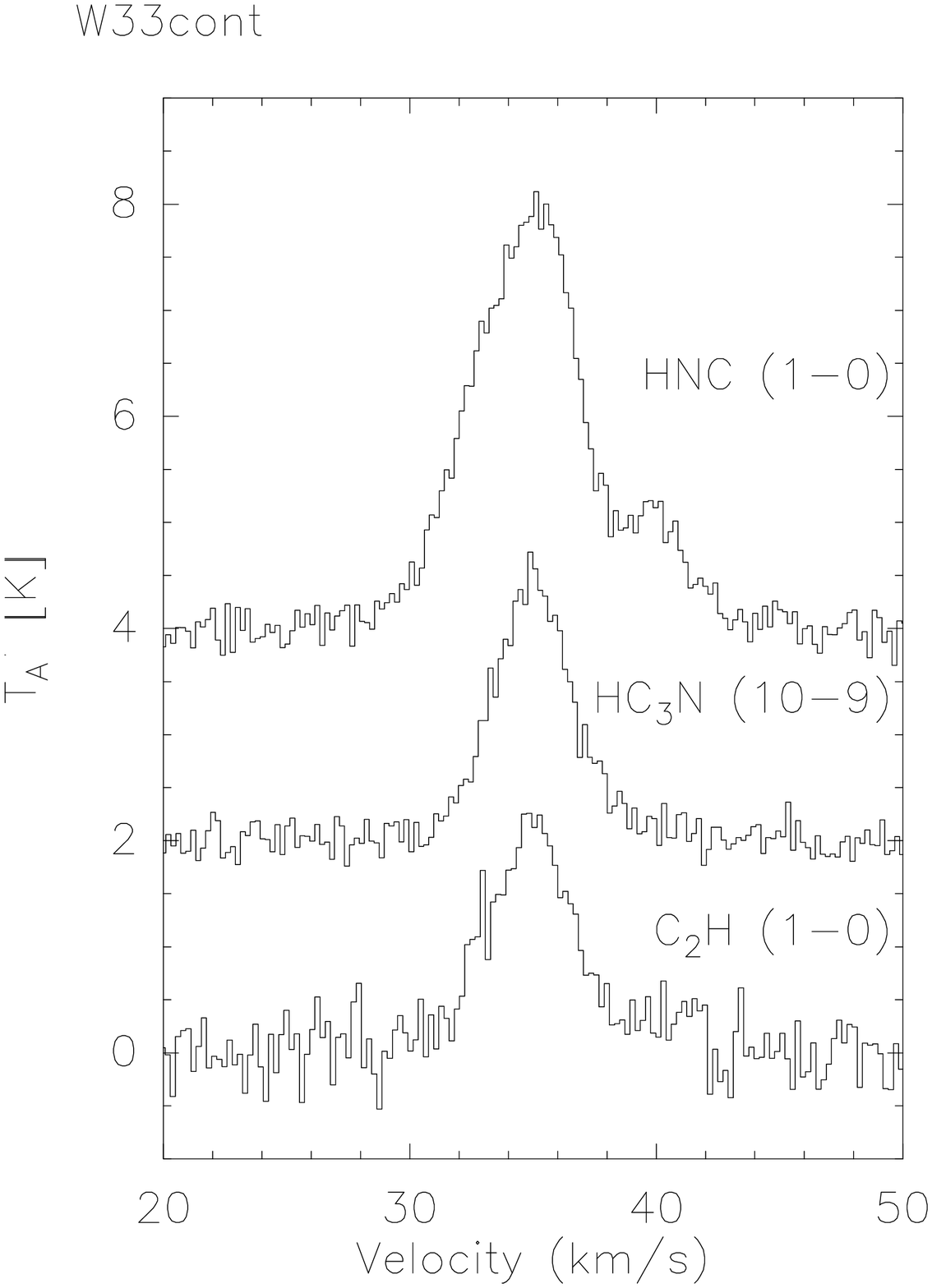}
\includegraphics[width=2.7in]{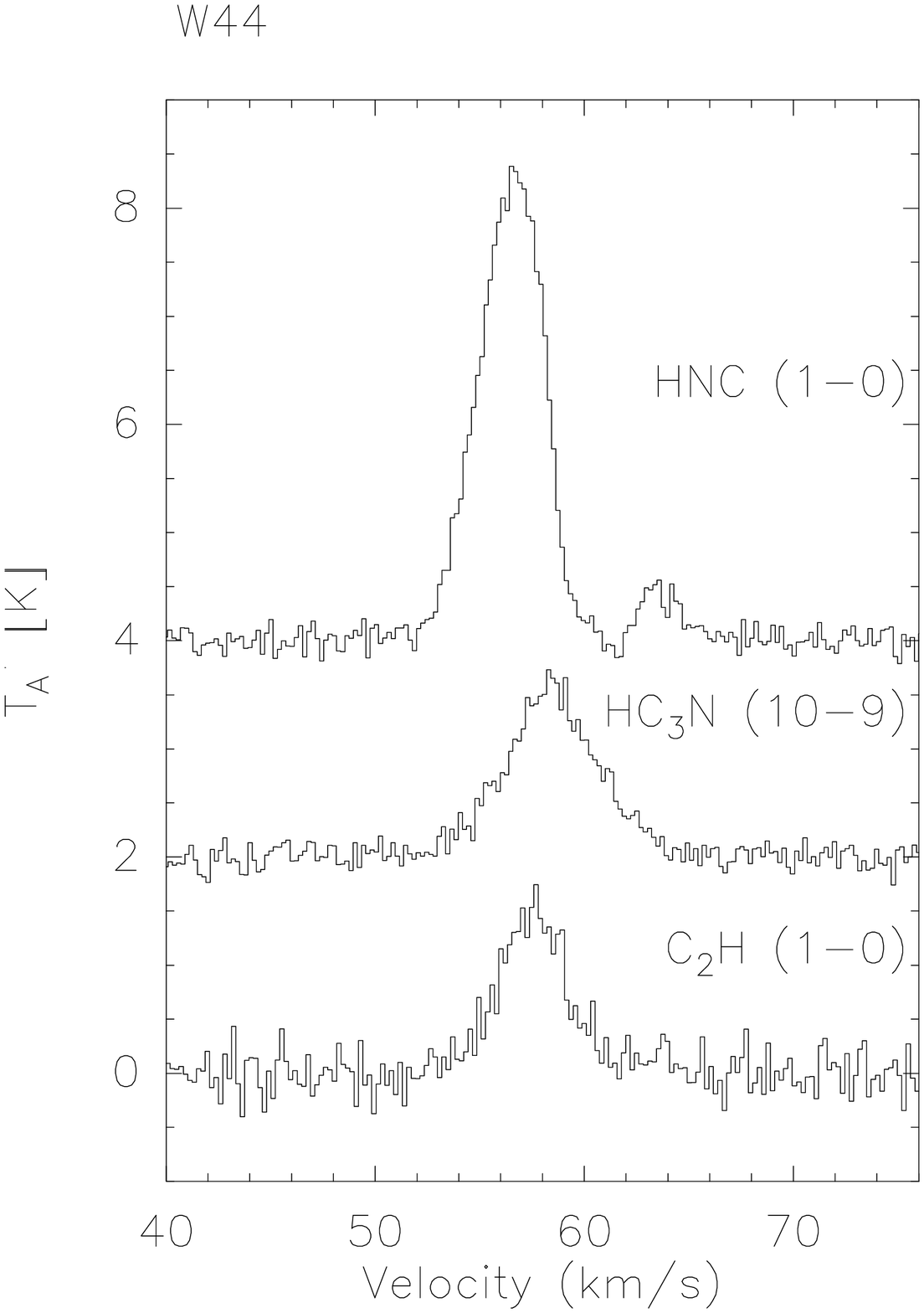}
\includegraphics[width=2.7in]{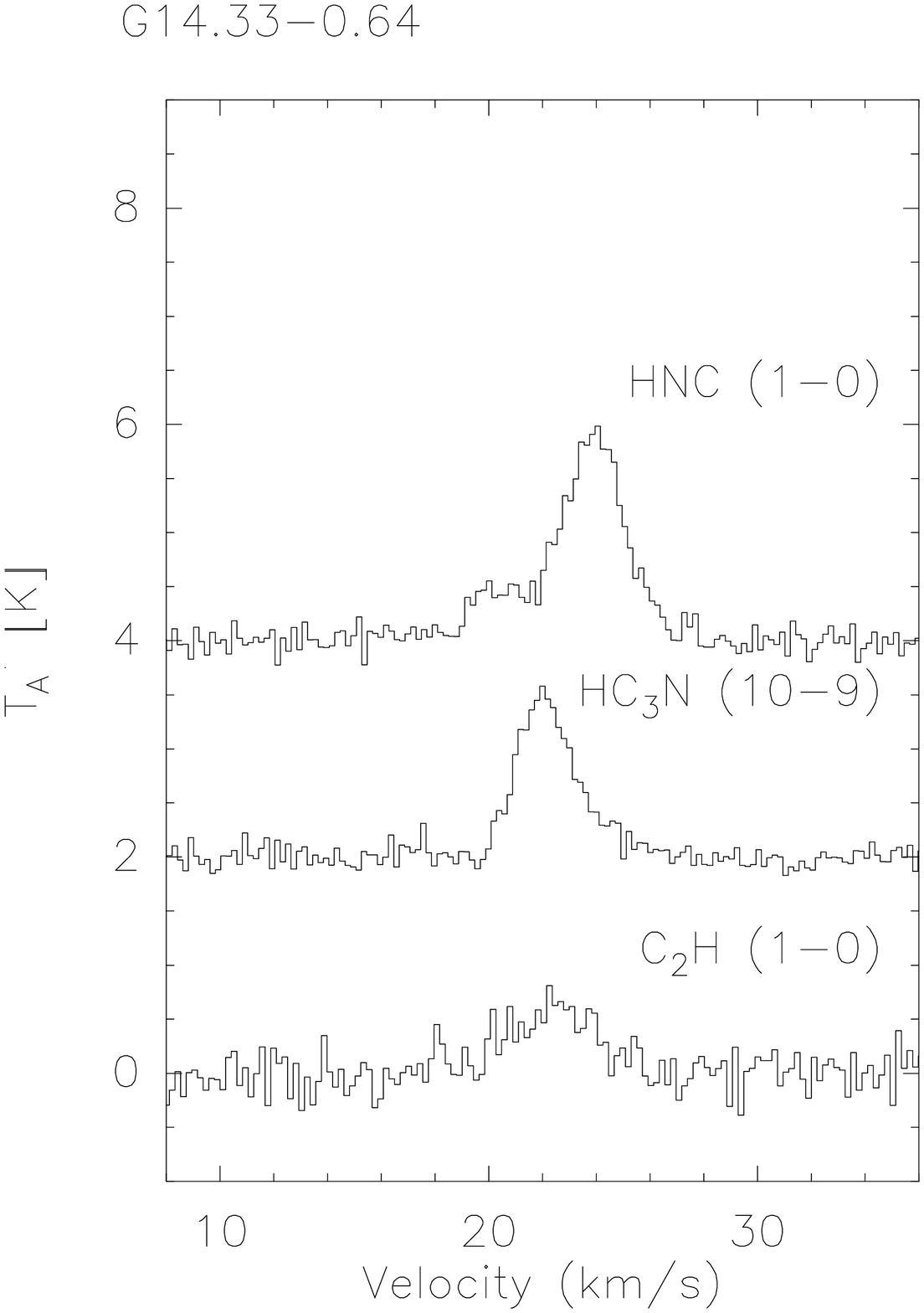}
\includegraphics[width=2.7in]{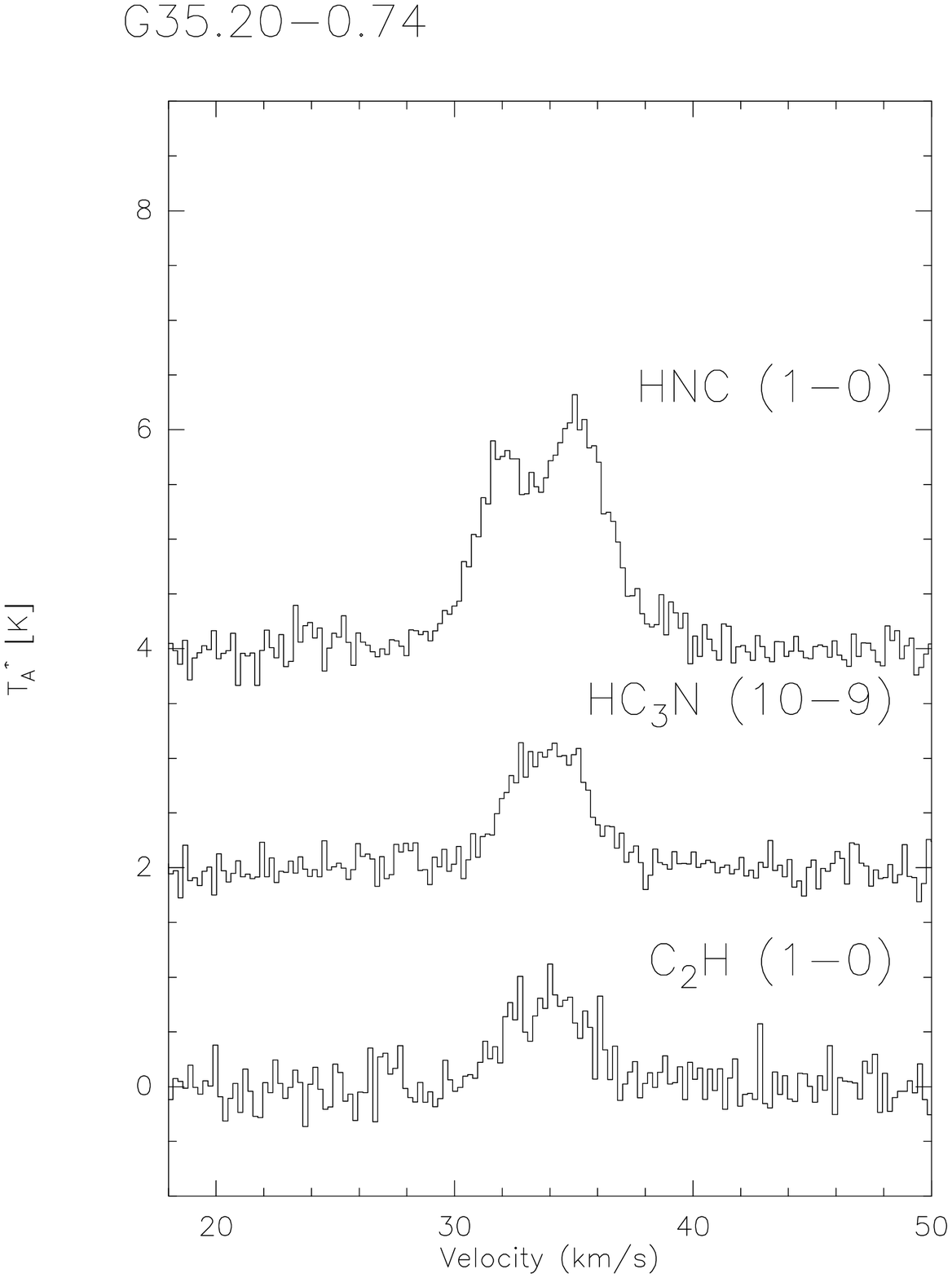}
\vspace*{-0.2 cm} \caption{\label{spec3} Spectra of \hcccn\ 10-9,
HNC 1-0 and \cch\ 1-0 (J = 3/2$\rightarrow$1/2, F =
2$\rightarrow$1) at the offset (0, 0) for W33cont, W44, G14.33-0.64
and G35.20-0.74. The identification of the transitions is given to
the right of each lines.}
\end{center}
\end{figure}

\clearpage

\begin{figure}[]
\begin{center}
\includegraphics[width=2.4in, angle=270]{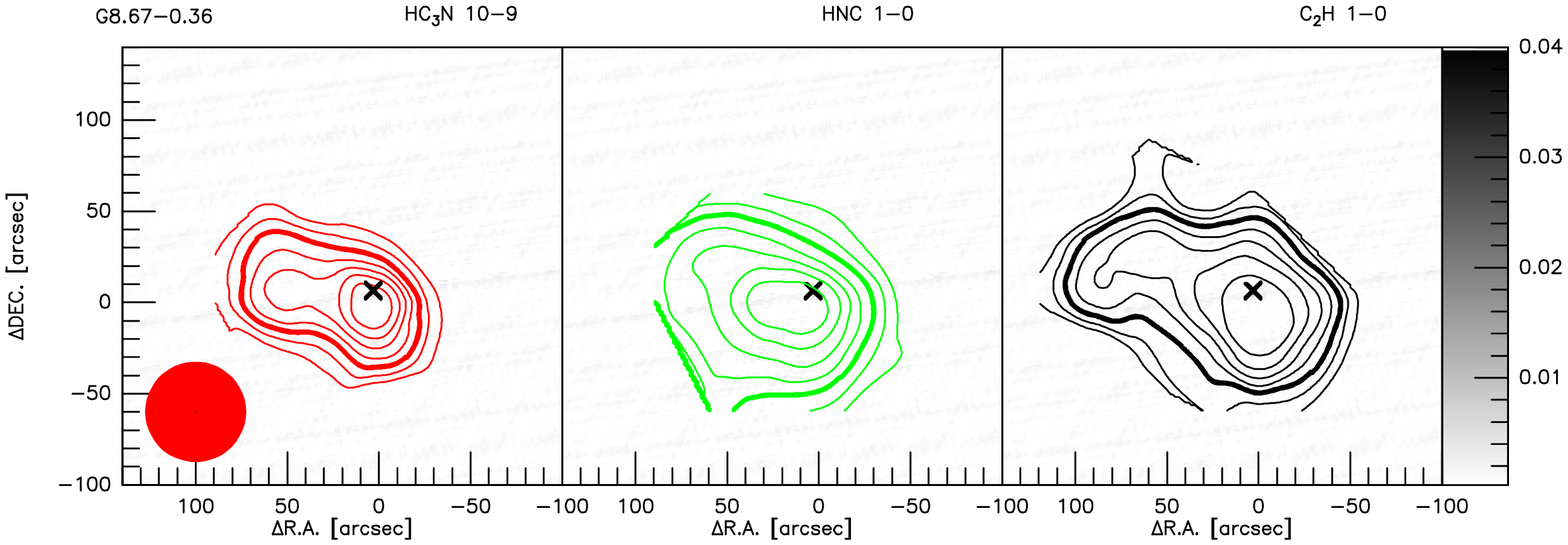}
\includegraphics[width=2.4in, angle=270]{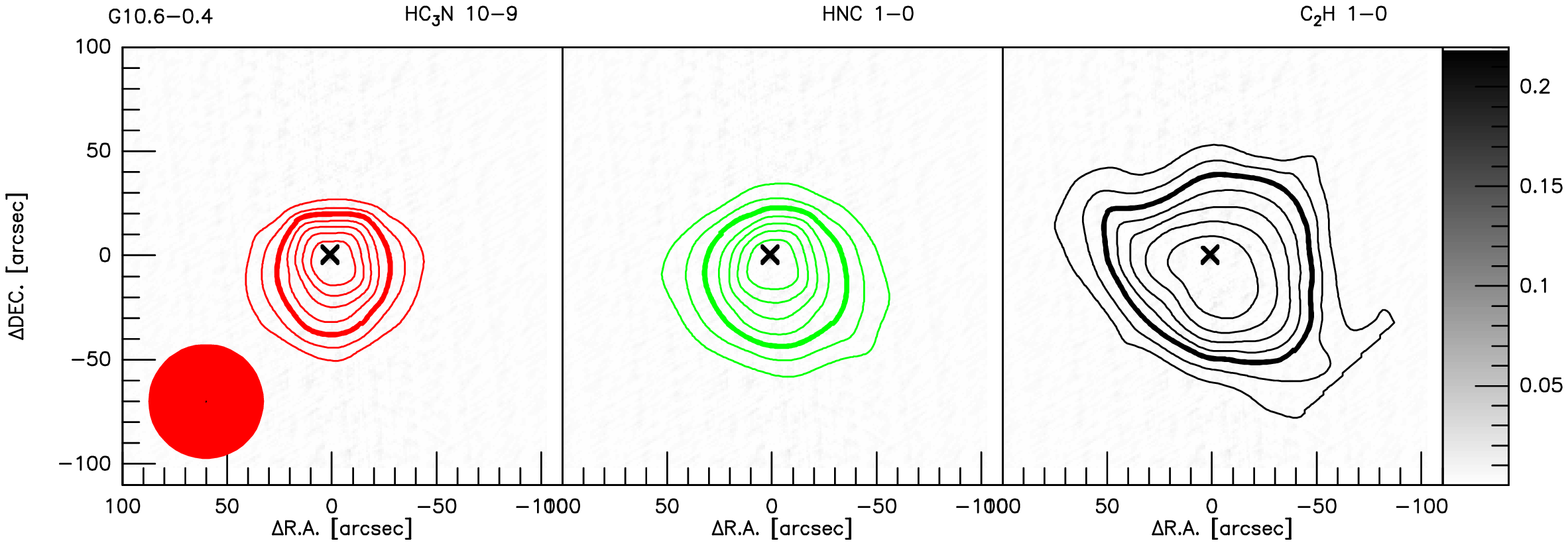}
\includegraphics[width=2.4in, angle=270]{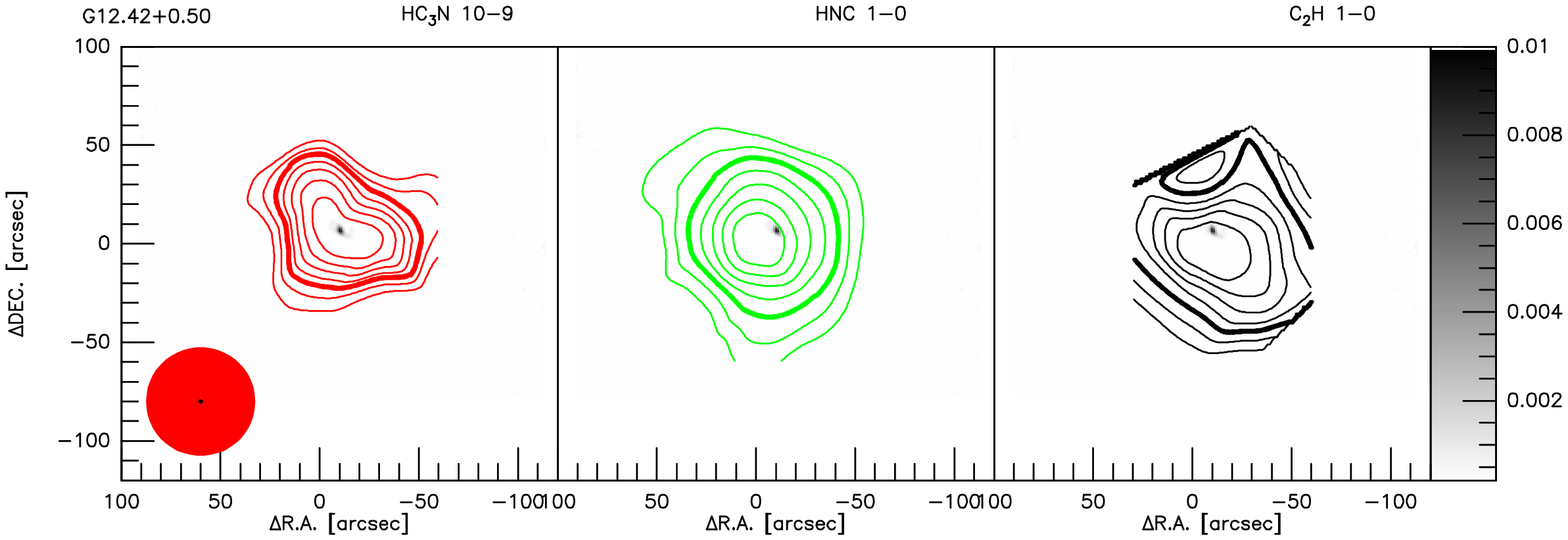}
\vspace*{-0.2 cm} \caption{\label{intensity1} Contour maps of
\hcccn\ 10-9 (left pannel), HNC 1-0 (middle pannel) and \cch\ 1-0 (F
= 3/2, 2 $\rightarrow$ 1/2, 1) (right pannel) integrated intensity
superimposed on its 8.4 or 4.8 GHz continuum map in gray scale. The
contour levels are 30\%, 40\%, 50\%, 60\%, 70\%, 80\% and 90\% of
the map peak, reported in Table 3 (see col. [3] for \hcccn, col. [7]
for HNC, and col. [11] for \cch). The heavy lines represent 50\% of
the map peak. The FWHM beam size for molecular lines (red big
circle) and continuum emission (black small ellipse) observations
are shown at the lower left of left map. The water maser is at (0,
0). Crosses are used to mark position of HII region in case that the
HII regions are too small to be recognized in the figure. The gray
scale range indicated in the wedge is in Jy/beam. (A color version
of this figure is available in the online journal.)}
\end{center}
\end{figure}

\clearpage

\begin{figure}[]
\begin{center}
\includegraphics[width=2.4in, angle=270]{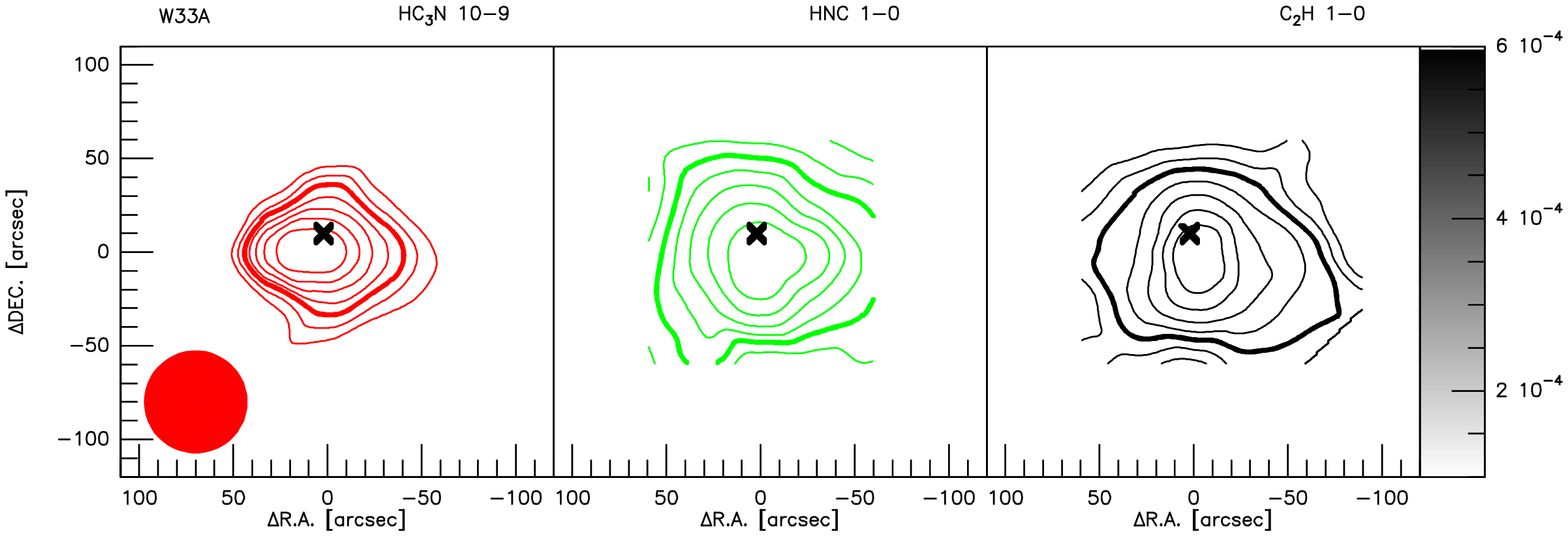}
\includegraphics[width=2.4in, angle=270]{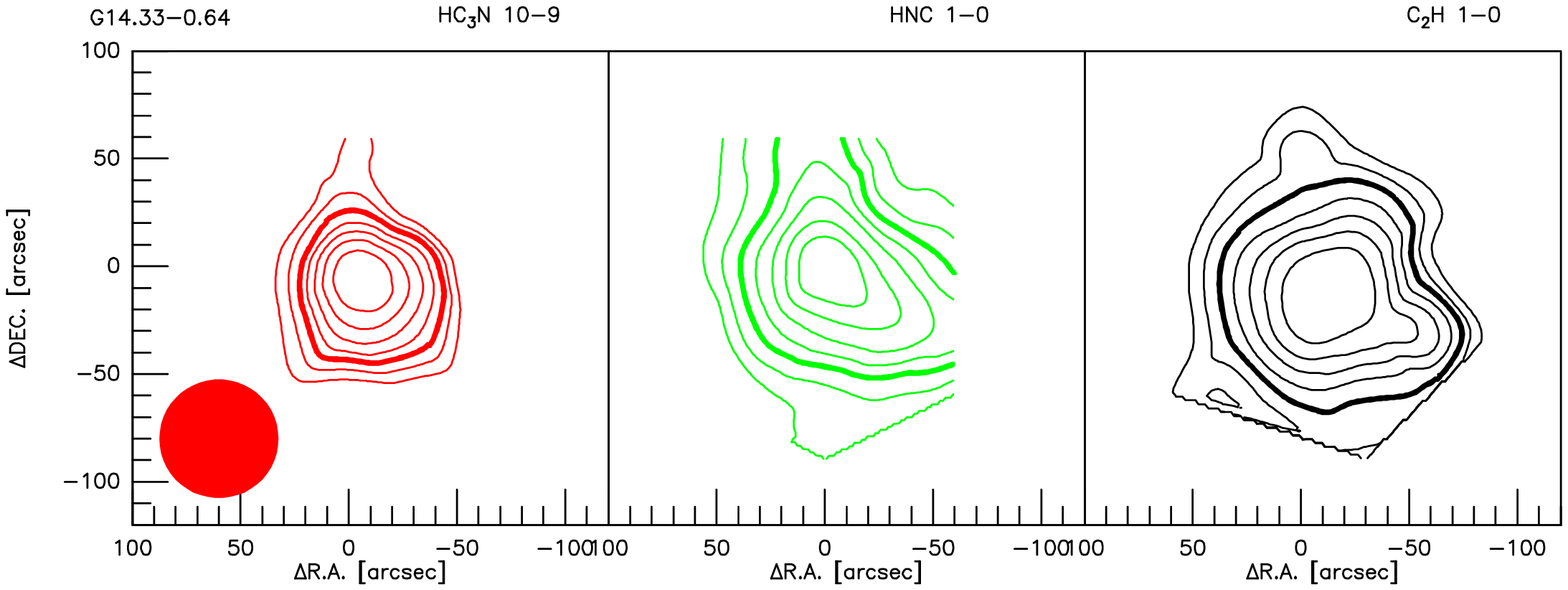}
\vspace*{-0.2 cm} \caption{\label{intensity2} Contour maps of
\hcccn\ 10-9 (left pannel), HNC 1-0 (middle pannel) and \cch\ 1-0 (F
= 3/2, 2 $\rightarrow$ 1/2, 1) (right pannel) integrated intensity
superimposed on its 8.4 or 4.8 GHz continuum map in gray scale. (A
color version of this figure is available in the online journal.)}
\end{center}
\end{figure}

\clearpage

\begin{figure}[]
\begin{center}
\includegraphics[width=2.4in, angle=270]{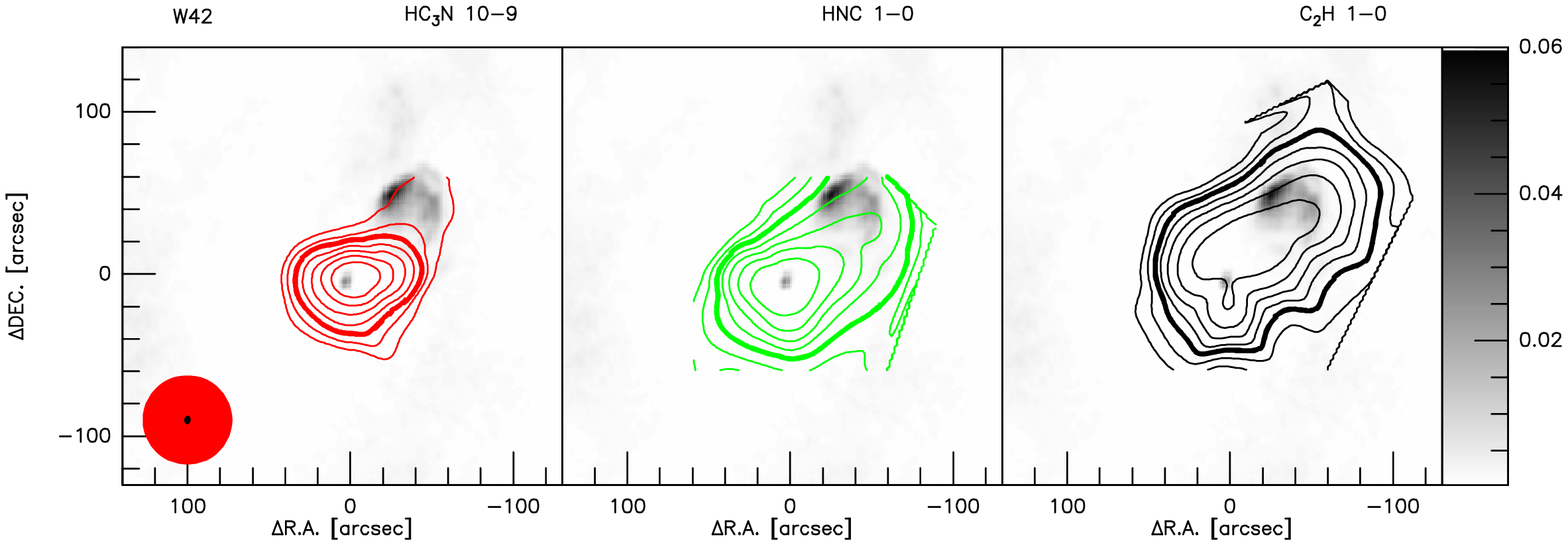}
\includegraphics[width=2.4in, angle=270]{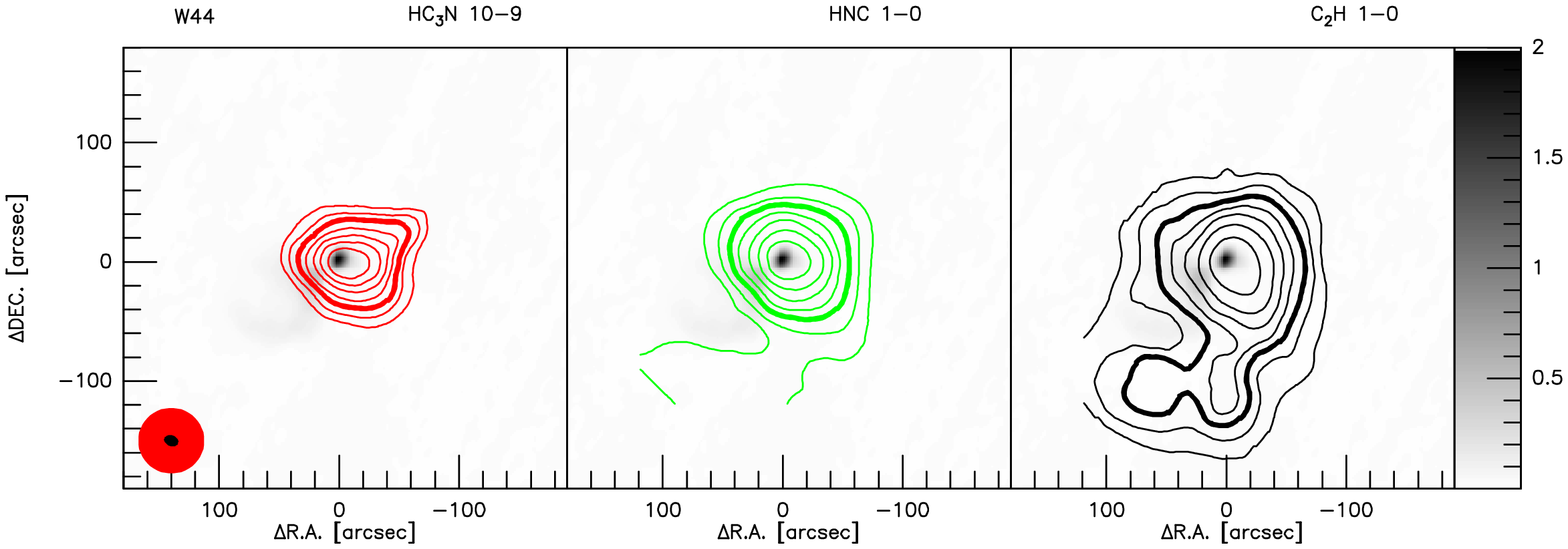}
\includegraphics[width=2.4in, angle=270]{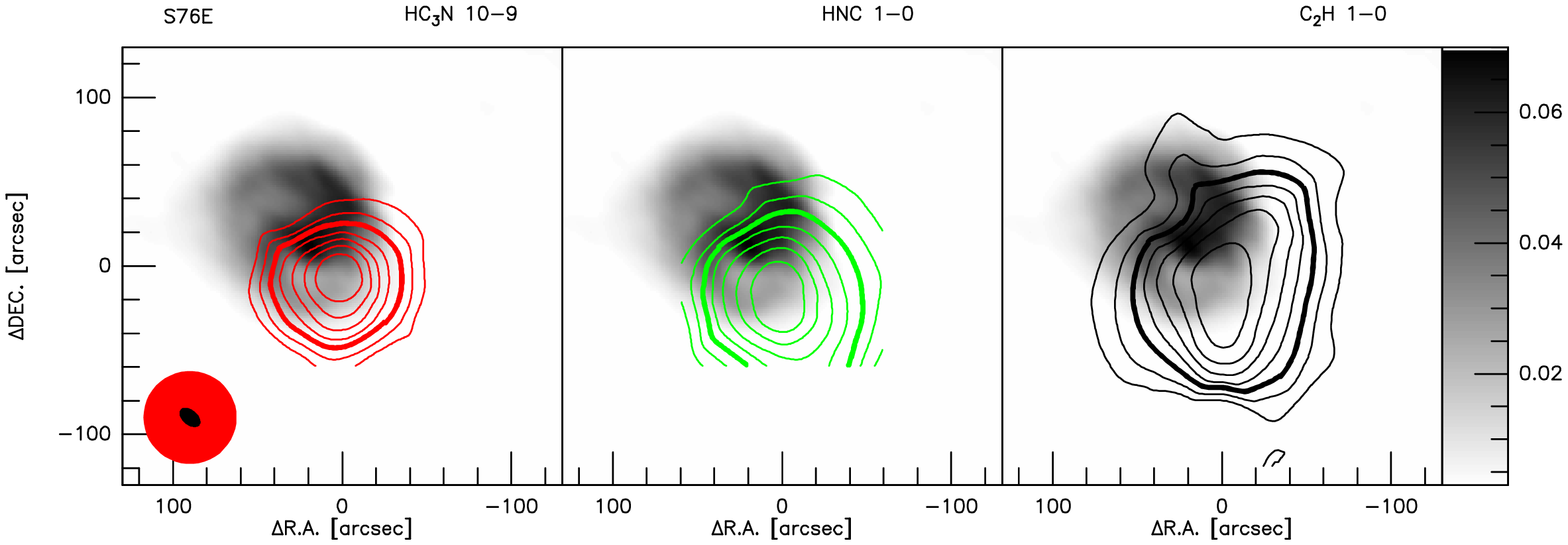}
\vspace*{-0.2 cm} \caption{\label{intensity3} Contour maps of
\hcccn\ 10-9 (left pannel), HNC 1-0 (middle pannel) and \cch\ 1-0 (F
= 3/2, 2 $\rightarrow$ 1/2, 1) (right pannel) integrated intensity
superimposed on its 8.4 or 4.8 GHz continuum map in gray scale. (A
color version of this figure is available in the online journal.)}
\end{center}
\end{figure}

\clearpage

\begin{figure}[]
\begin{center}
\includegraphics[width=2.4in, angle=270]{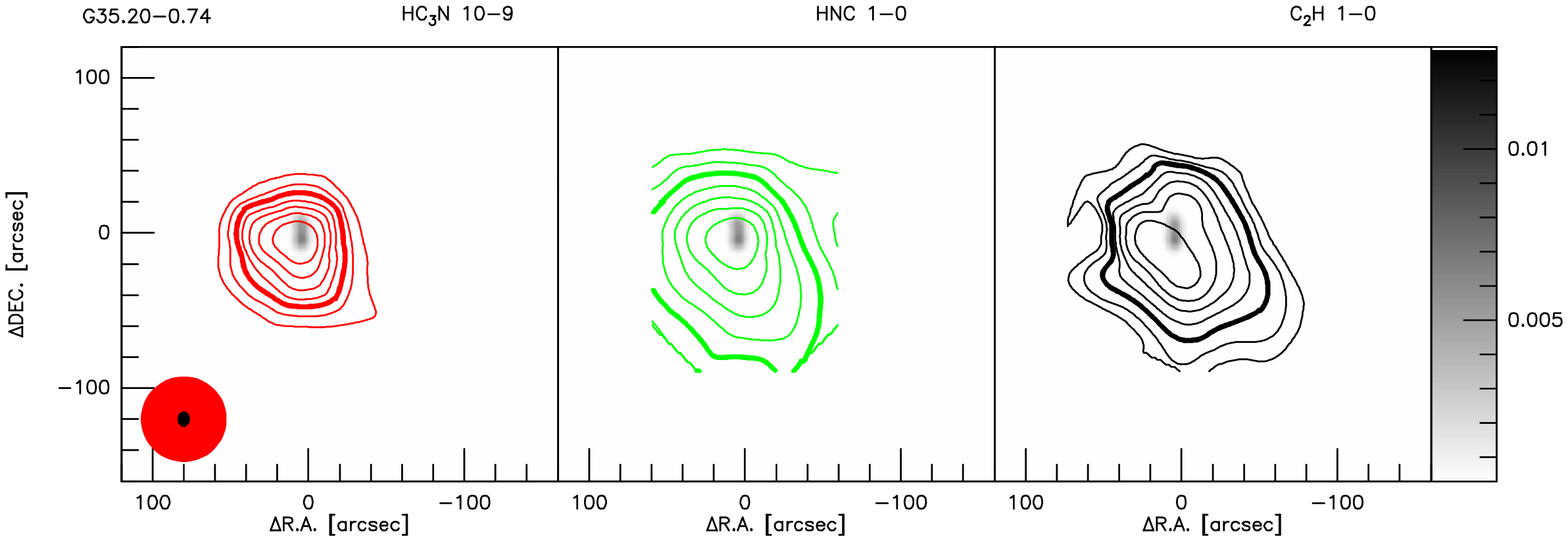}
\includegraphics[width=2.4in, angle=270]{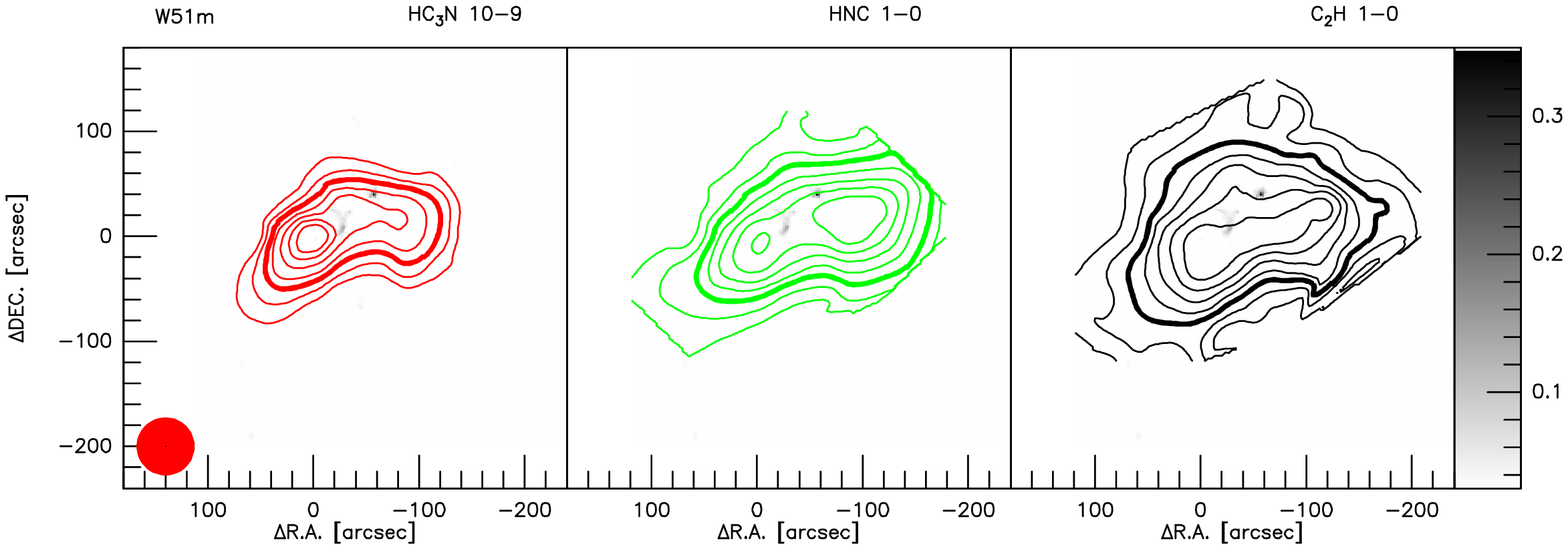}
\includegraphics[width=2.4in, angle=270]{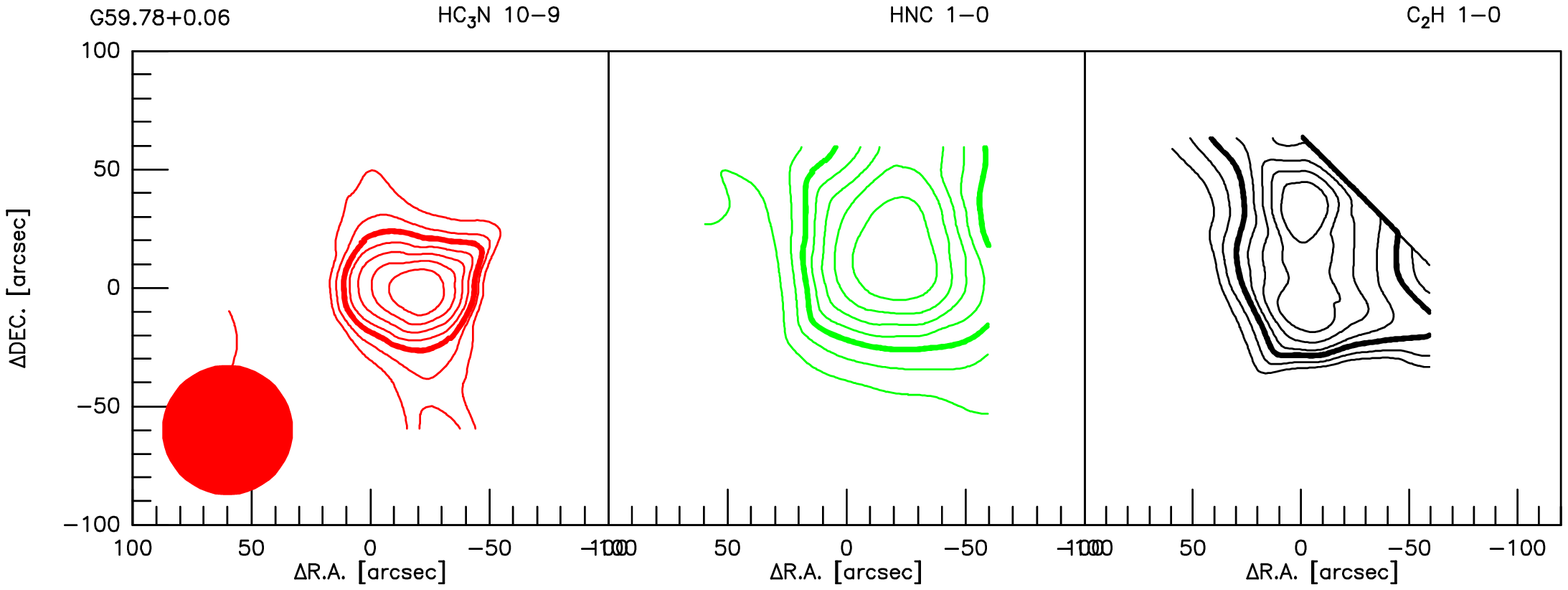}
\vspace*{-0.2 cm} \caption{\label{intensity4} Contour maps of
\hcccn\ 10-9 (left pannel), HNC 1-0 (middle pannel) and \cch\ 1-0 (F
= 3/2, 2 $\rightarrow$ 1/2, 1) (right pannel) integrated intensity
superimposed on its 8.4 or 4.8 GHz continuum map in gray scale. (A
color version of this figure is available in the online journal.)}
\end{center}
\end{figure}

\clearpage

\begin{figure}[]
\begin{center}
\includegraphics[width=2.4in, angle=270]{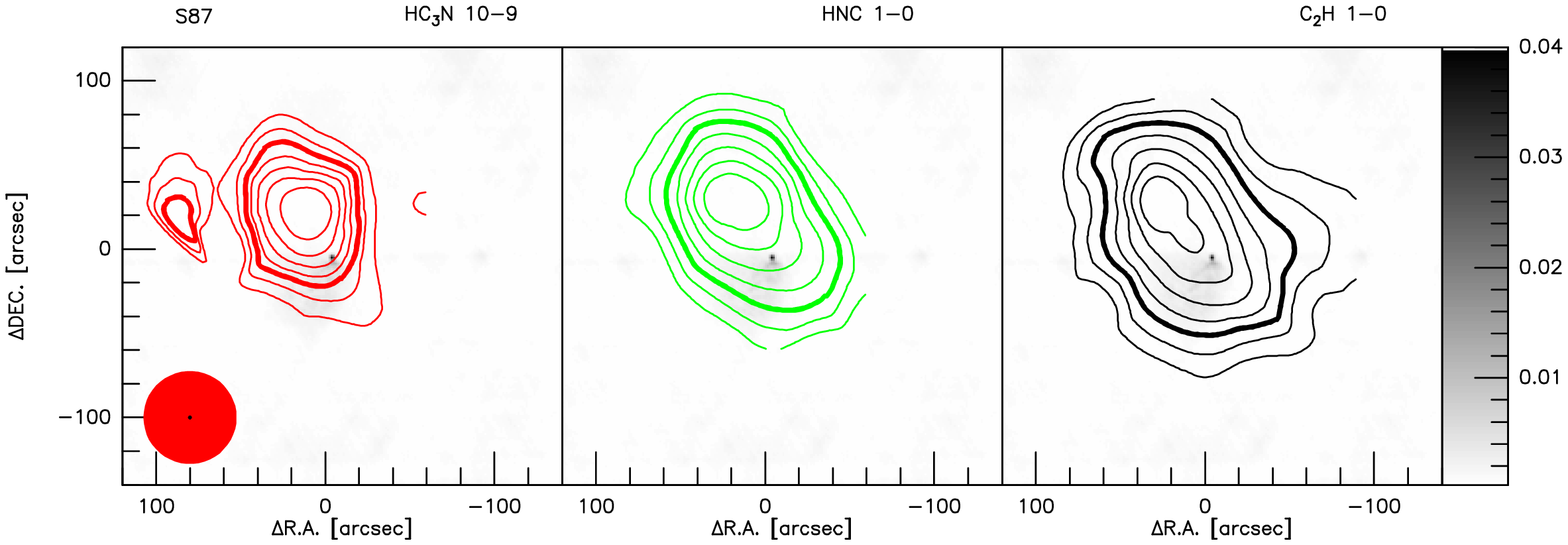}
\includegraphics[width=2.4in, angle=270]{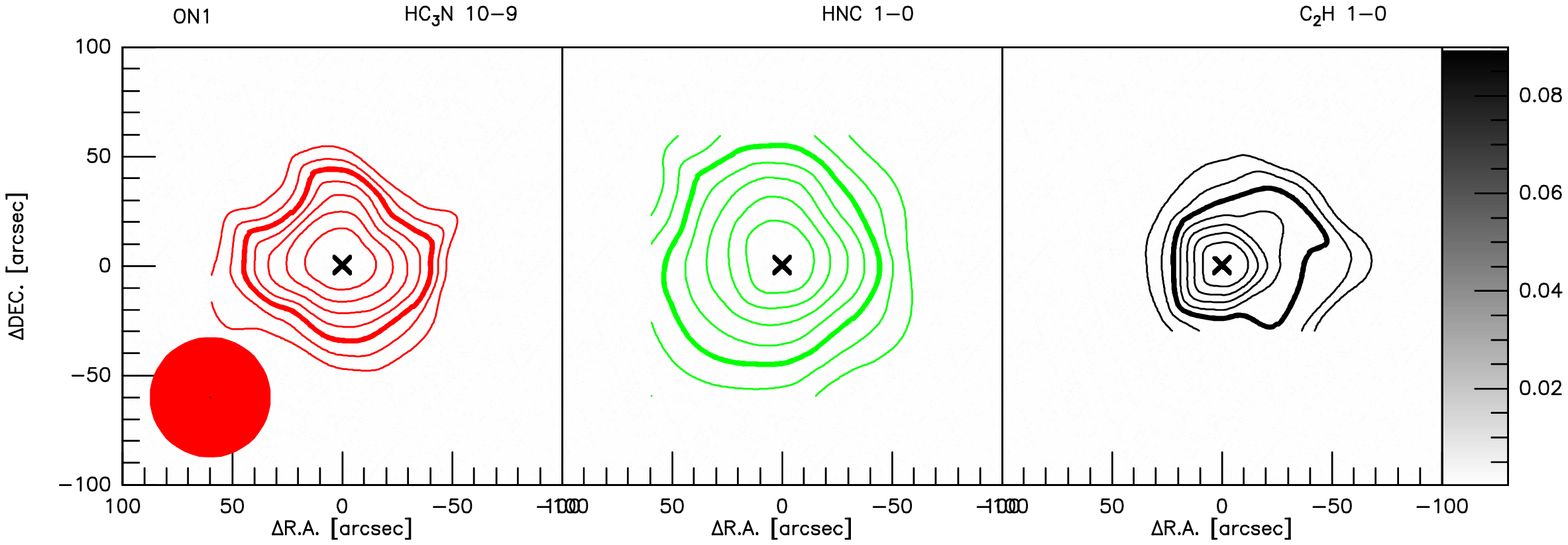}
\includegraphics[width=2.4in, angle=270]{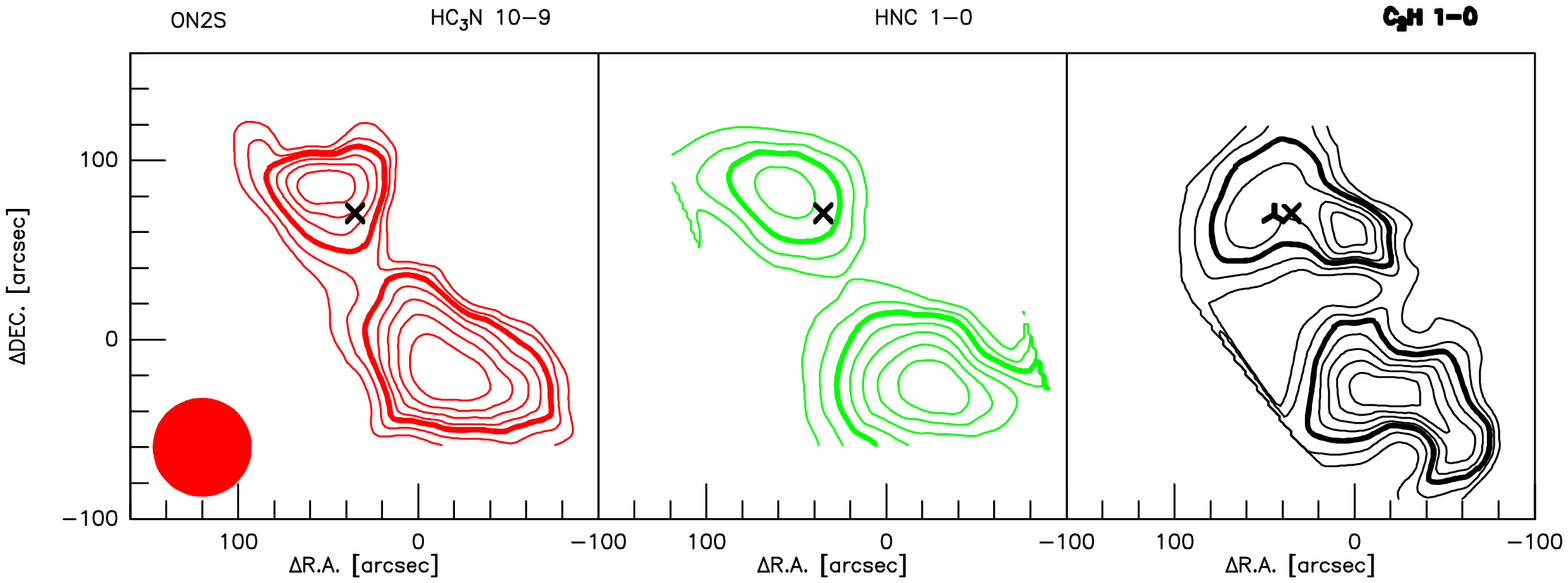}
\vspace*{-0.2 cm} \caption{\label{intensity5} Contour maps of
\hcccn\ 10-9 (left pannel), HNC 1-0 (middle pannel) and \cch\ 1-0 (F
= 3/2, 2 $\rightarrow$ 1/2, 1) (right pannel) integrated intensity
superimposed on its 8.4 or 4.8 GHz continuum map in gray scale. (A
color version of this figure is available in the online journal.)}
\end{center}
\end{figure}

\clearpage

\begin{figure}[]
\begin{center}
\includegraphics[width=2.4in, angle=270]{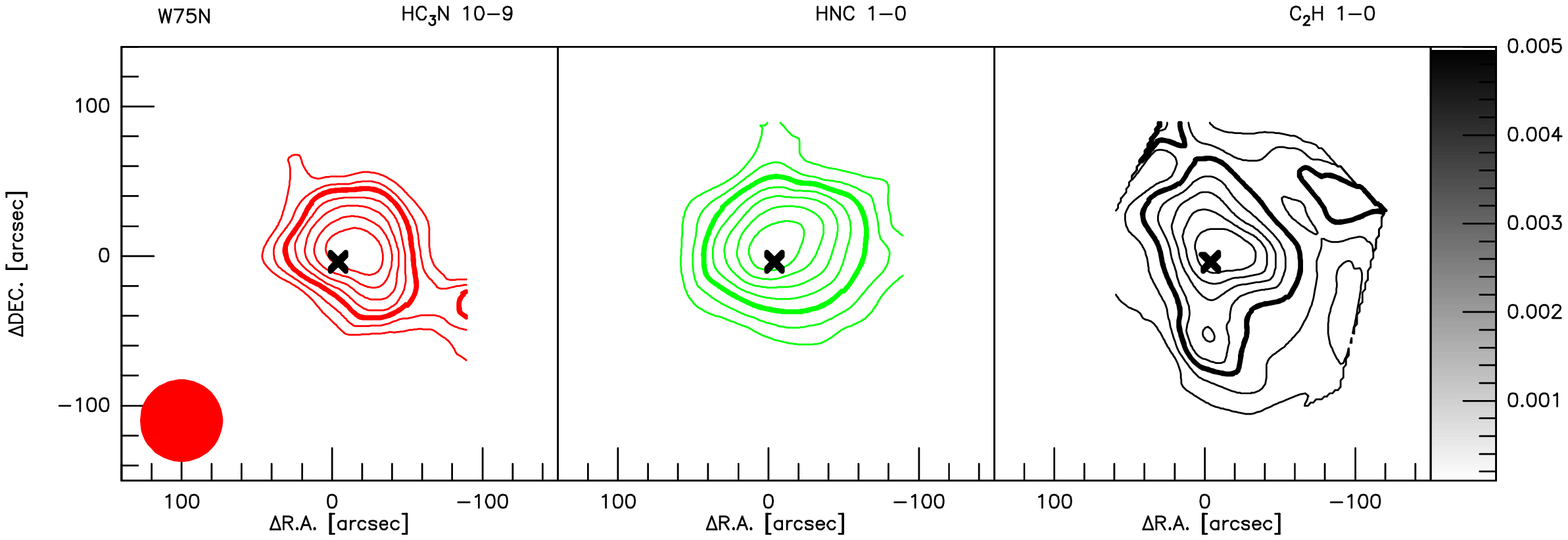}
\includegraphics[width=2.4in, angle=270]{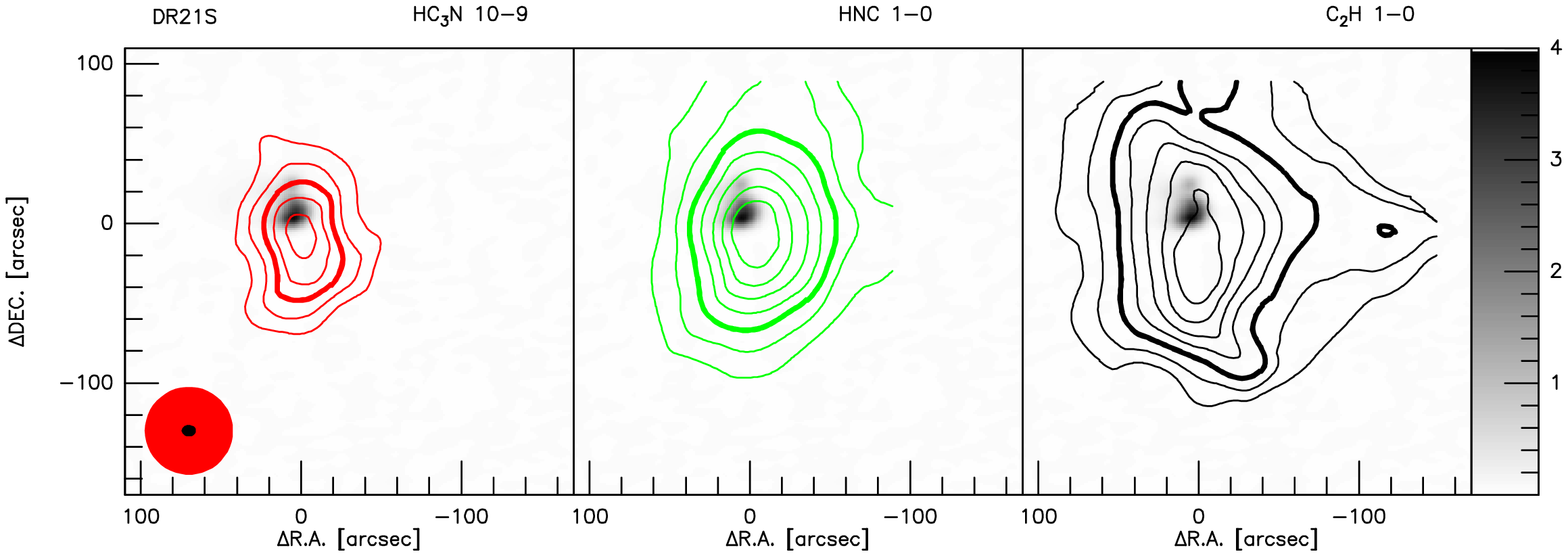}
\includegraphics[width=2.4in, angle=270]{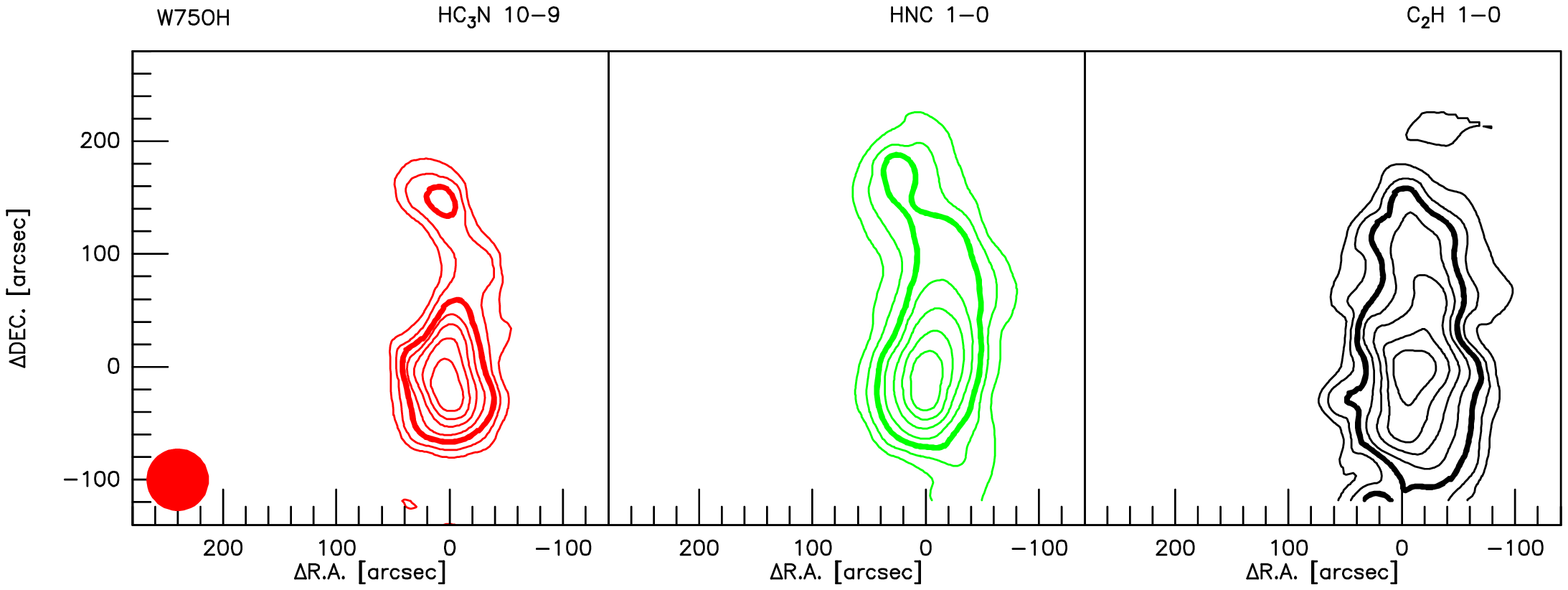}
\vspace*{-0.2 cm} \caption{\label{intensity6} Contour maps of
\hcccn\ 10-9 (left pannel), HNC 1-0 (middle pannel) and \cch\ 1-0 (F
= 3/2, 2 $\rightarrow$ 1/2, 1) (right pannel) integrated intensity
superimposed on its 8.4 or 4.8 GHz continuum map in gray scale. (A
color version of this figure is available in the online journal.)}
\end{center}
\end{figure}

\clearpage

\begin{figure}[]
\begin{center}
\includegraphics[width=2.4in, angle=270]{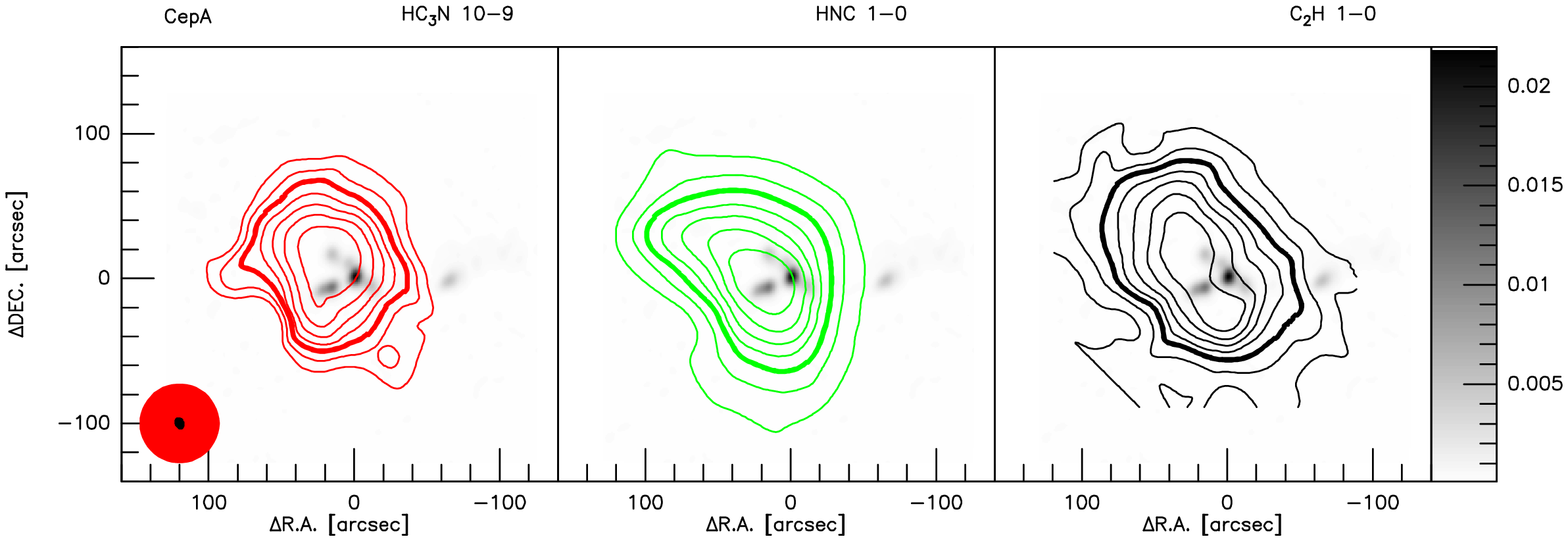}
\includegraphics[width=2.4in, angle=270]{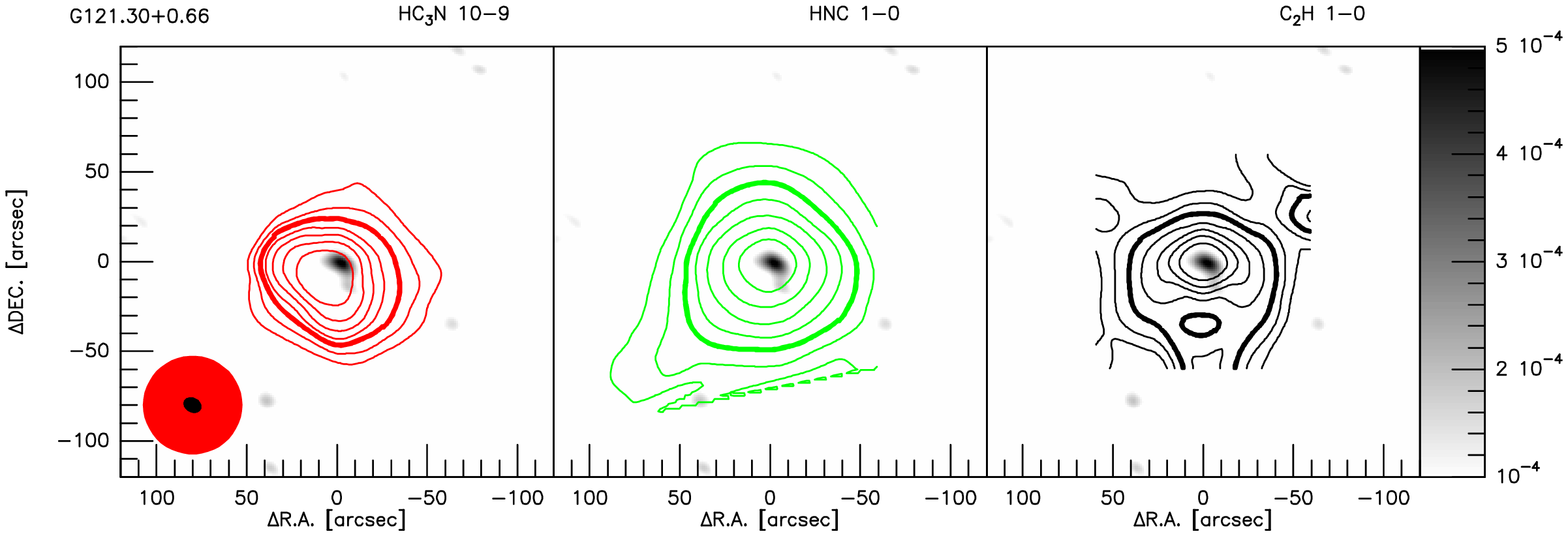}
\includegraphics[width=2.4in, angle=270]{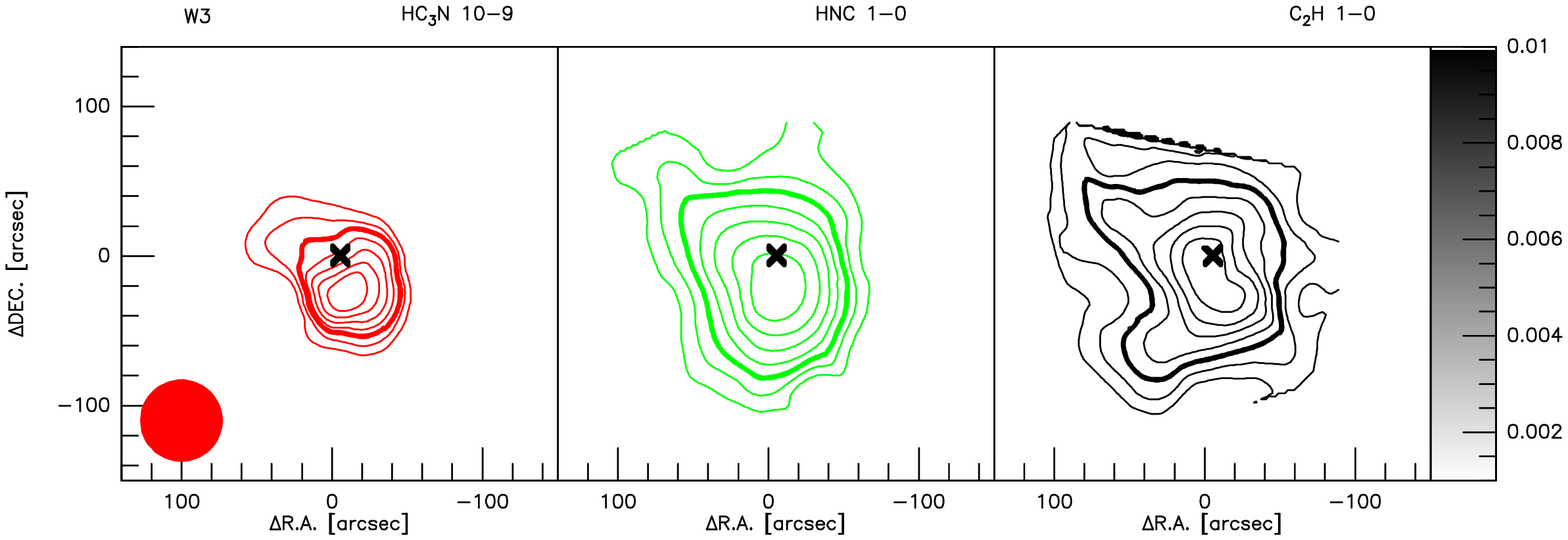}
\vspace*{-0.2 cm} \caption{\label{intensity7} Contour maps of
\hcccn\ 10-9 (left pannel), HNC 1-0 (middle pannel) and \cch\ 1-0 (F
= 3/2, 2 $\rightarrow$ 1/2, 1) (right pannel) integrated intensity
superimposed on its 8.4 or 4.8 GHz continuum map in gray scale. (A
color version of this figure is available in the online journal.)}
\end{center}
\end{figure}

\clearpage

\begin{figure}[]
\begin{center}
\includegraphics[width=2.4in, angle=270]{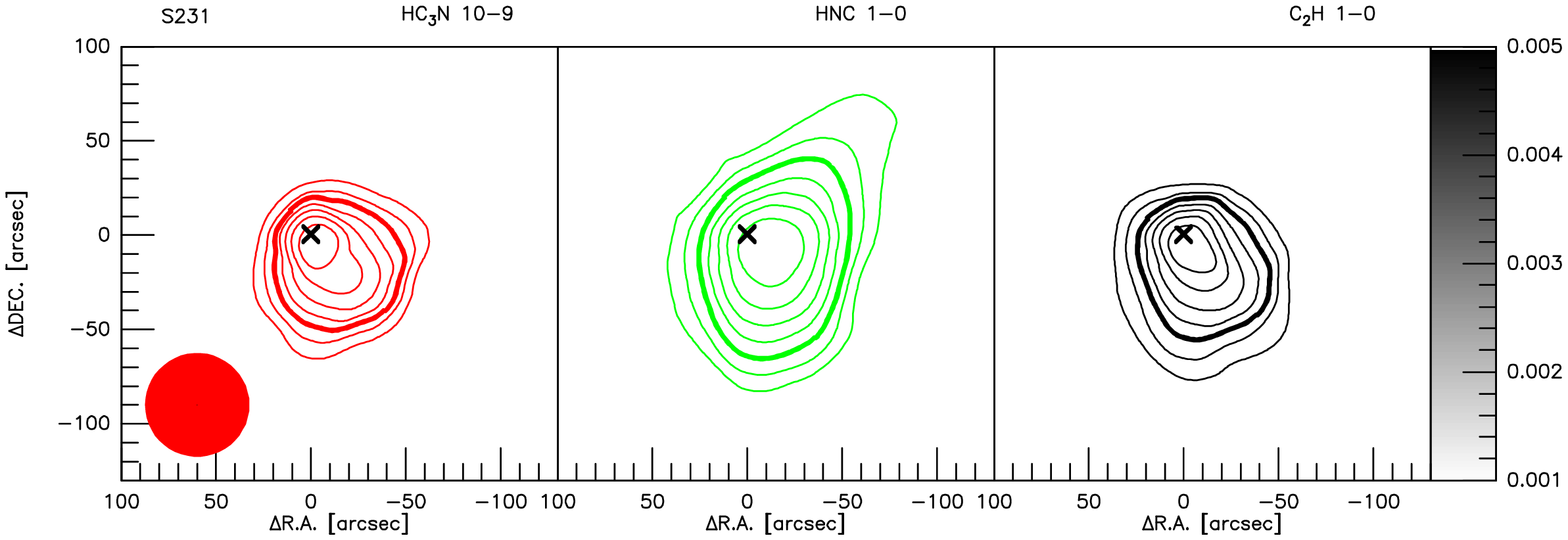}
\includegraphics[width=2.4in, angle=270]{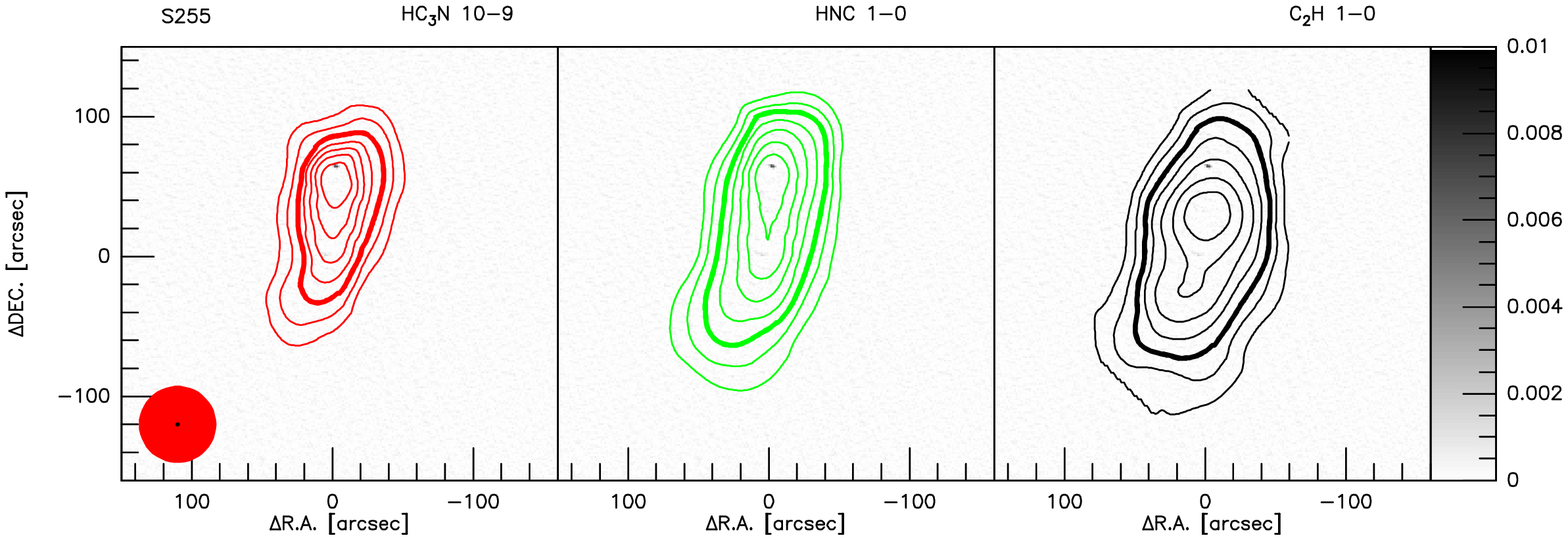}
\vspace*{-0.2 cm} \caption{\label{intensity8} Contour maps of
\hcccn\ 10-9 (left pannel), HNC 1-0 (middle pannel) and \cch\ 1-0 (F
= 3/2, 2 $\rightarrow$ 1/2, 1) (right pannel) integrated intensity
superimposed on its 8.4 or 4.8 GHz continuum map in gray scale. (A
color version of this figure is available in the online journal.)}
\end{center}
\end{figure}

\clearpage

\begin{figure}[]
\begin{center}
\includegraphics[width=2.4in, angle=270]{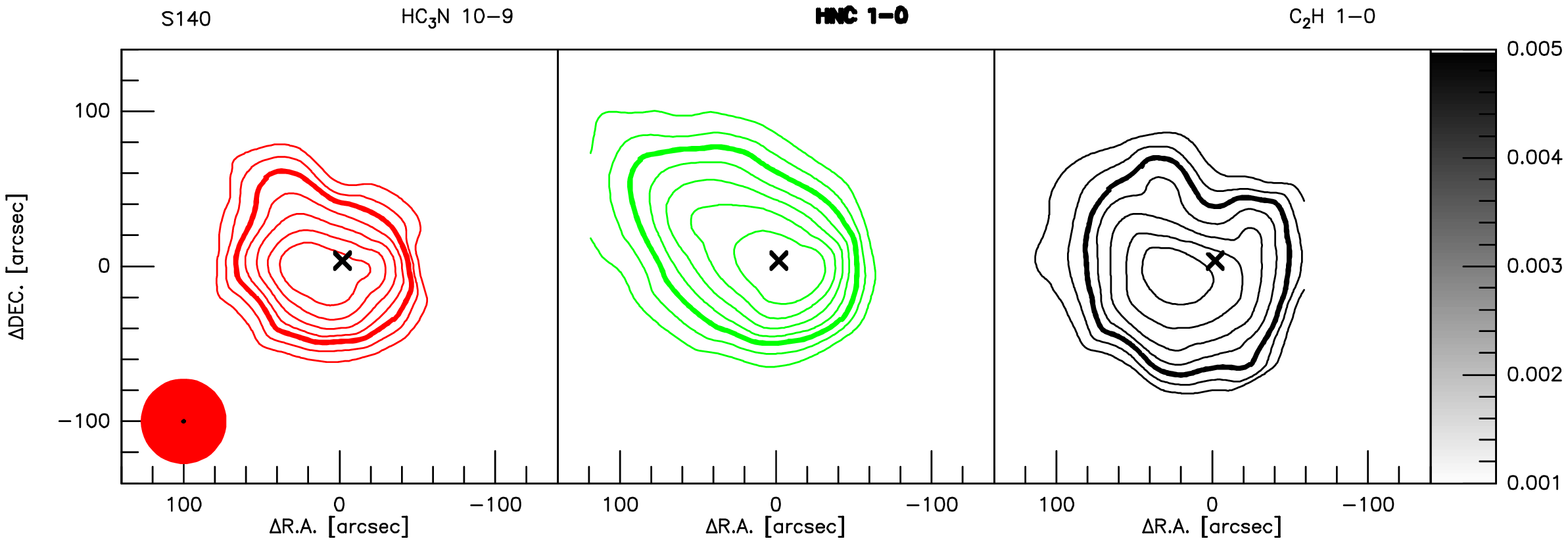}
\includegraphics[width=2.4in, angle=270]{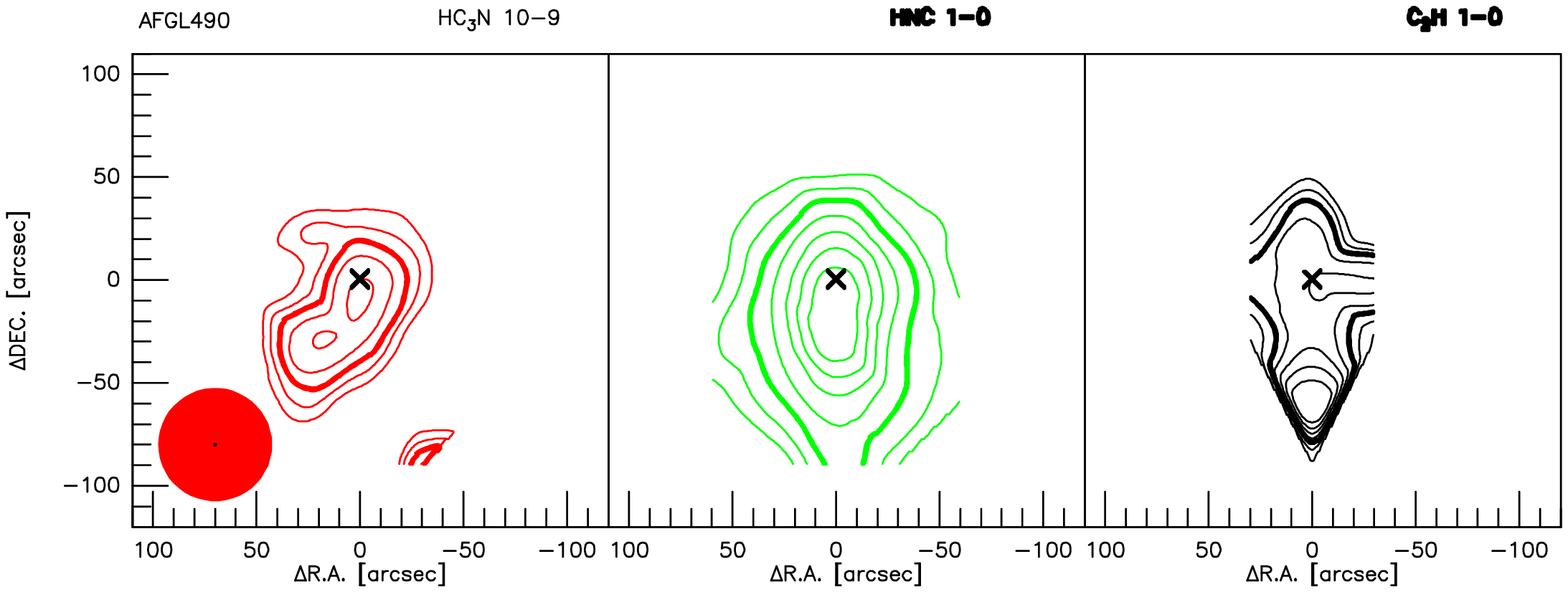}
\includegraphics[width=2.4in, angle=270]{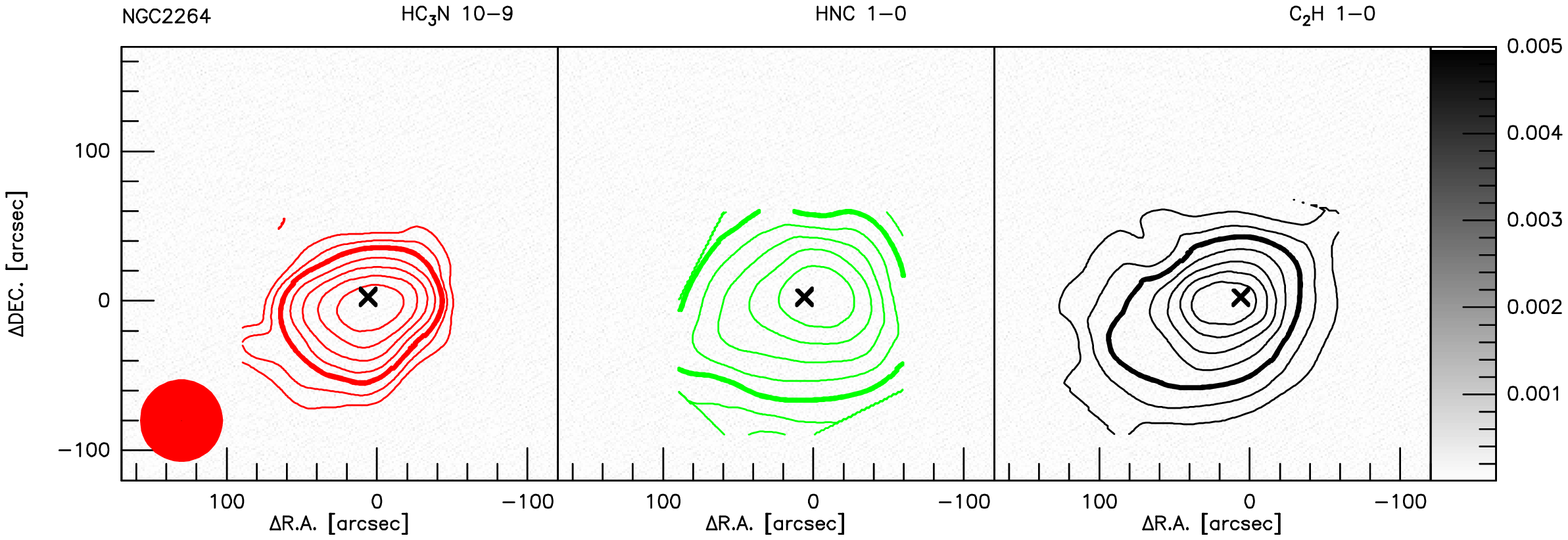}
\vspace*{-0.2 cm} \caption{\label{intensity9} Figure
\ref{intensity1} continued. (A color version of this figure is
available in the online journal.)}
\end{center}
\end{figure}

\clearpage

\begin{figure}[]
\begin{center}
\includegraphics[width=3.0in]{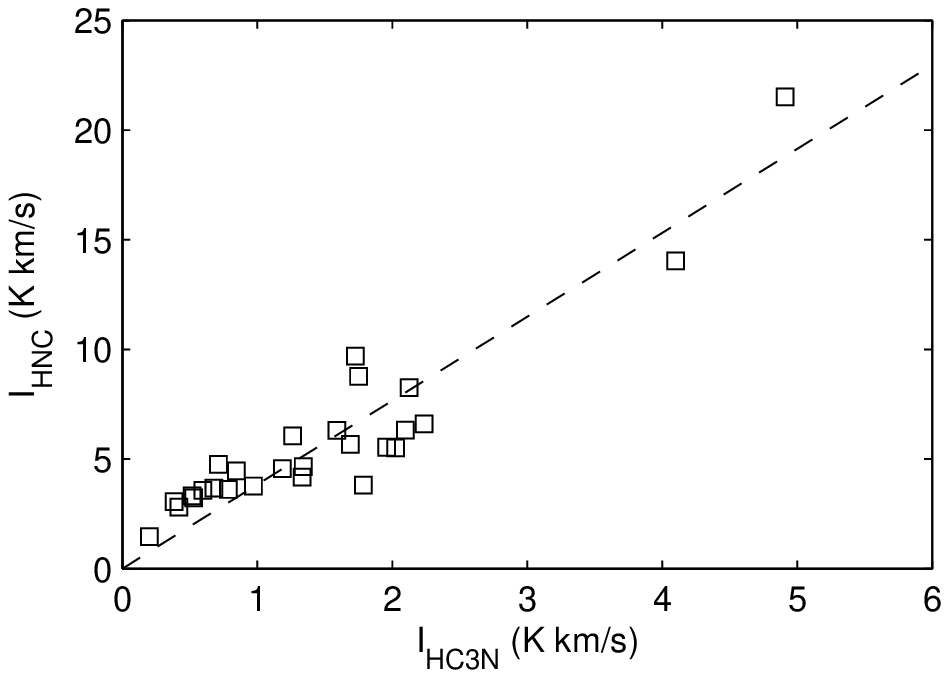}
\includegraphics[width=3.0in]{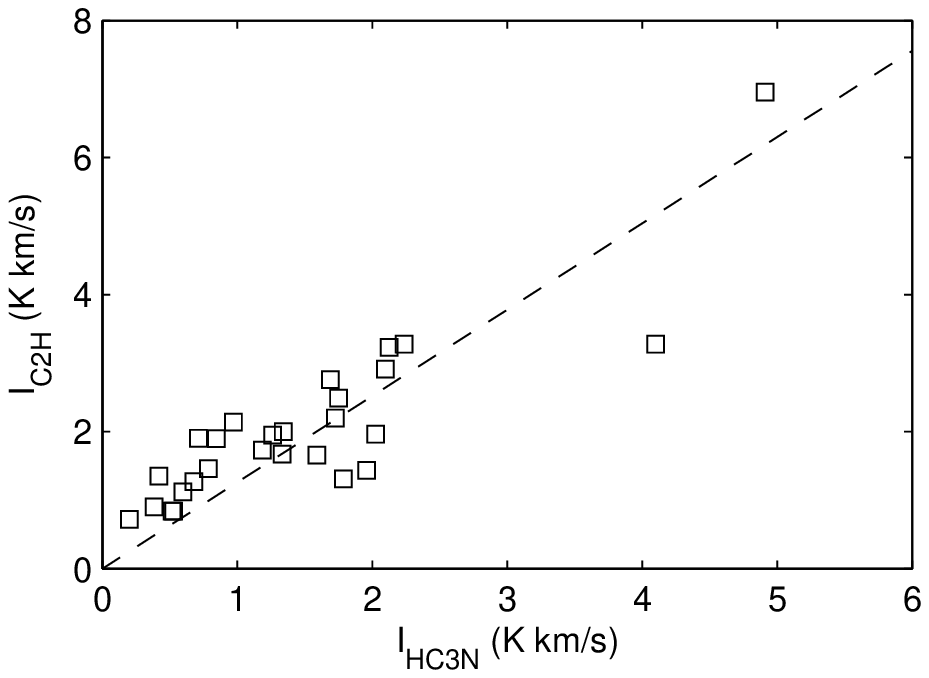}
\vspace*{-0.2 cm} \caption{\label{intencomp} Left: Plot of
integrated intensity of HNC averaged over the entire regions versus
those of \hcccn. Right: same for \cch\ vs. \hcccn. The dashed lines
represent the least-square fitting results, with slopes of 3.8 and
1.3. }
\end{center}
\end{figure}

\clearpage

\begin{figure}[]
\begin{center}
\includegraphics[width=3.0in]{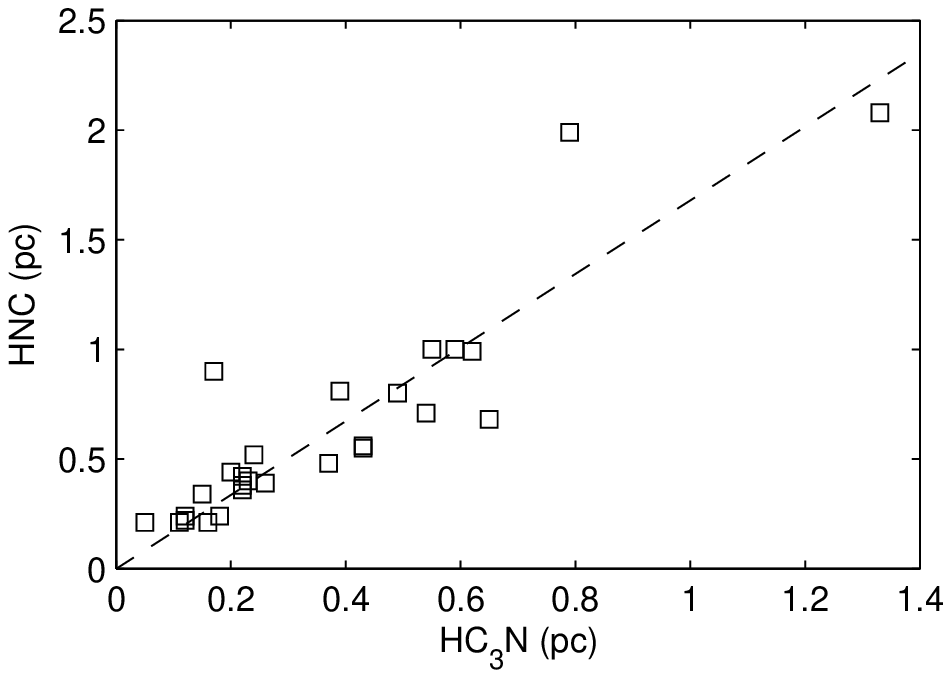}
\includegraphics[width=3.0in]{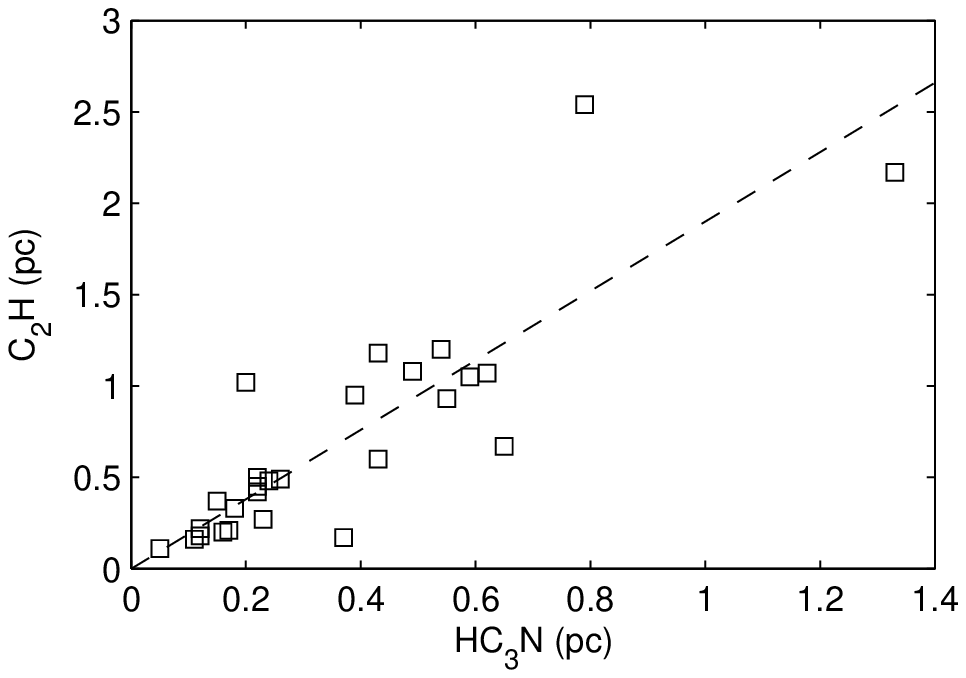}
\vspace*{-0.2 cm} \caption{\label{sizecomp} Plot of FWHM size of HNC
versus \hcccn\ (left) and \cch\ versus \hcccn\ (right), and the
dashed line represents the least-squares fit result to \cch\ vs.
\hcccn.}
\end{center}
\end{figure}

\clearpage

%%\begin{figure}[]            %  not necessary
%%\begin{center}
%%\includegraphics[width=7.0in]{historadius.eps}
%%\vspace*{-0.2 cm} \caption{\label{historadius} Distribution of FWHM
%%radius of \hcccn, HNC and \cch. The dotted vertical line represent
%%the median of FWHM radius.}
%%\end{center}
%%\end{figure}
%%
%%\clearpage

%\begin{figure}[]
%\begin{center}
%\includegraphics[width=3.0in]{offpeakhc3n.eps}
%\vspace*{-0.2 cm} \caption{\label{offhc3n} Distribution of angular
%separation between peak of radio continuum emission and \hcccn\
%emission.}
%\end{center}
%\end{figure}
%
%\clearpage

\begin{figure}[]
\begin{center}
\includegraphics[width=3.5in, angle=270]{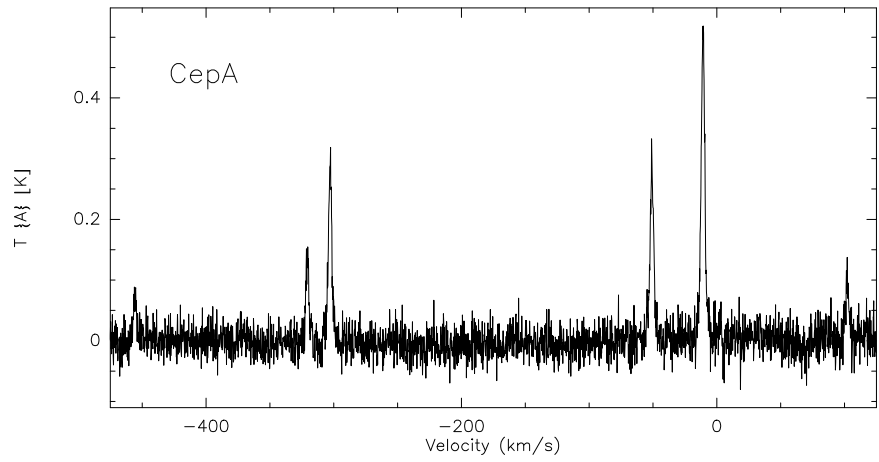}
\vspace*{-0.2 cm} \caption{\label{specc2h} Observed spectra of \cch\
1-0 toward Cep A, averaged over the whole emitting region.}
\end{center}
\end{figure}

\clearpage

\begin{figure}[]
\begin{center}
\includegraphics[width=2.0in, angle=270]{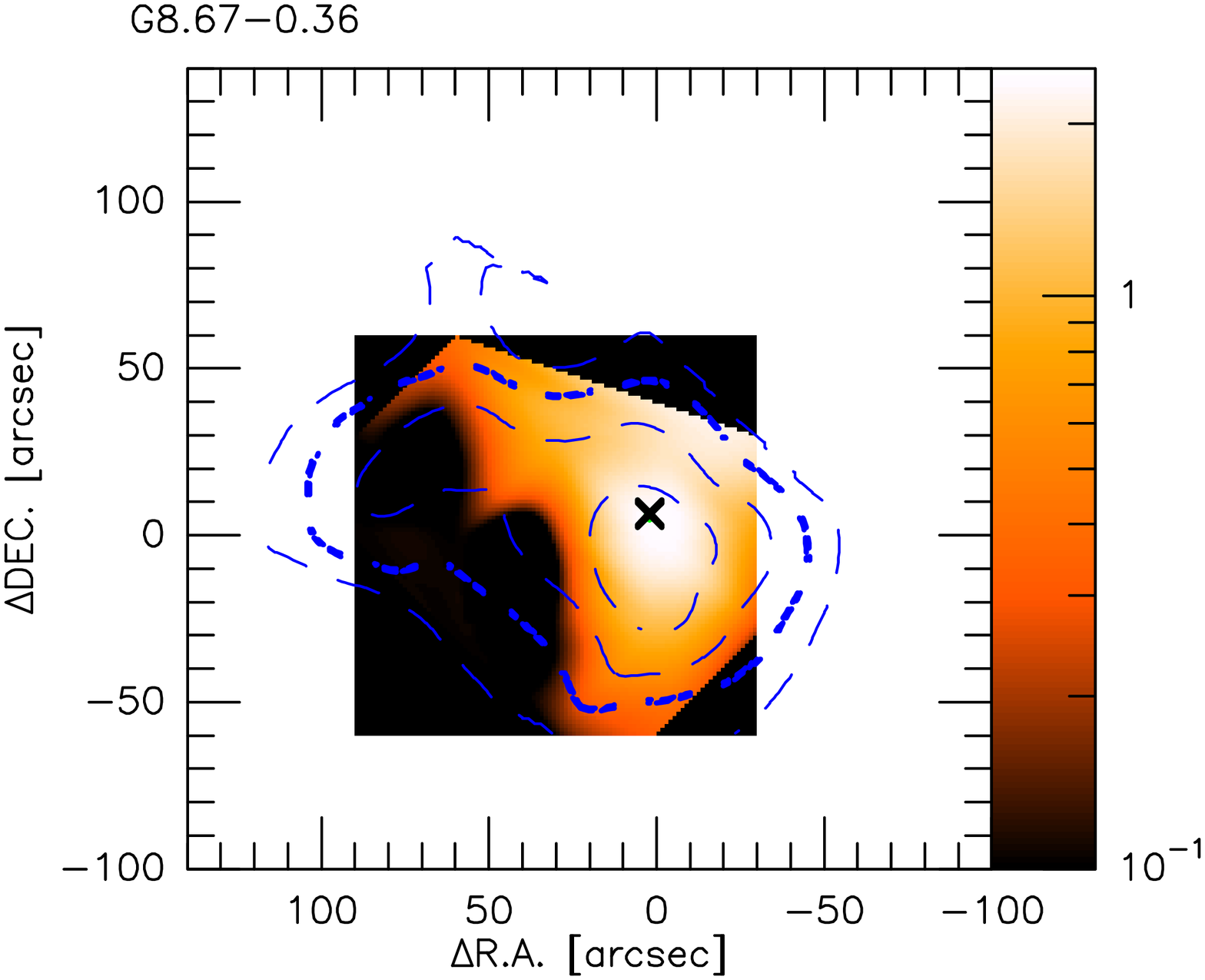}
\includegraphics[width=2.0in, angle=270]{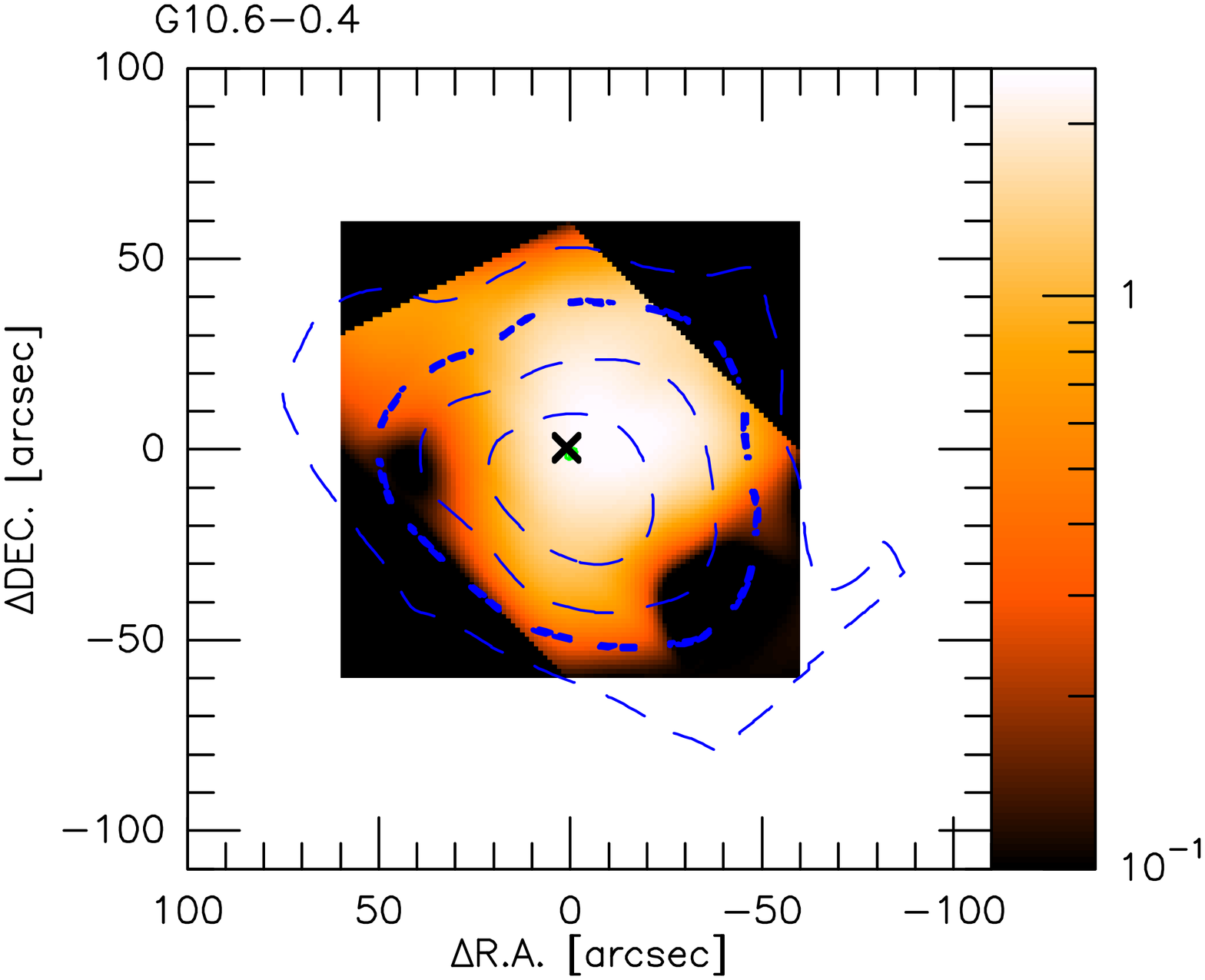}
\includegraphics[width=2.0in, angle=270]{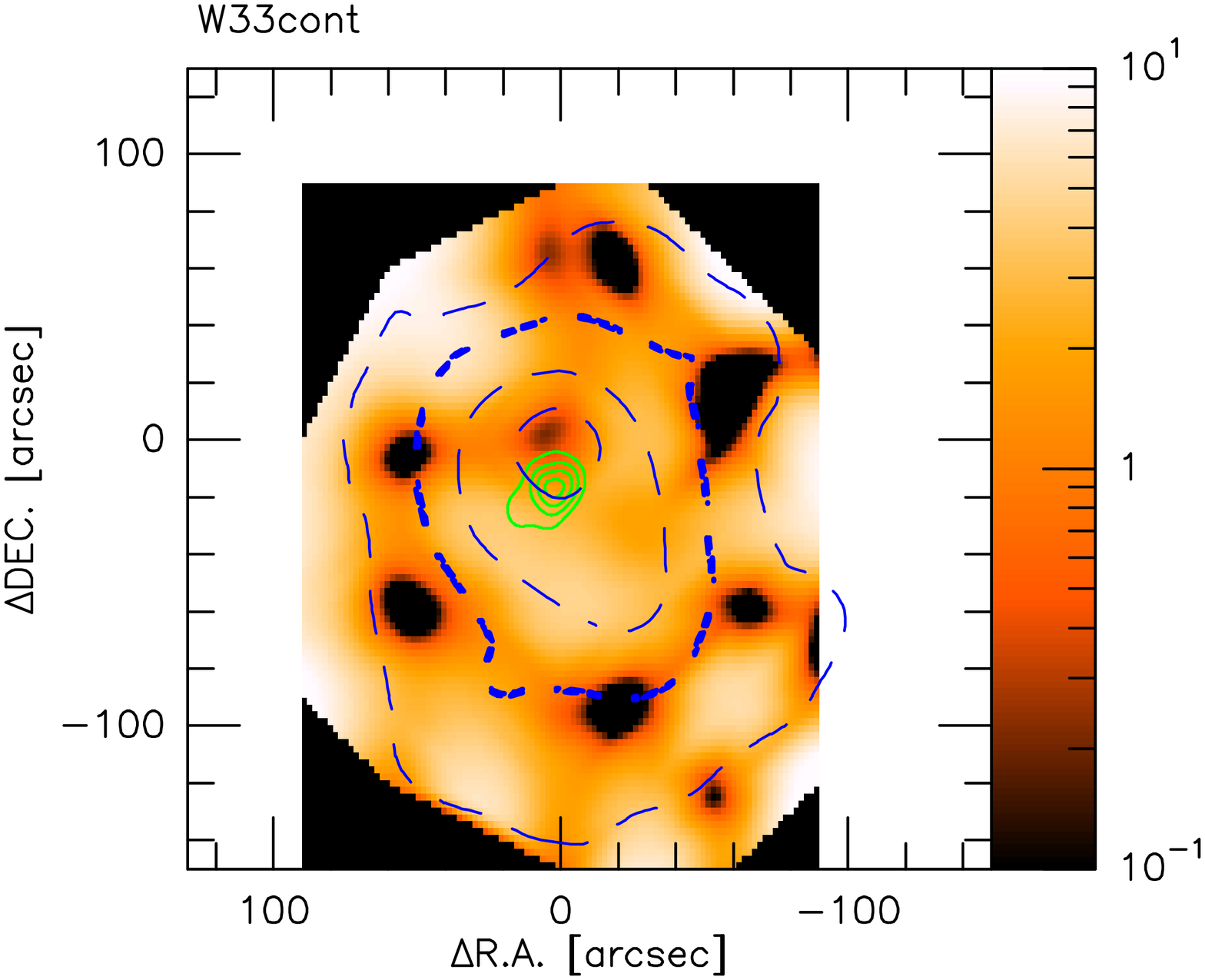}
\includegraphics[width=2.0in, angle=270]{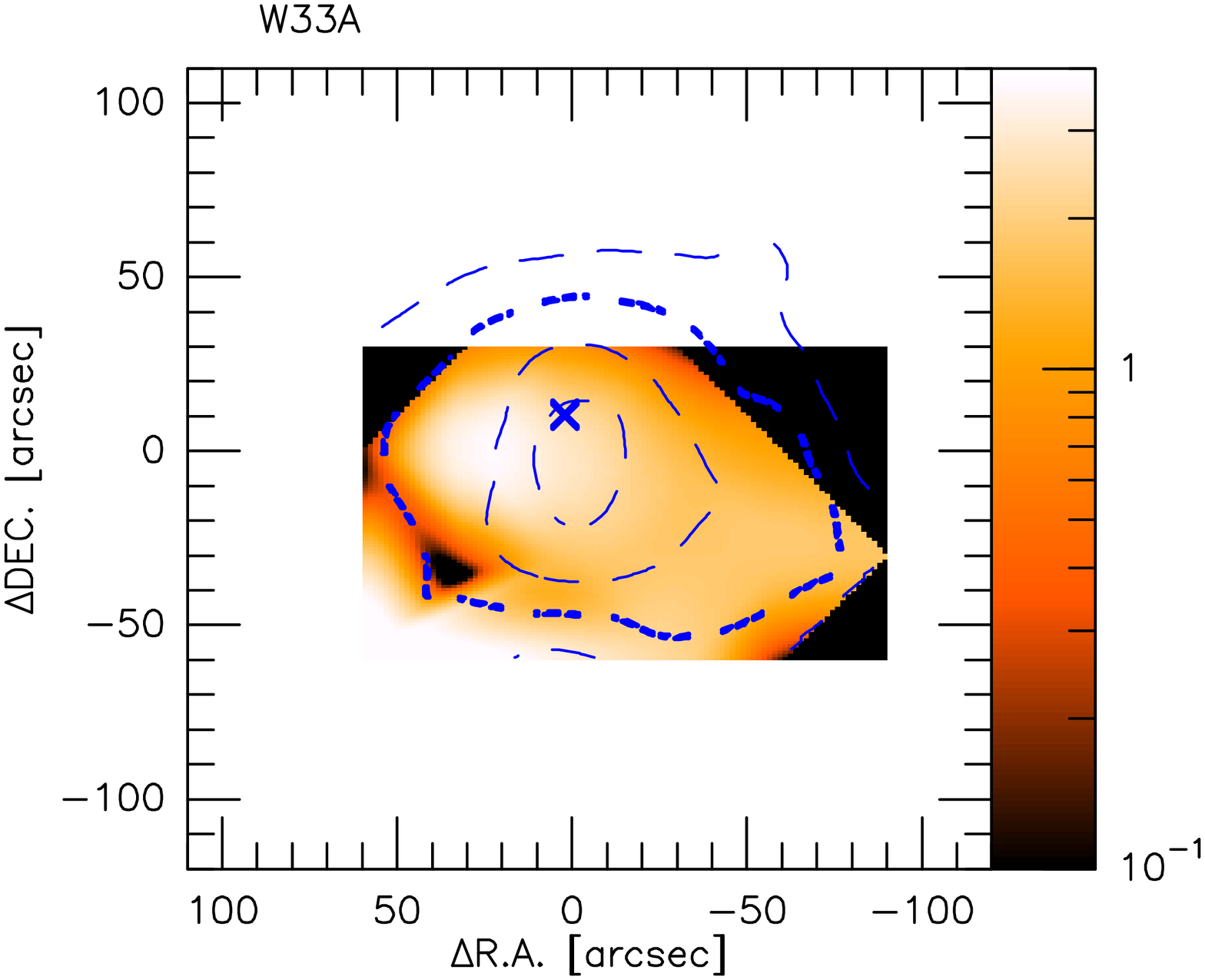}
\includegraphics[width=2.0in, angle=270]{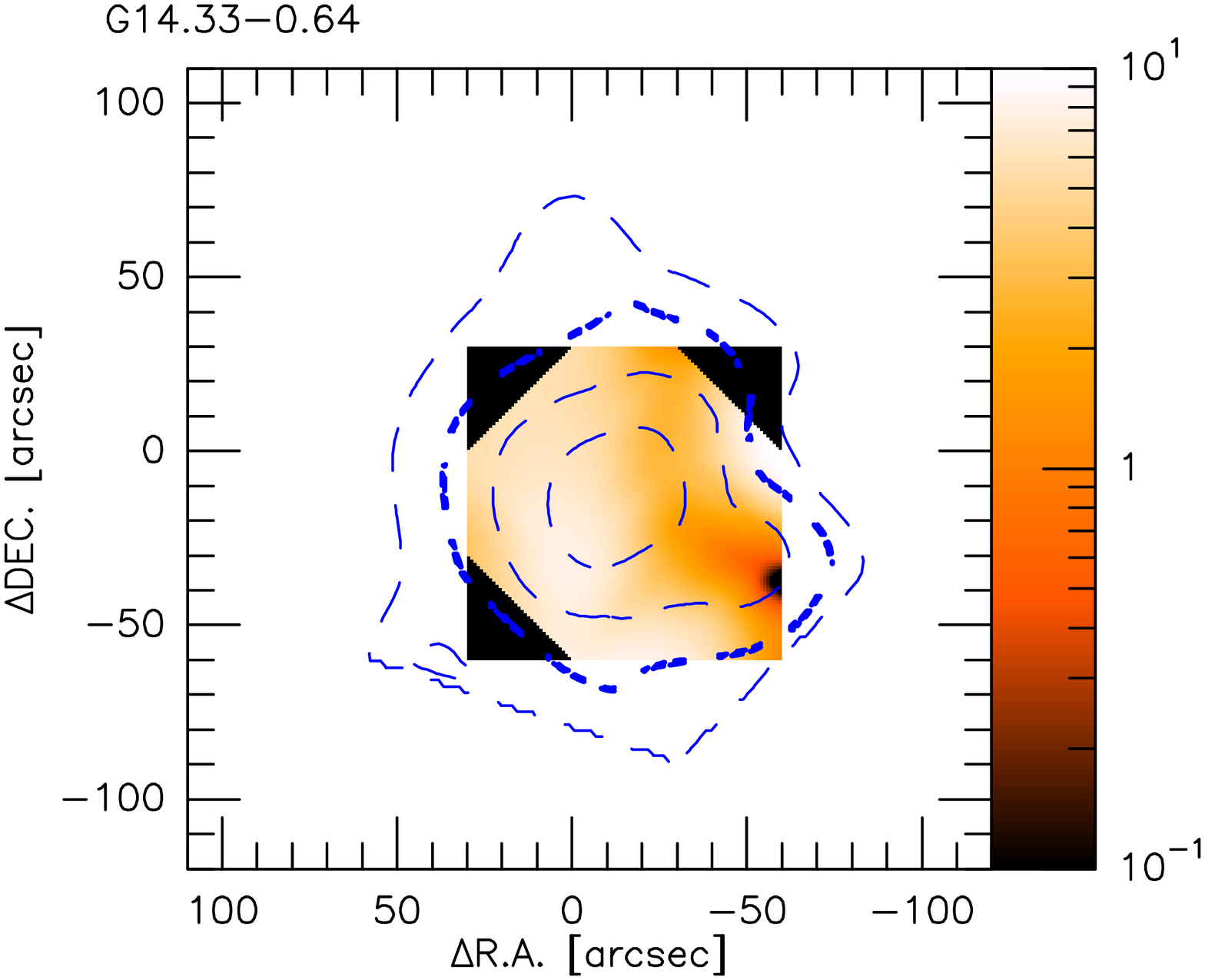}
\includegraphics[width=2.0in, angle=270]{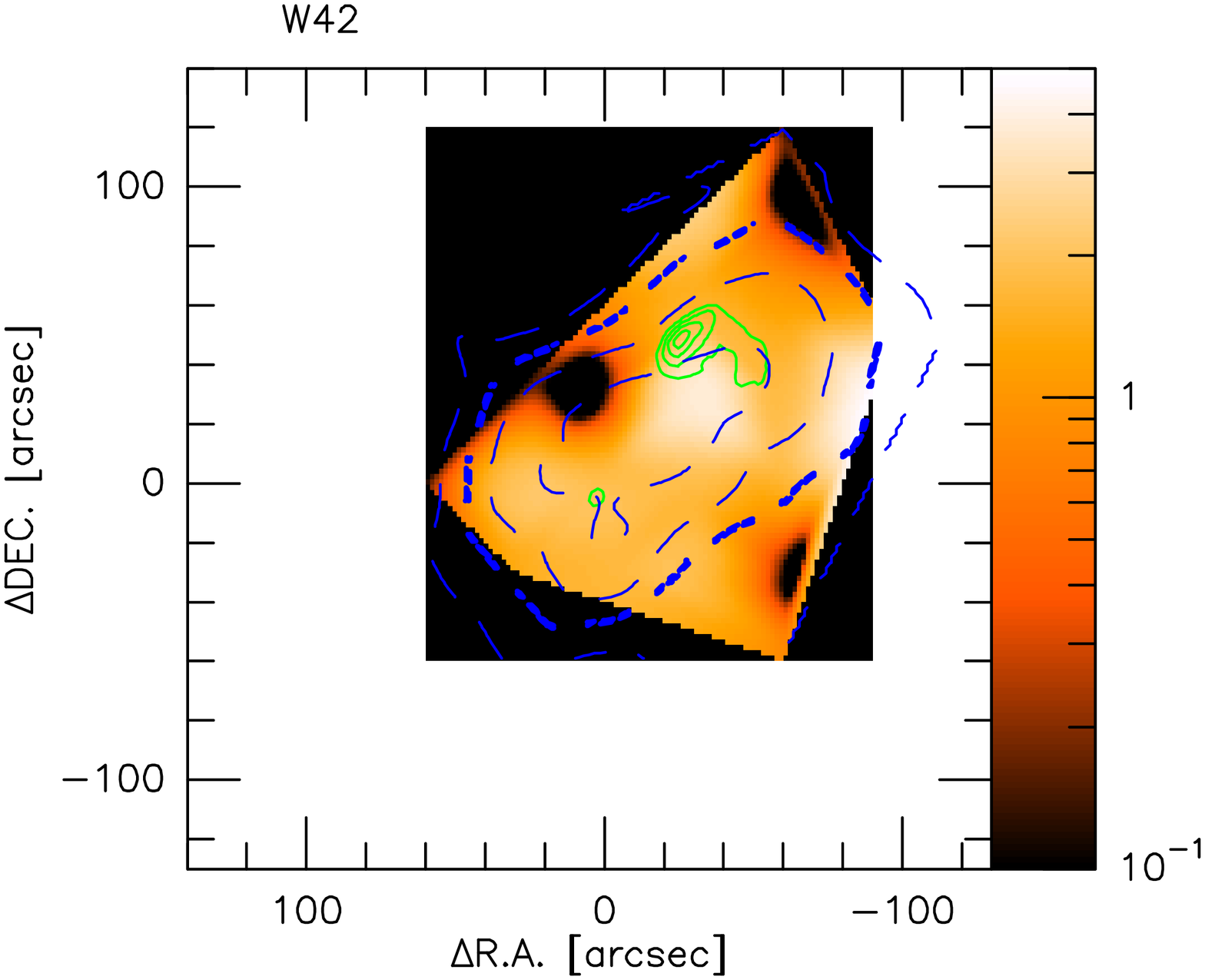}
\includegraphics[width=2.0in, angle=270]{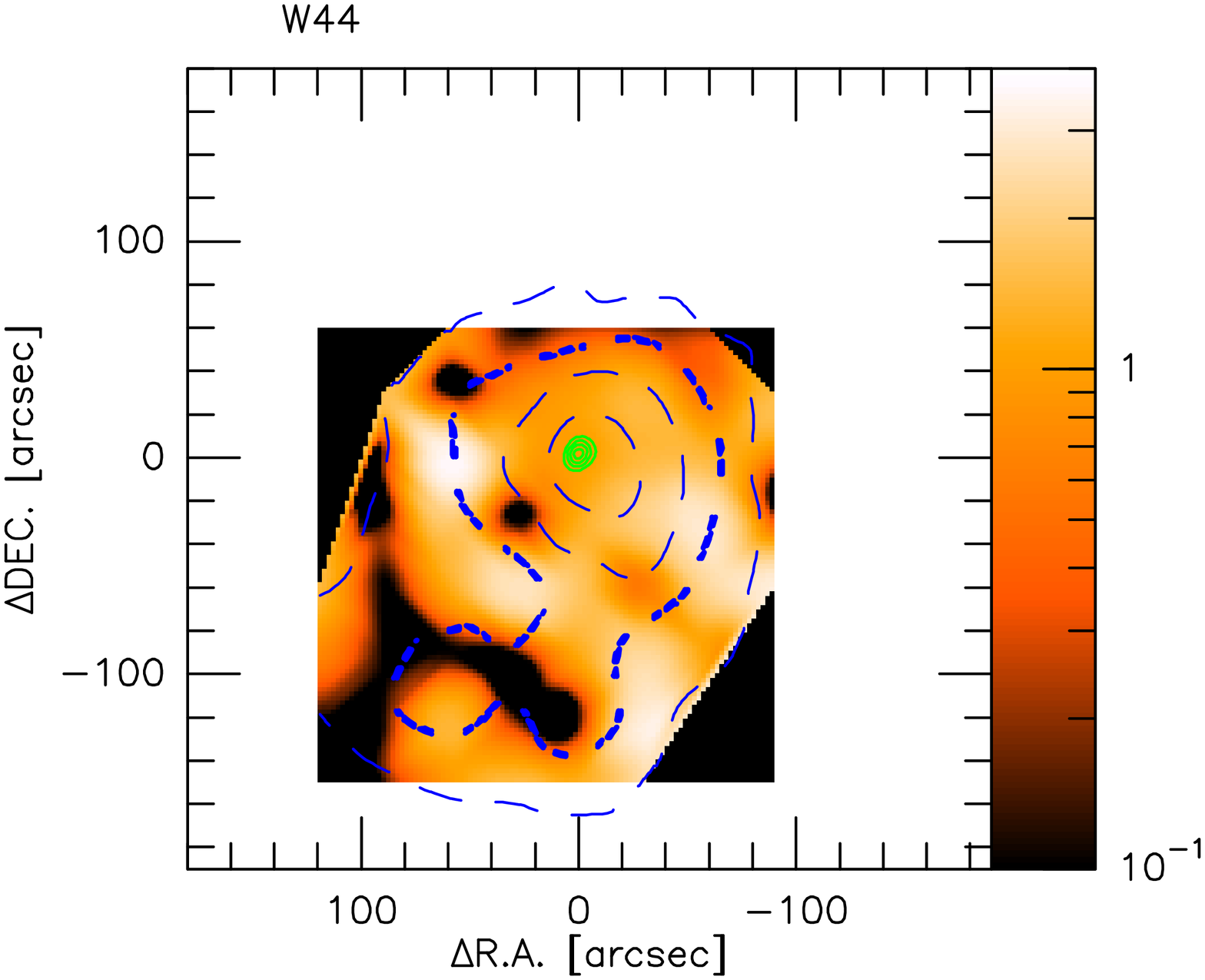}
\includegraphics[width=2.0in, angle=270]{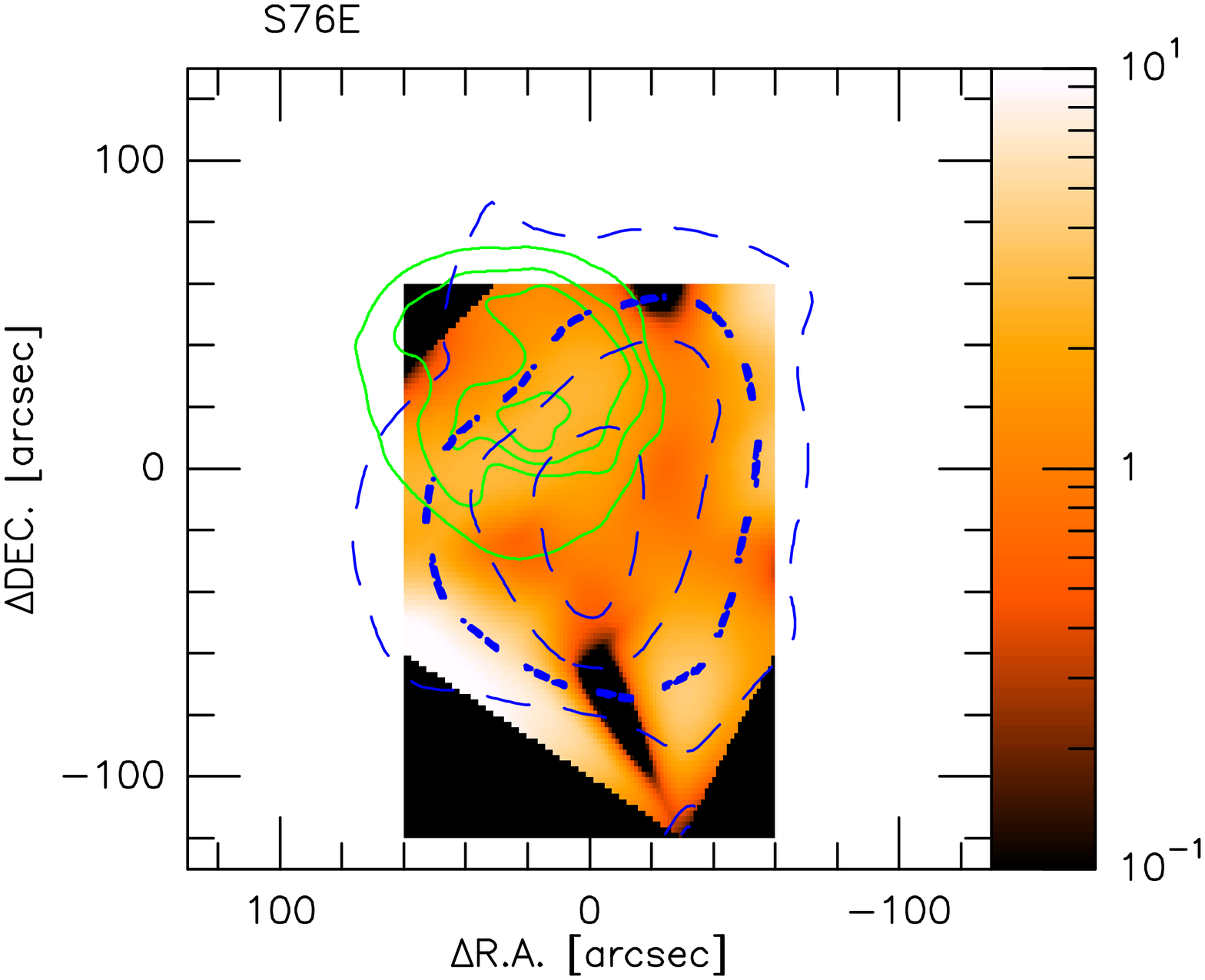}
\vspace*{-0.2 cm} \caption{\label{opdepth1} Contour maps \cch\ 1-0
(J = 3/2$\rightarrow$1/2, F = 2$\rightarrow$1) (dashed line)
integrated intensity and radio continuum emission (solid line)
superimposed on the optical depth of \cch\ 1-0 (J =
3/2$\rightarrow$1/2, F = 2$\rightarrow$1) in gray scale. Dashed
lines represent contour levels of 30\%, 50\%, 70\% and 90\% of the
peak intensity of \cch\ emission, reported in Table 3. Solid lines
represent contour levels of 30\%, 50\%, 70\% and 90\% of the peak
intensity of radio continuum emission, reported in Table 5. Crosses
are used to mark position of HII regions in case that the HII
regions are too small to be recognized in the figure. (A color
version of this figure is available in the online journal.)}
\end{center}
\end{figure}

\clearpage

\begin{figure}[]
\begin{center}
\includegraphics[width=2.0in, angle=270]{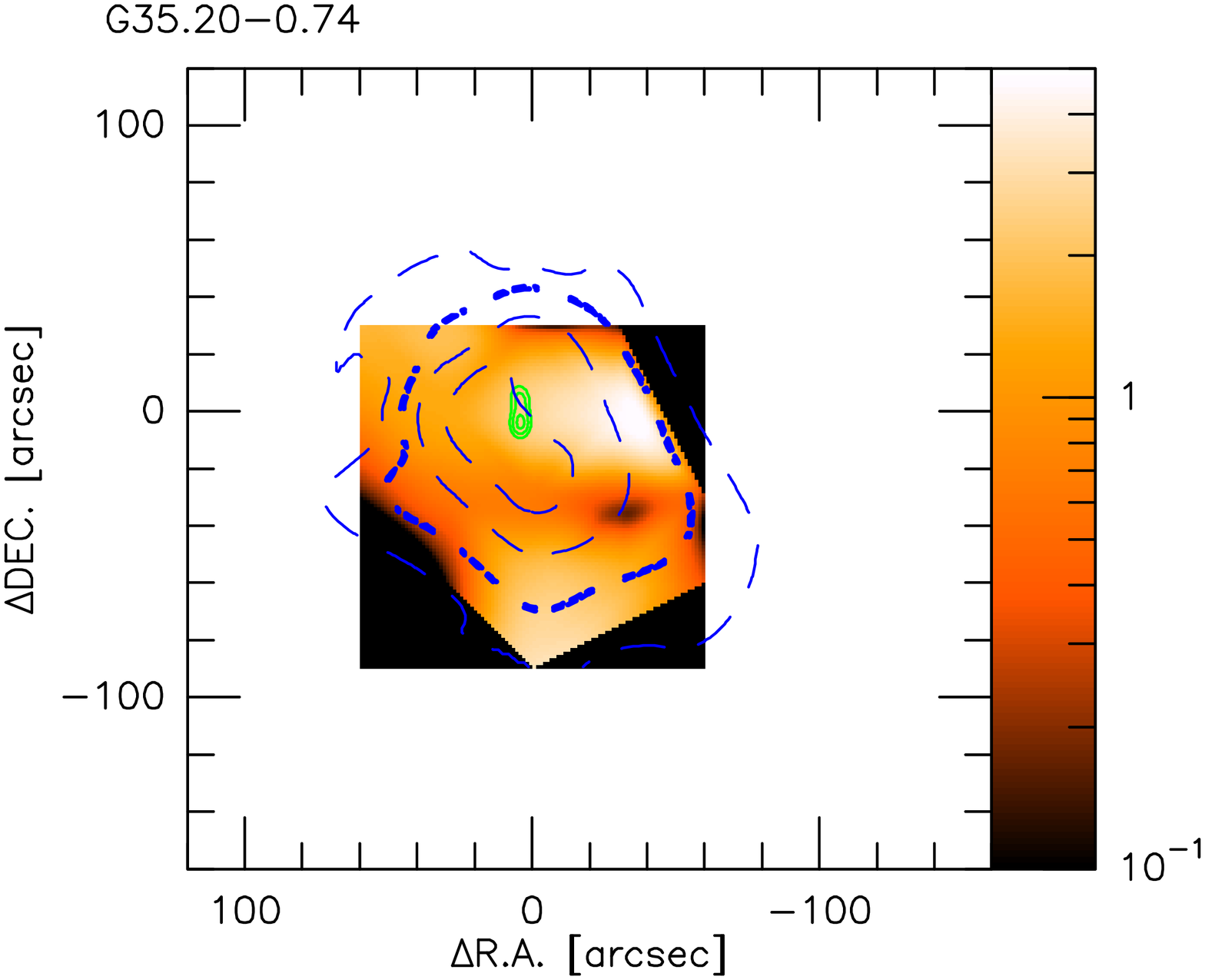}
\includegraphics[width=2.0in, angle=270]{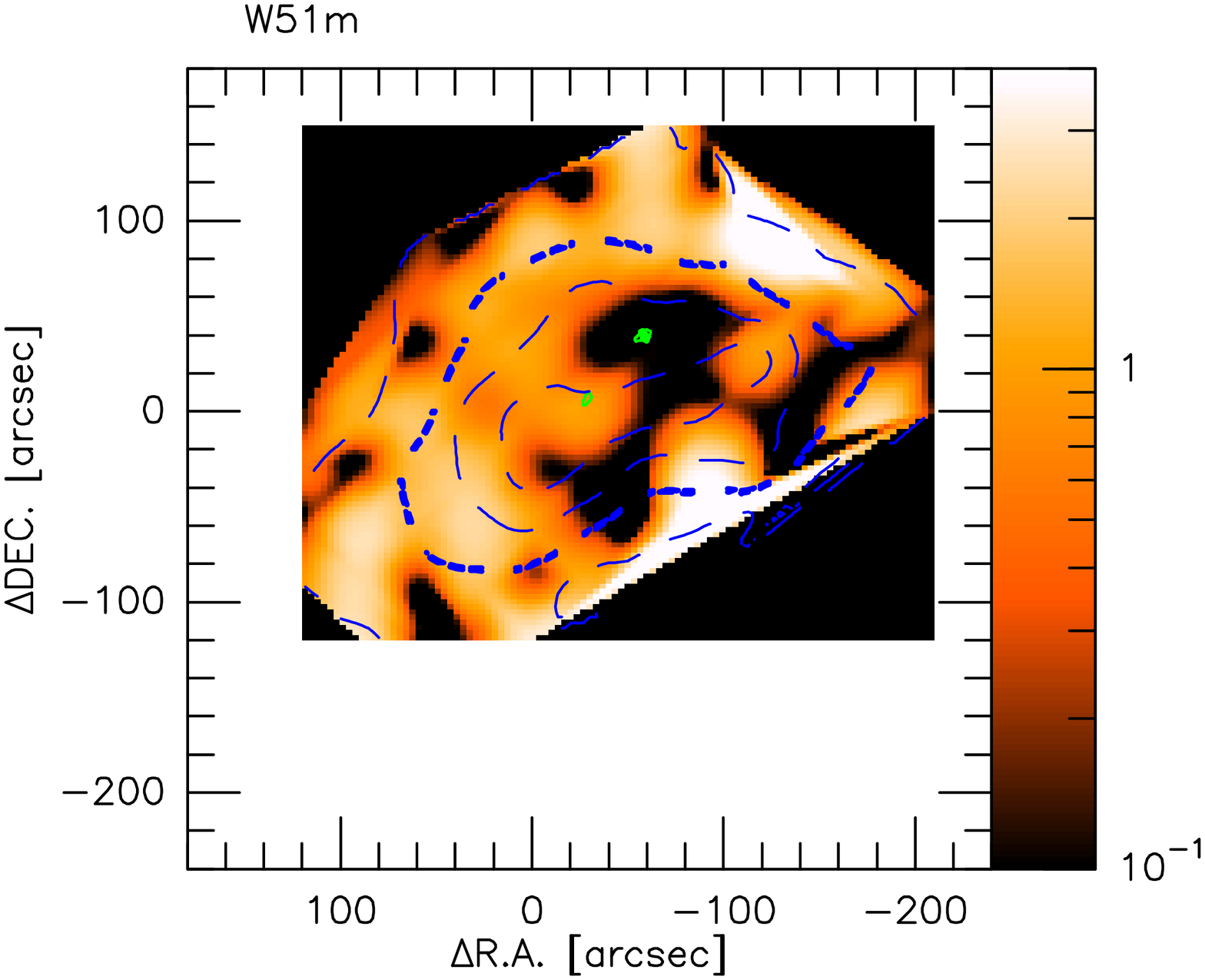}
\includegraphics[width=2.0in, angle=270]{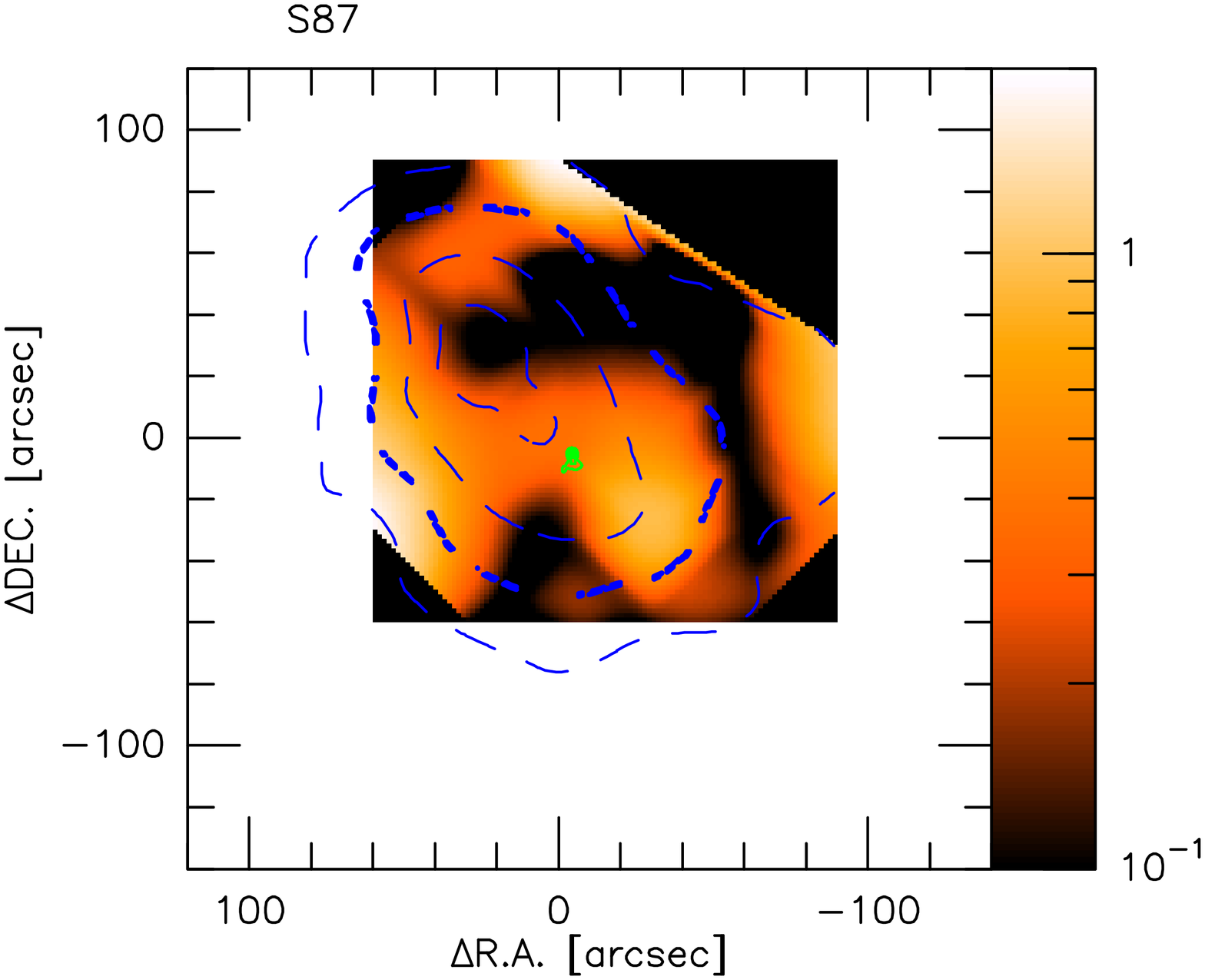}
\includegraphics[width=2.0in, angle=270]{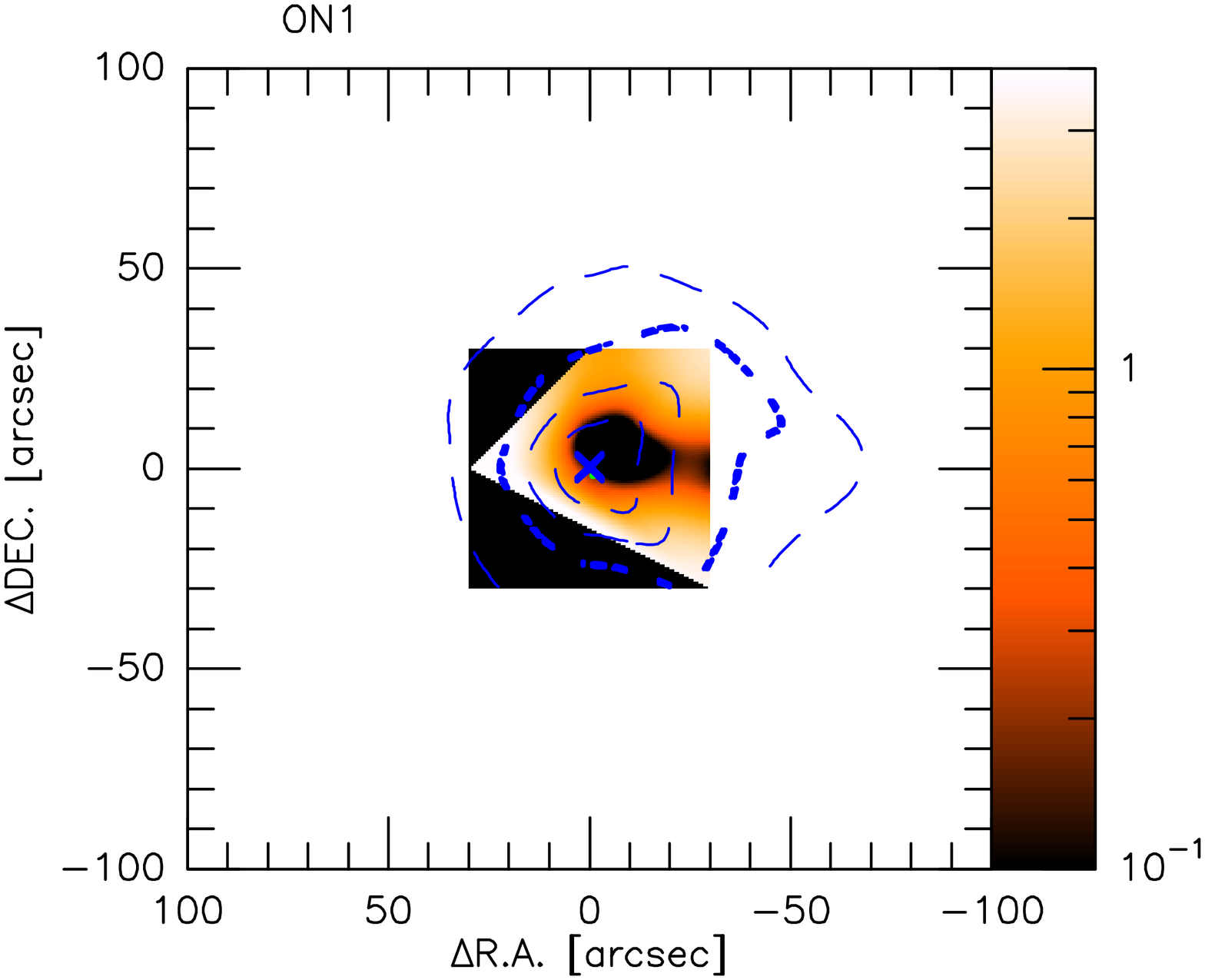}
\includegraphics[width=2.0in, angle=270]{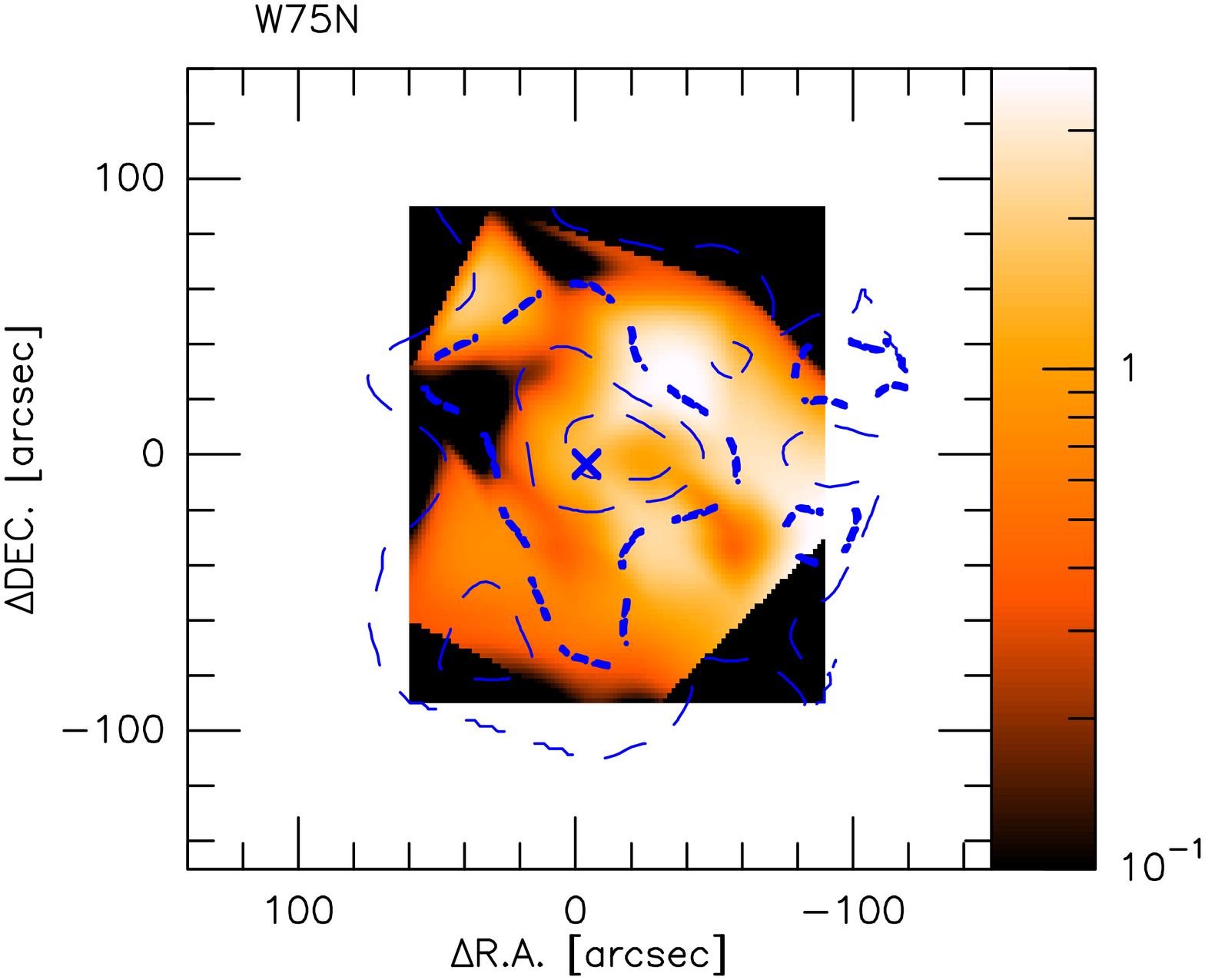}
\includegraphics[width=2.0in, angle=270]{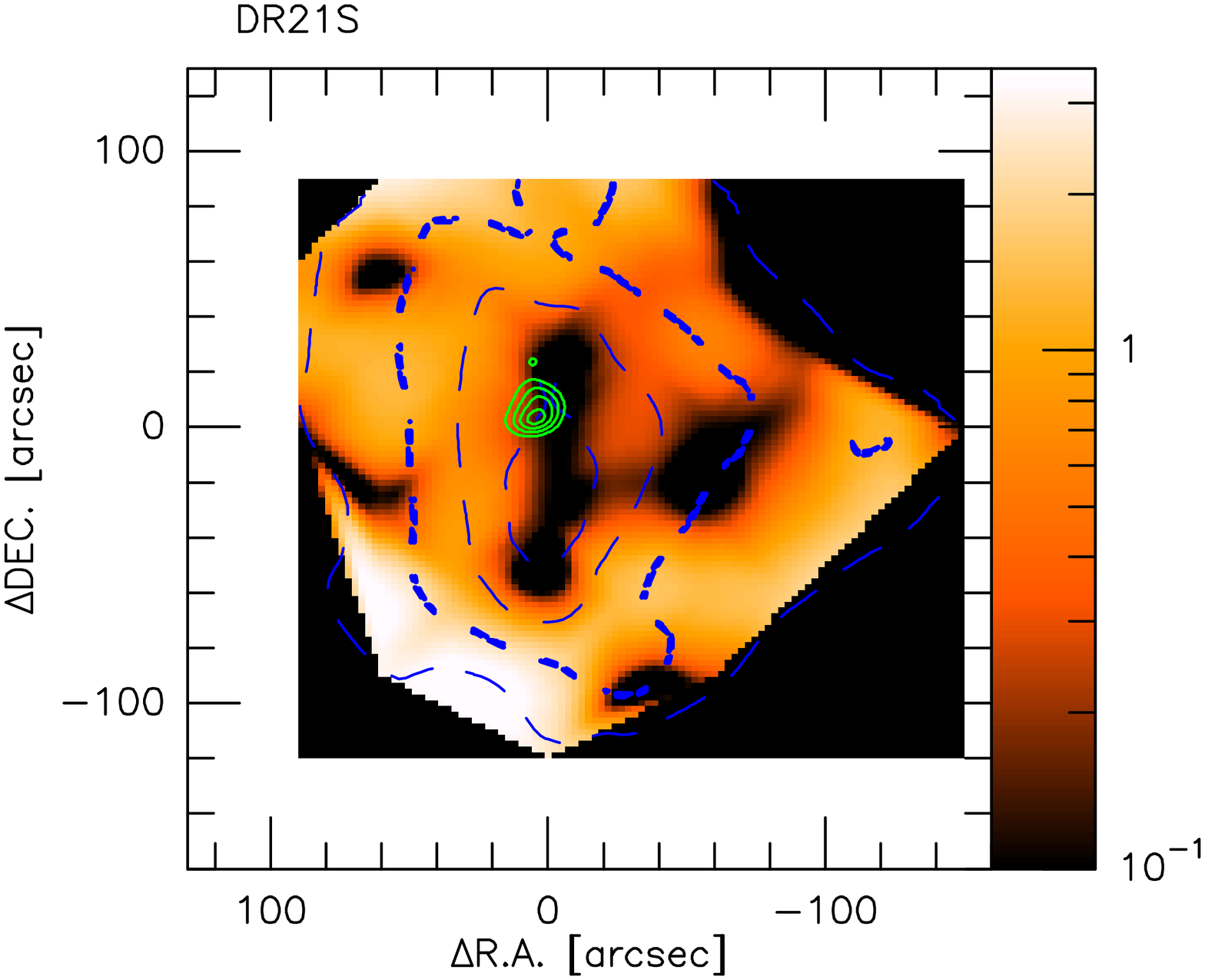}
\includegraphics[width=2.0in, angle=270]{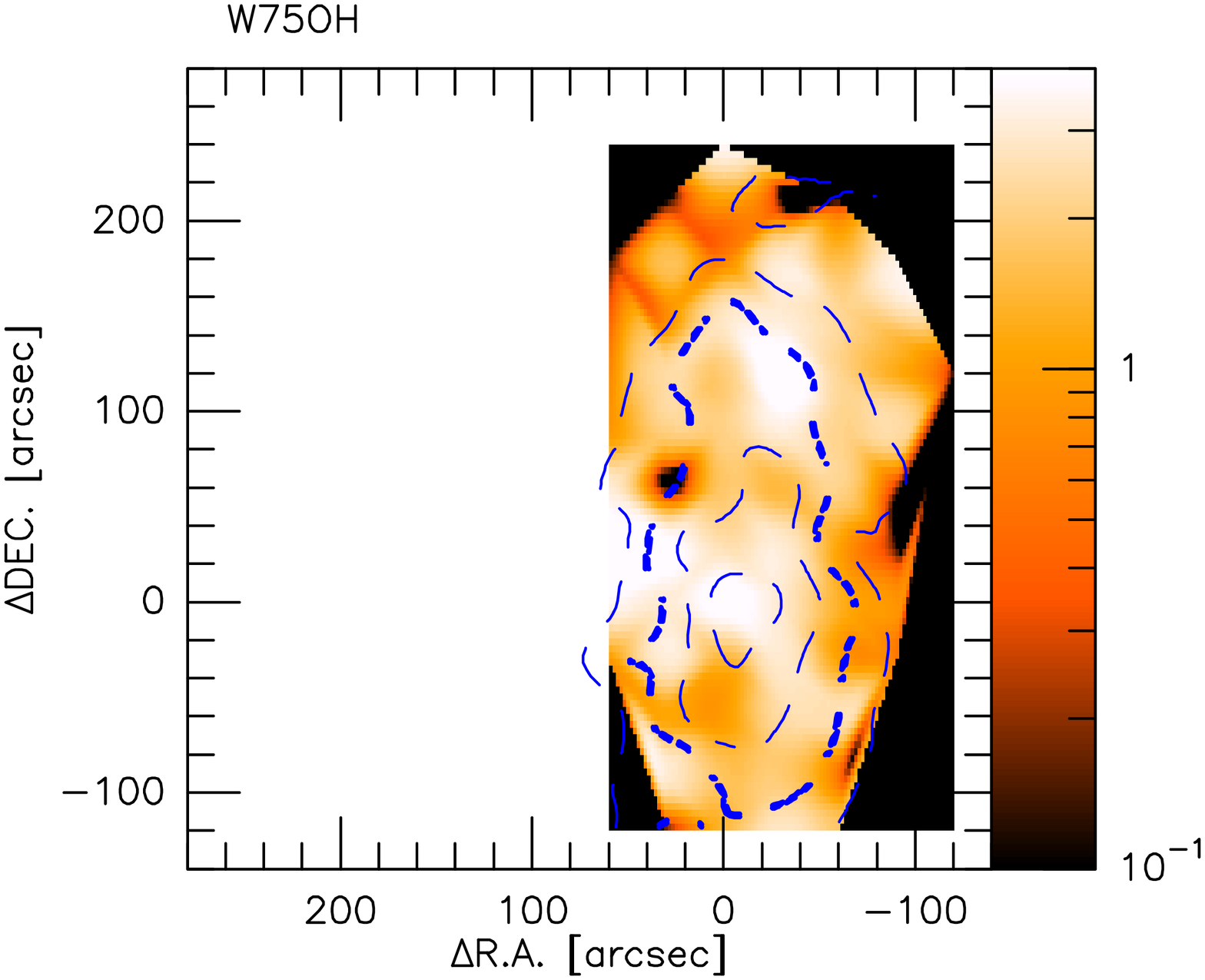}
\includegraphics[width=2.0in, angle=270]{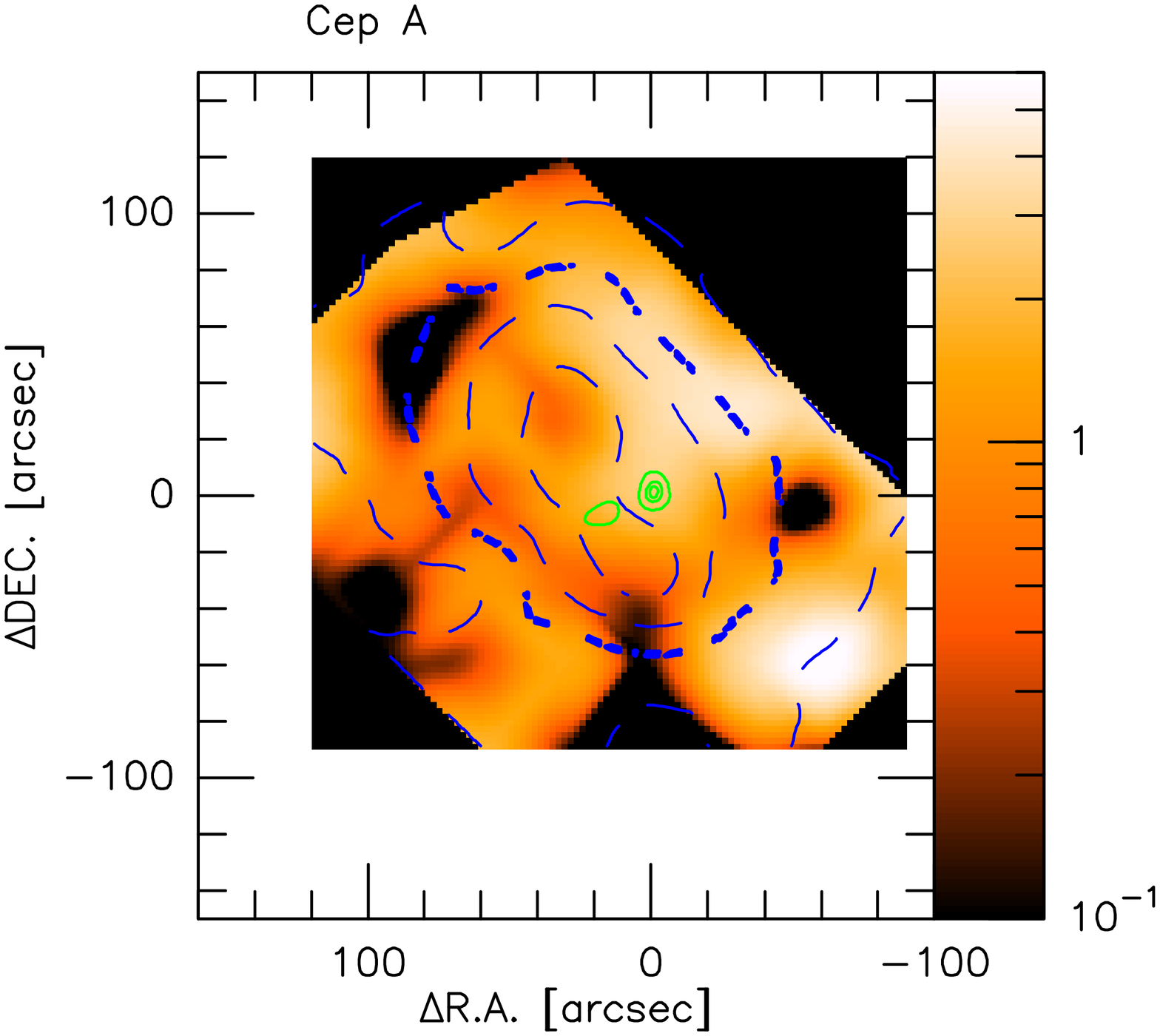}
\vspace*{-0.2 cm} \caption{\label{opdepth2} Contour maps \cch\ 1-0
(J = 3/2$\rightarrow$1/2, F = 2$\rightarrow$1) (dashed line)
integrated intensity and radio continuum emission (solid line)
superimposed on the optical depth of \cch\ 1-0 (J =
3/2$\rightarrow$1/2, F = 2$\rightarrow$1) in gray scale. (A color
version of this figure is available in the online journal.)}
\end{center}
\end{figure}

\clearpage

\begin{figure}[]
\begin{center}
\includegraphics[width=2.0in, angle=270]{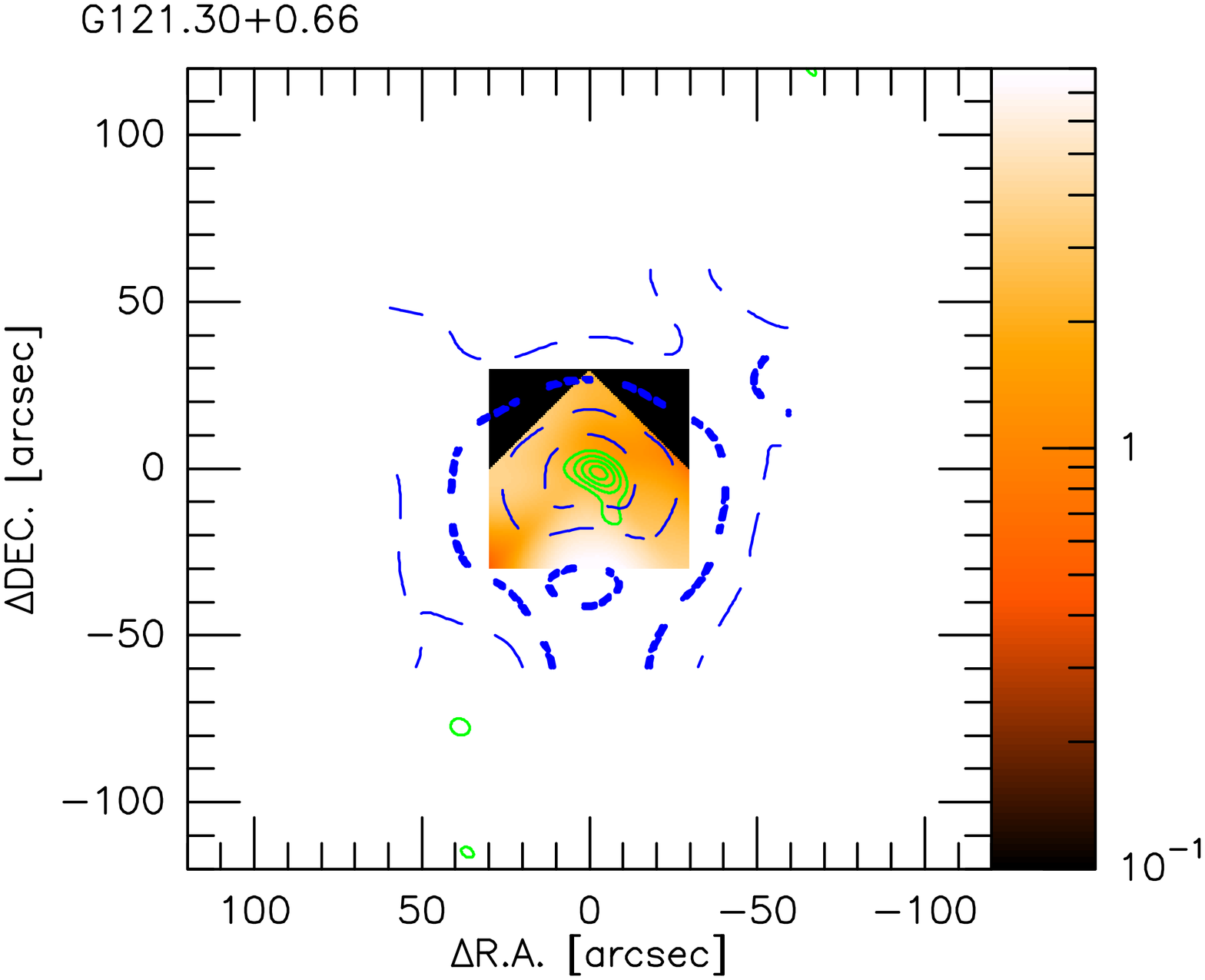}
\includegraphics[width=2.0in, angle=270]{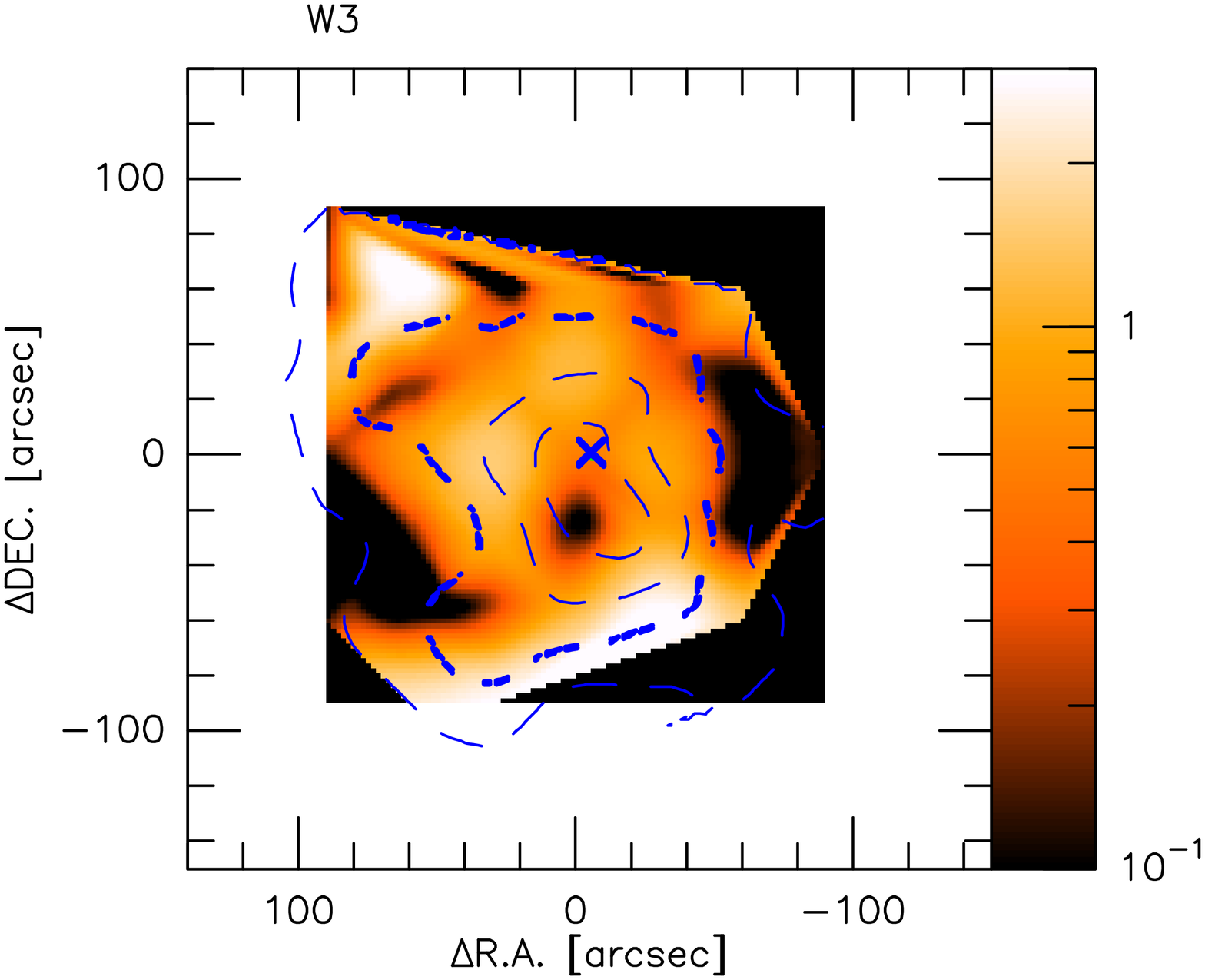}
\includegraphics[width=2.0in, angle=270]{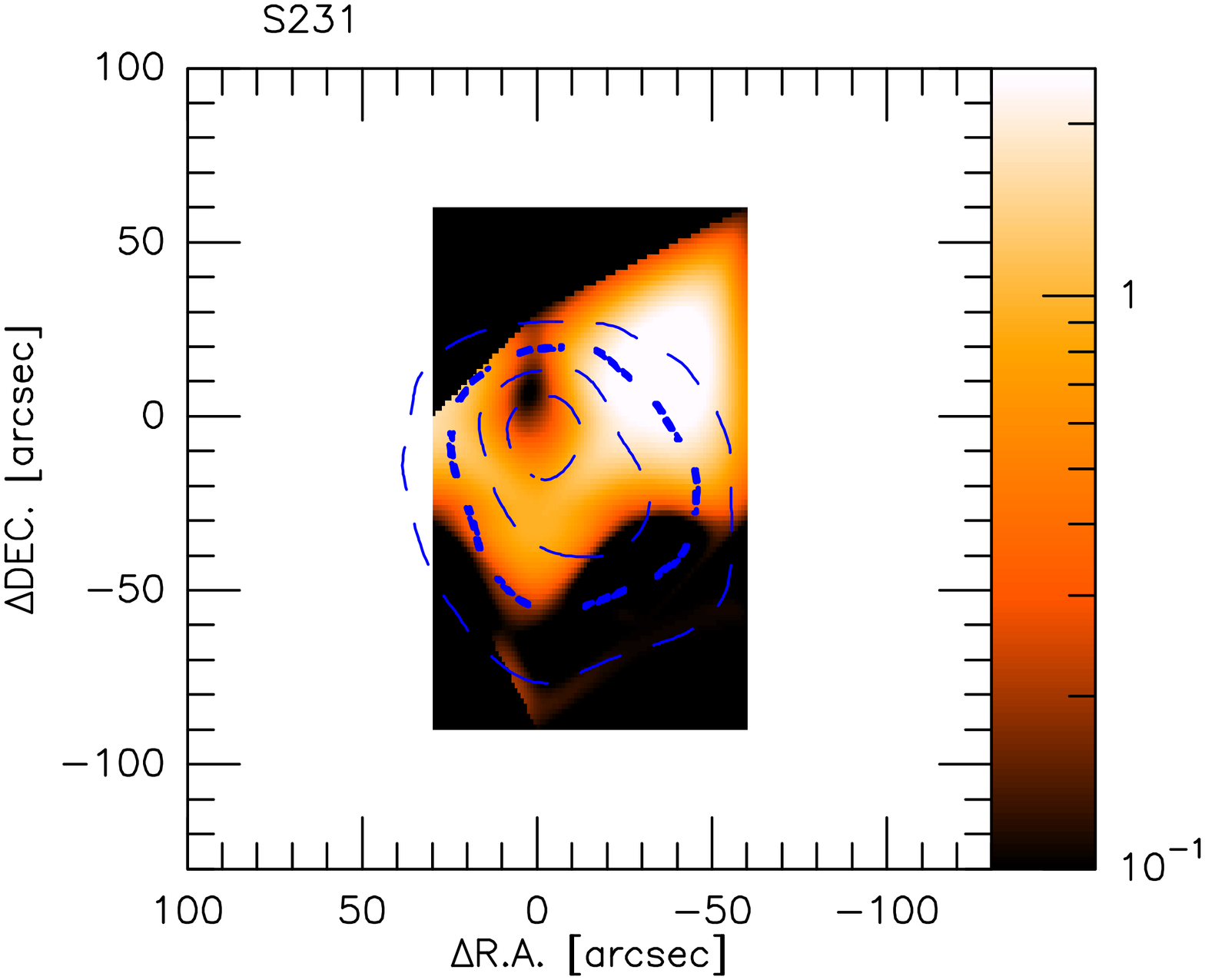}
\includegraphics[width=2.0in, angle=270]{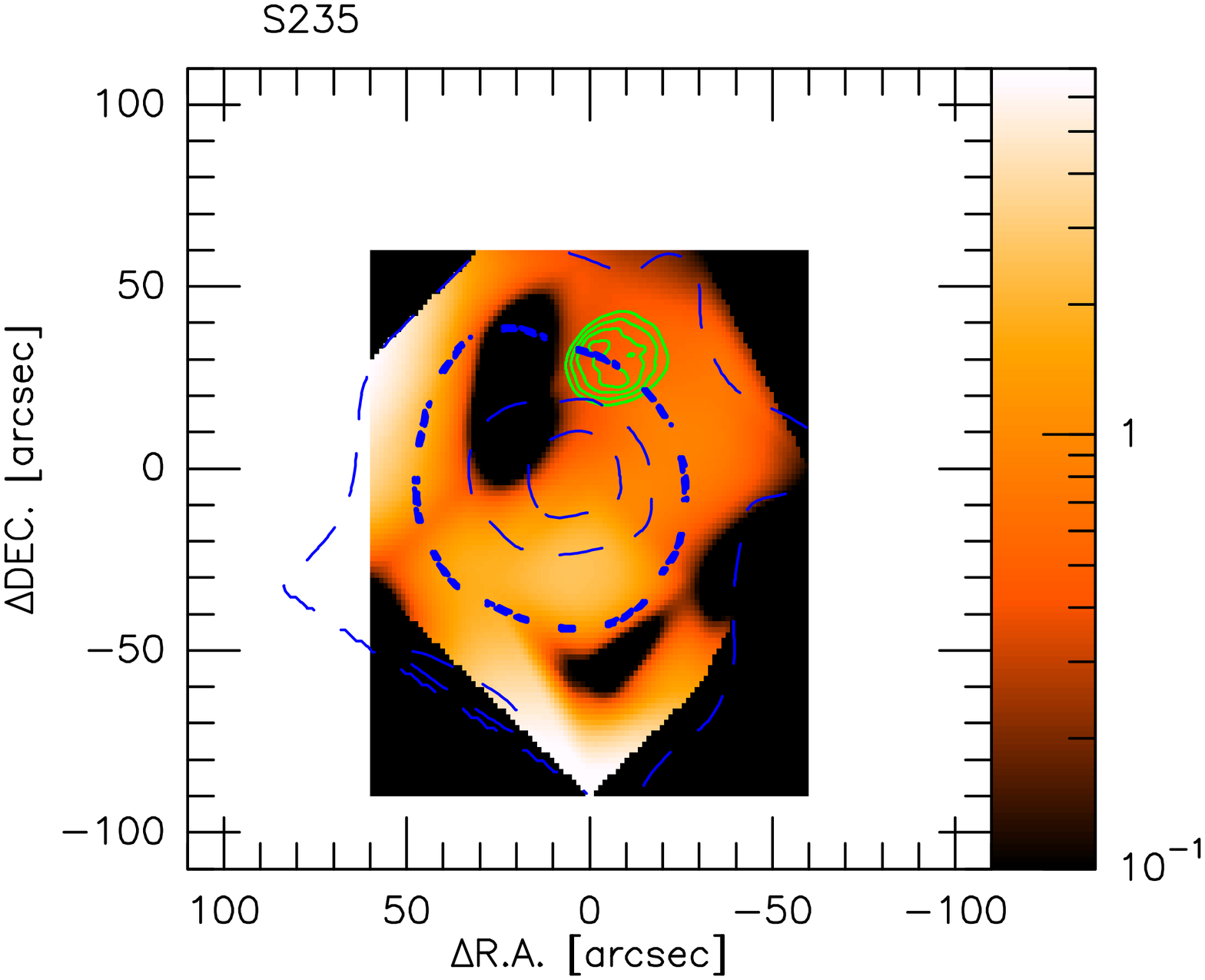}
\includegraphics[width=2.0in, angle=270]{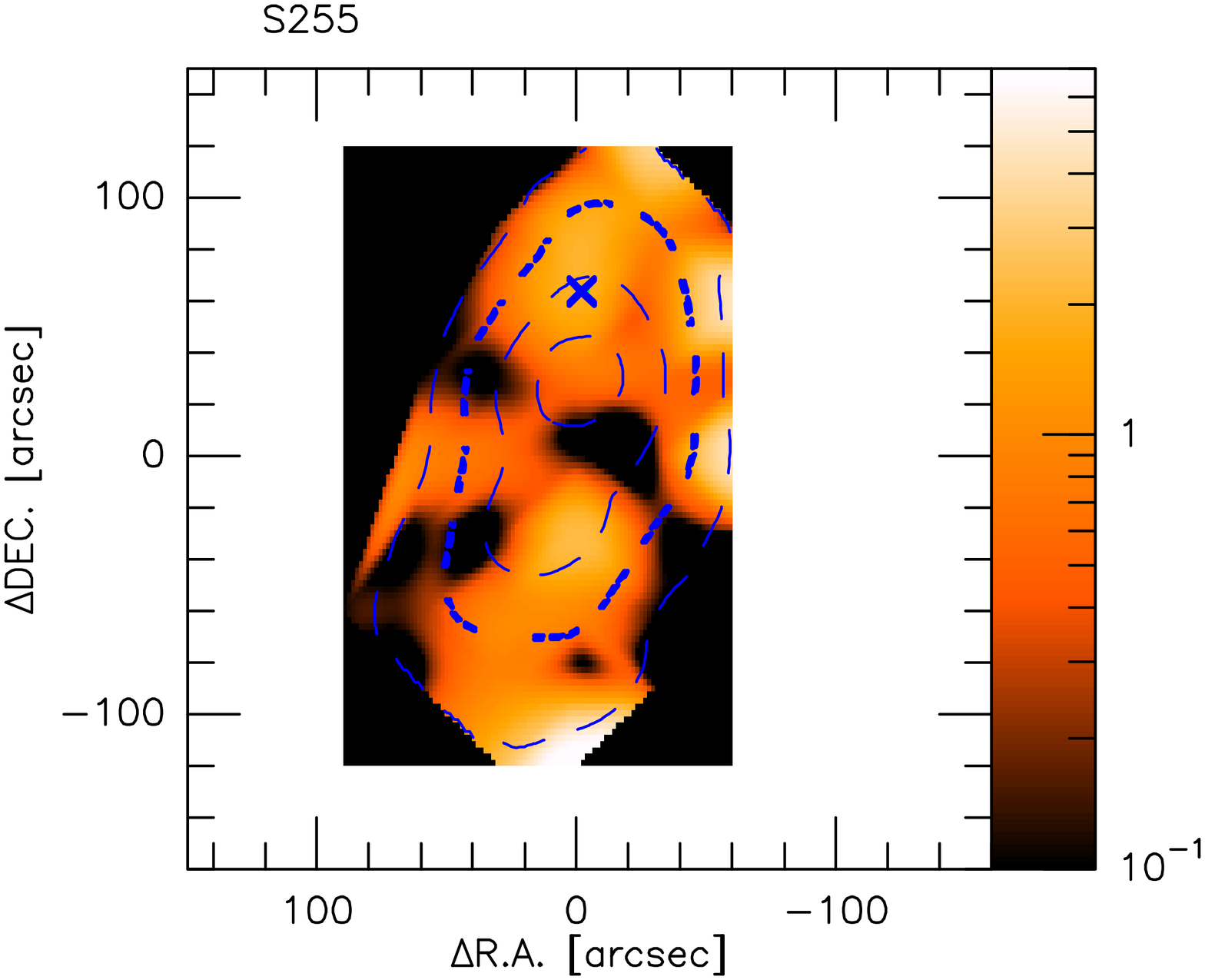}
\includegraphics[width=2.0in, angle=270]{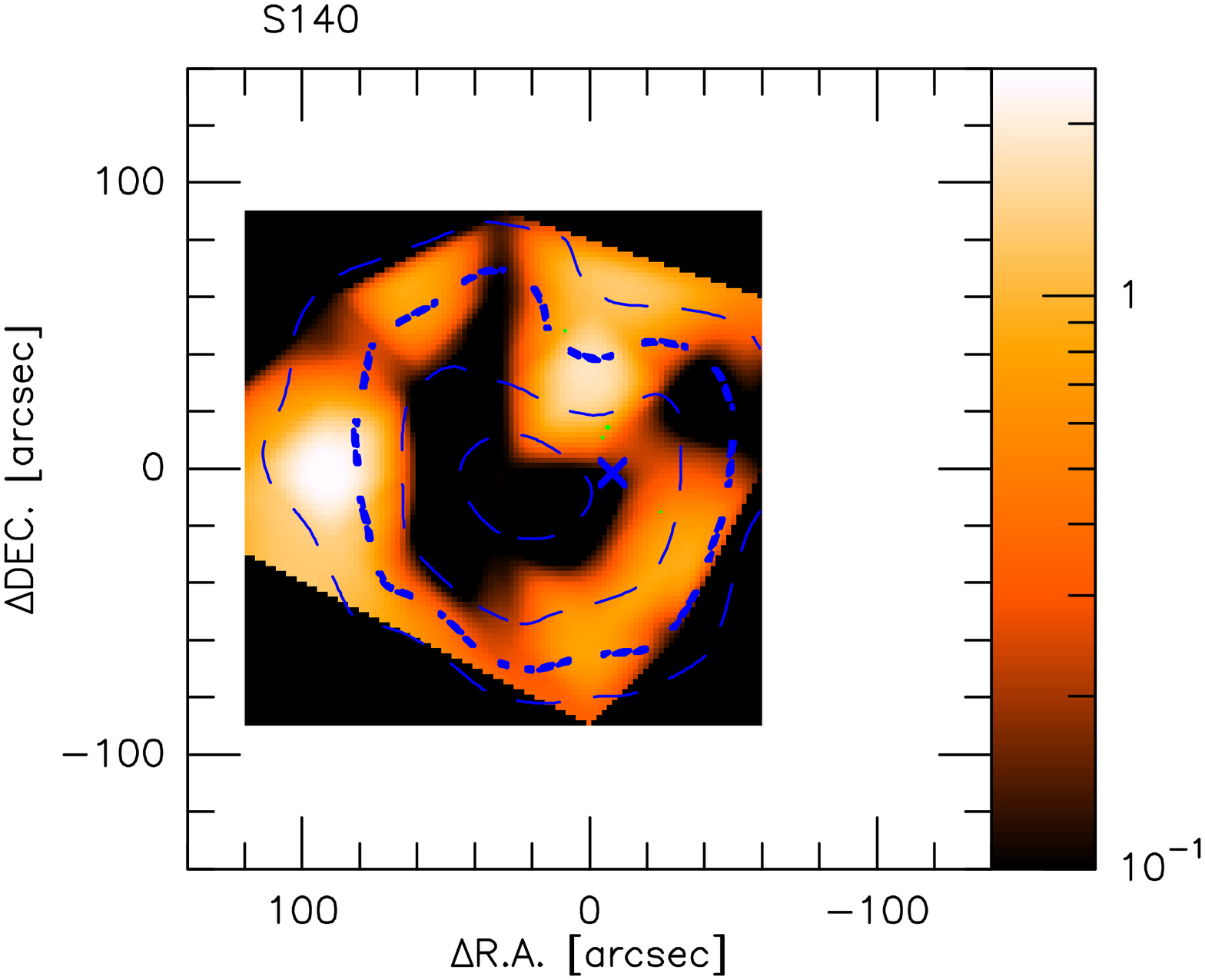}
\includegraphics[width=2.0in, angle=270]{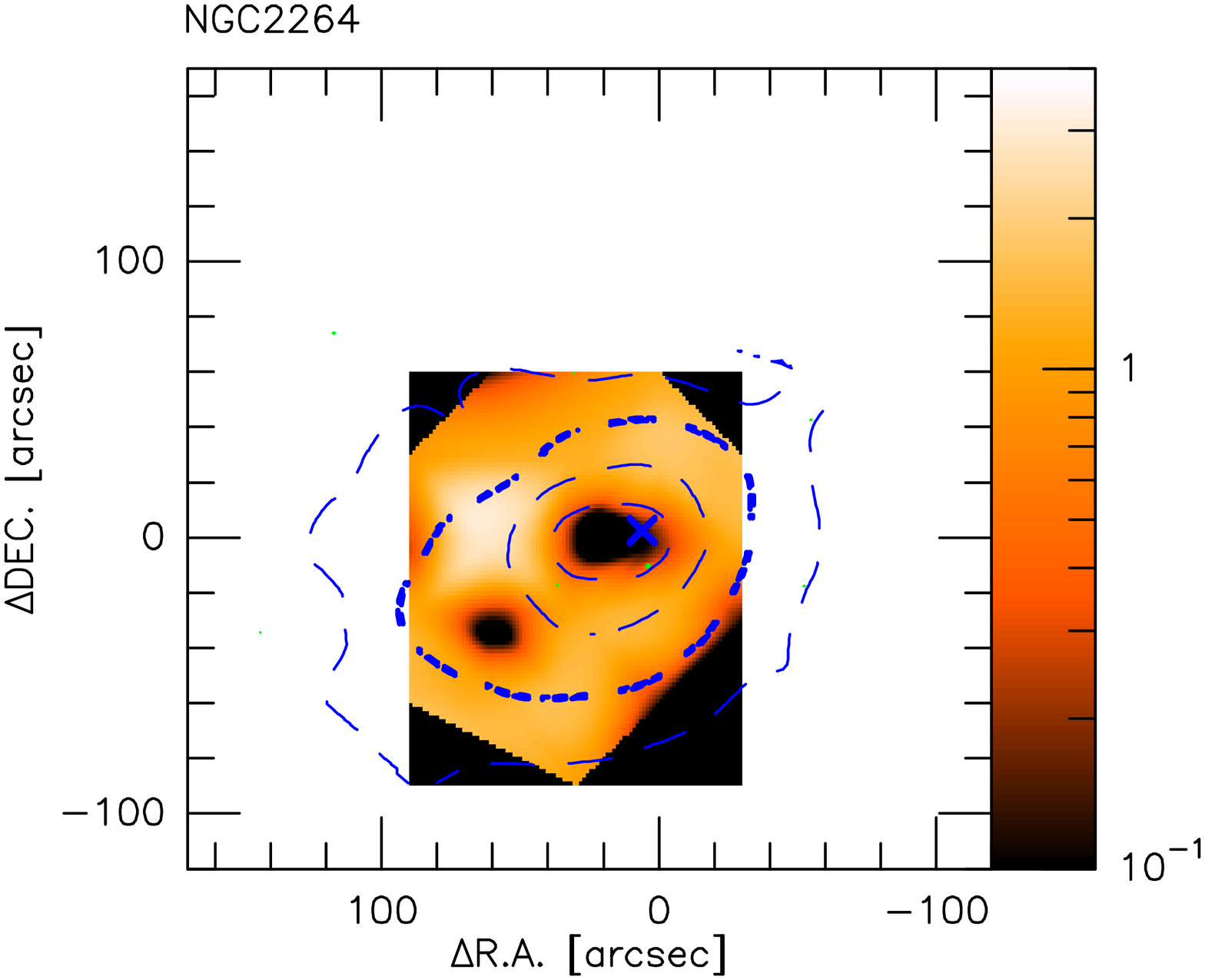}
\vspace*{-0.2 cm} \caption{\label{opdepth3} Contour maps \cch\ 1-0
(J = 3/2$\rightarrow$1/2, F = 2$\rightarrow$1) (dashed line)
integrated intensity and radio continuum emission (solid line)
superimposed on the optical depth of \cch\ 1-0 (J =
3/2$\rightarrow$1/2, F = 2$\rightarrow$1) in gray scale. (A color
version of this figure is available in the online journal.)}
\end{center}
\end{figure}

\clearpage

\begin{figure}[]
\begin{center}
\includegraphics[width=5.1in]{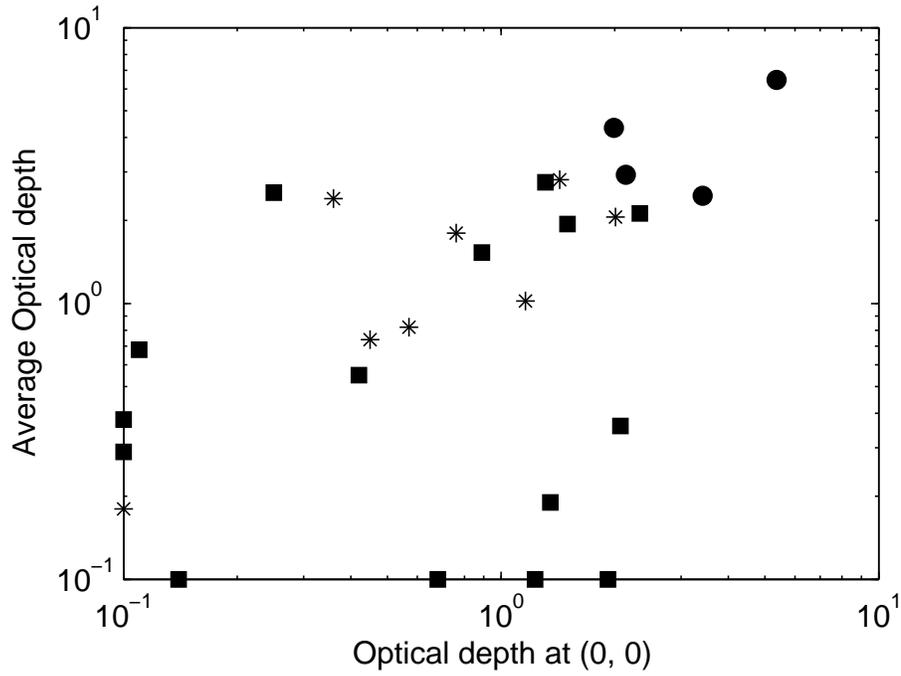}
\vspace*{-0.2 cm} \caption{\label{opdistribution} Plot of optical
depth for \cch\ 1-0 main line at (0, 0)
versus optical depth of \cch\ 1-0 main line averaging over the whole
emitting regions. Sources not associated with a
detectable HII regions and sources associated with HCHII regions are
plotted as circles. Sources associated with UCHII regions are
plotted as squares, while sources associated with CHII or extended
HII regions are shown in stars. (A color version of this figure is
available in the online journal.)}
\end{center}
\end{figure}

\clearpage

\begin{figure}[]
\begin{center}
\includegraphics[width=3.1in]{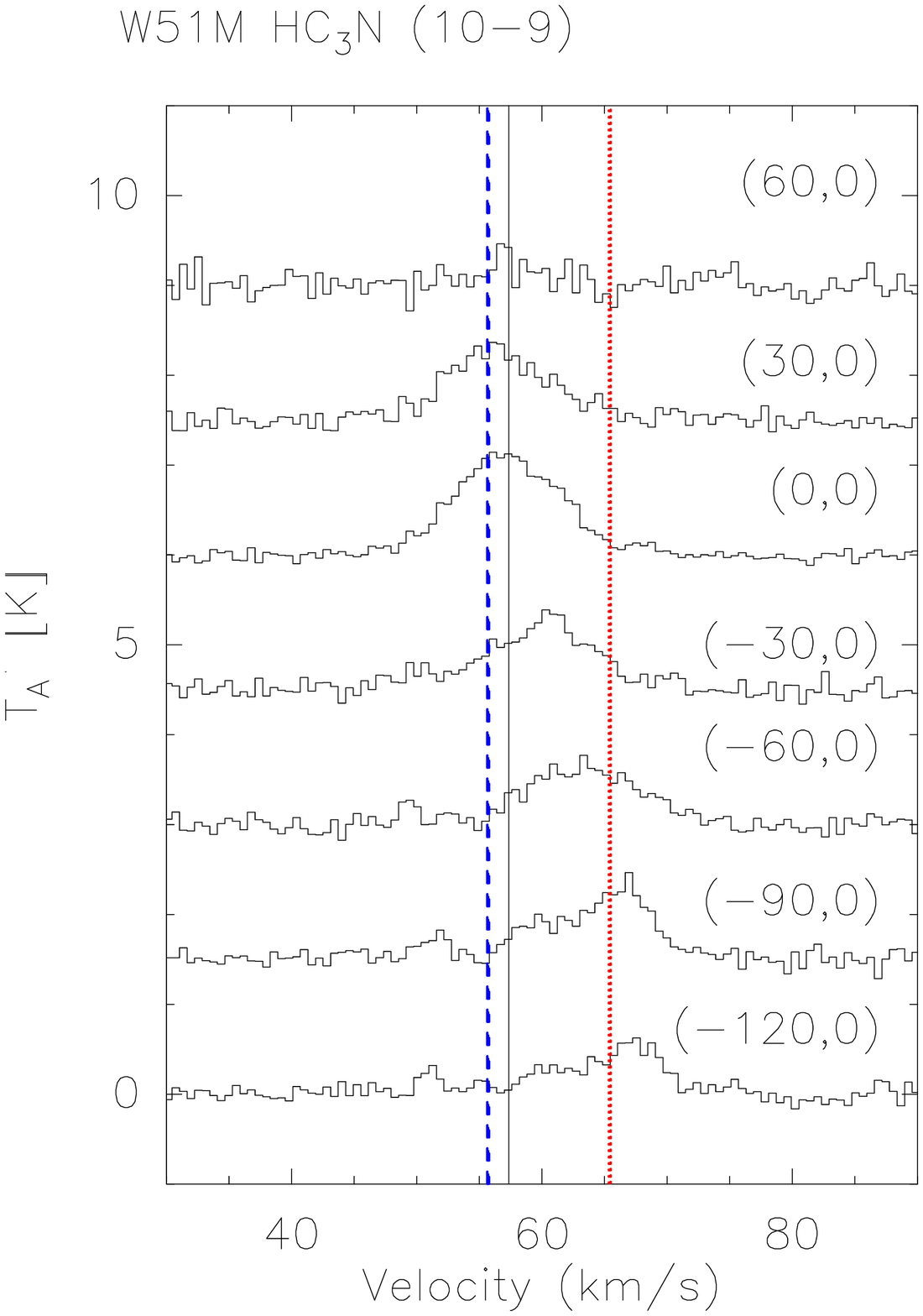}
\includegraphics[width=3.1in]{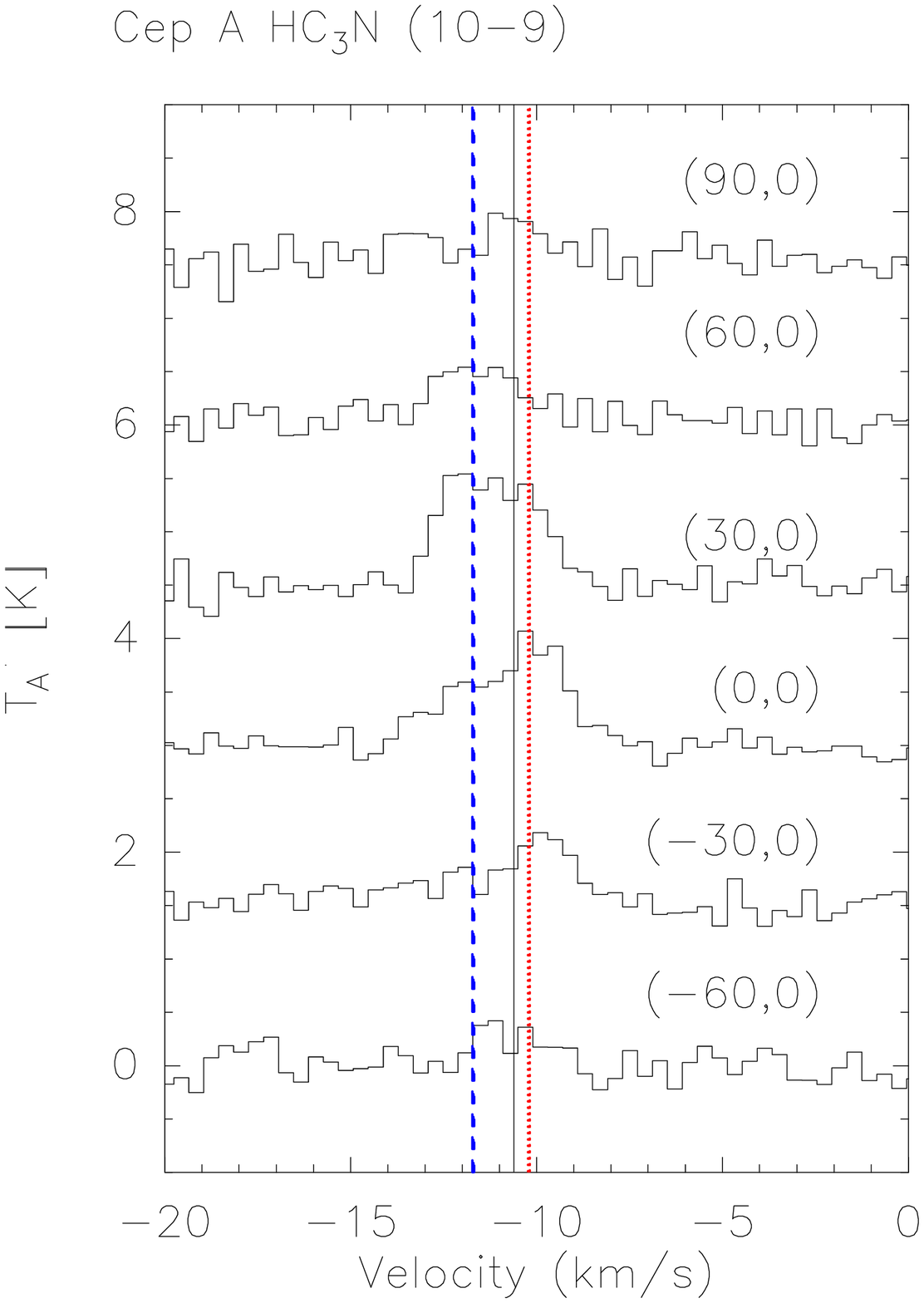}
\vspace*{-0.2 cm} \caption{\label{velspec} Left: Spectra of \hcccn\
(10-9) at the offset position of $\triangle Dec = 0$ for W51M. The
blue dashed line mark centroid velocity for spectra line at (30,
0), while the solid line marks centroid velocity of spectral line at (0, 0),
and the red dotted line marks centroid velocity for
spectra line at (-120, 0). Right: Spectra of \hcccn\ (10-9) at the
offset position of $\triangle Dec = 0$ for Cep A. The blue dashed
line mark centroid velocity of spectra line at (60, 0), while the solid
line marks centroid veloicty of spectra line at (0, 0), and the
red dotted line marks centroid velocity of spectra line at (-30,
0). Centroid velocities are derived from Gaussian fit. (A color version of
this figure is available on line.)}
\end{center}
\end{figure}

\clearpage

%%\begin{figure}[]                    %%%%% not necessary
%%\begin{center}
%%\includegraphics[width=2.0in]{gradhc3n.eps}
%%\includegraphics[width=2.0in]{gradhnc.eps}
%%\includegraphics[width=2.0in]{gradc2h.eps}
%%\vspace*{-0.2 cm} \caption{\label{grad} Distribution of gradient
%%magnitude ($Grad \geq 3 \sigma_{Grad}$) for \hcccn, HNC and \cch.
%%The dotted vertical lines in each panel represent the median of
%%$Grad$.}
%%\end{center}
%%\end{figure}
%%
%%\clearpage

\begin{figure}[]
\begin{center}
\includegraphics[width=2.0in]{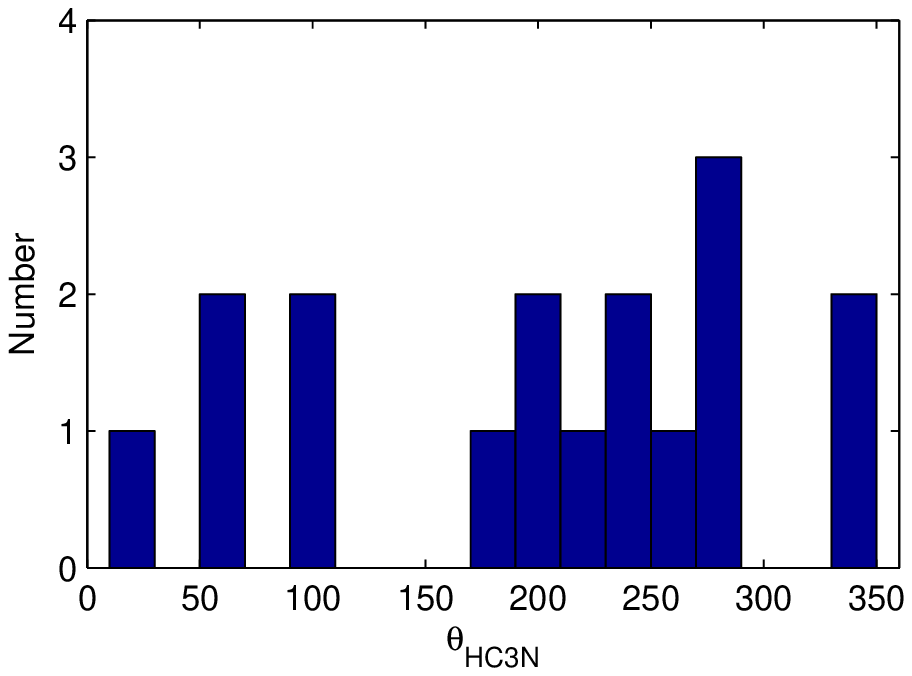}
\includegraphics[width=2.0in]{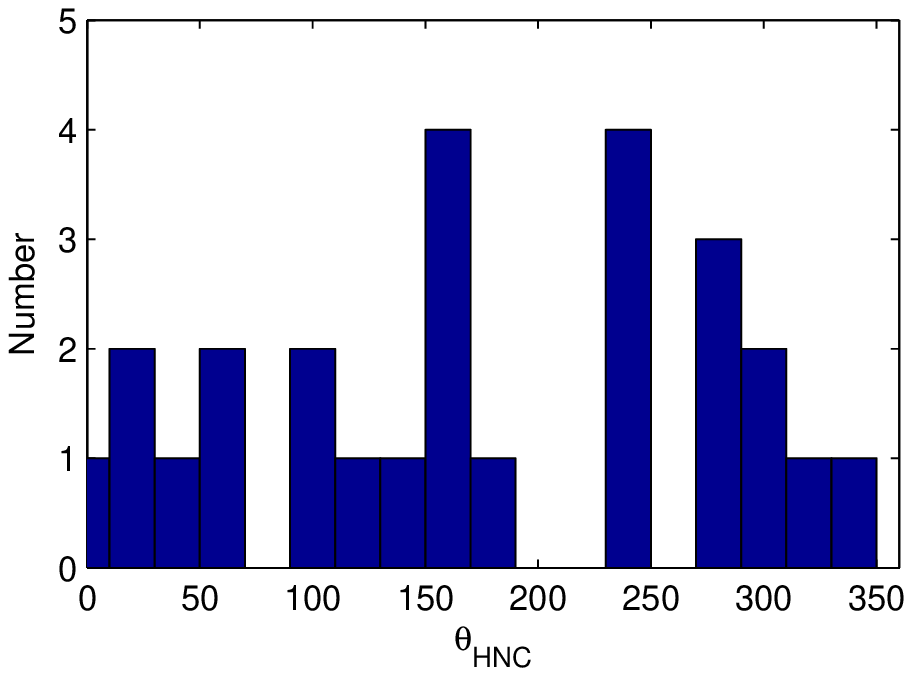}
\includegraphics[width=2.0in]{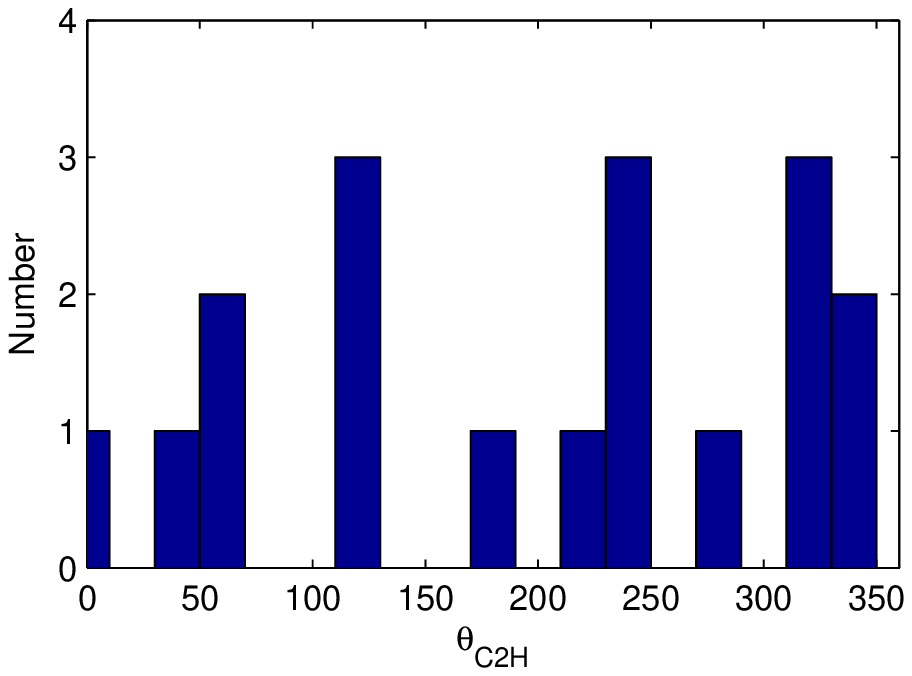}
\vspace*{-0.2 cm} \caption{\label{graddirec} Distribution of
gradient direction ($Grad \geq 3 \sigma_{Grad}$) for \hcccn, HNC and
\cch.}
\end{center}
\end{figure}

\clearpage

\begin{figure}[]
\begin{center}
\includegraphics[width=3.0in]{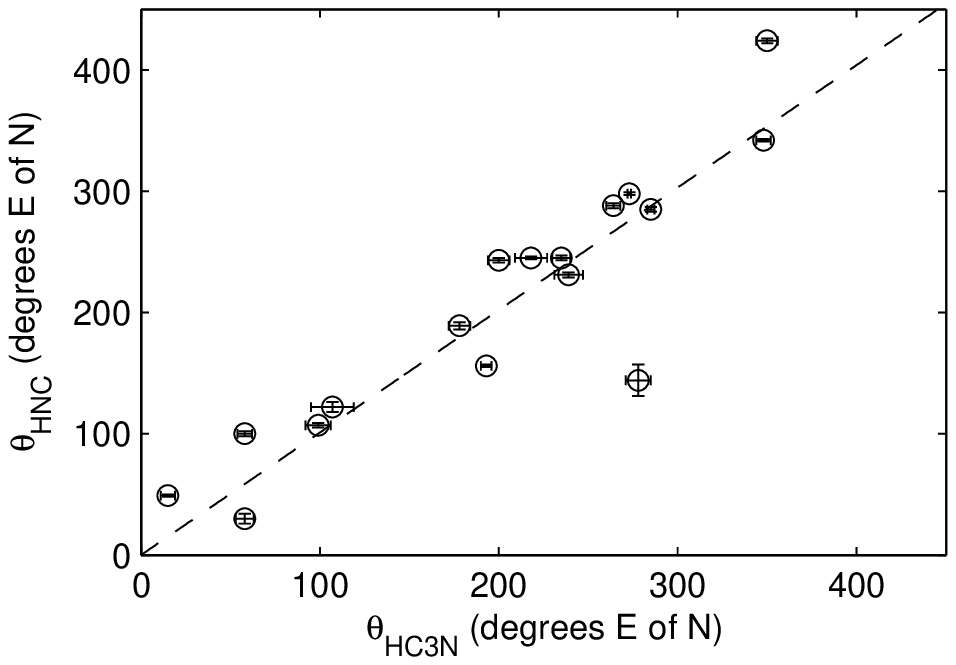}
\includegraphics[width=3.0in]{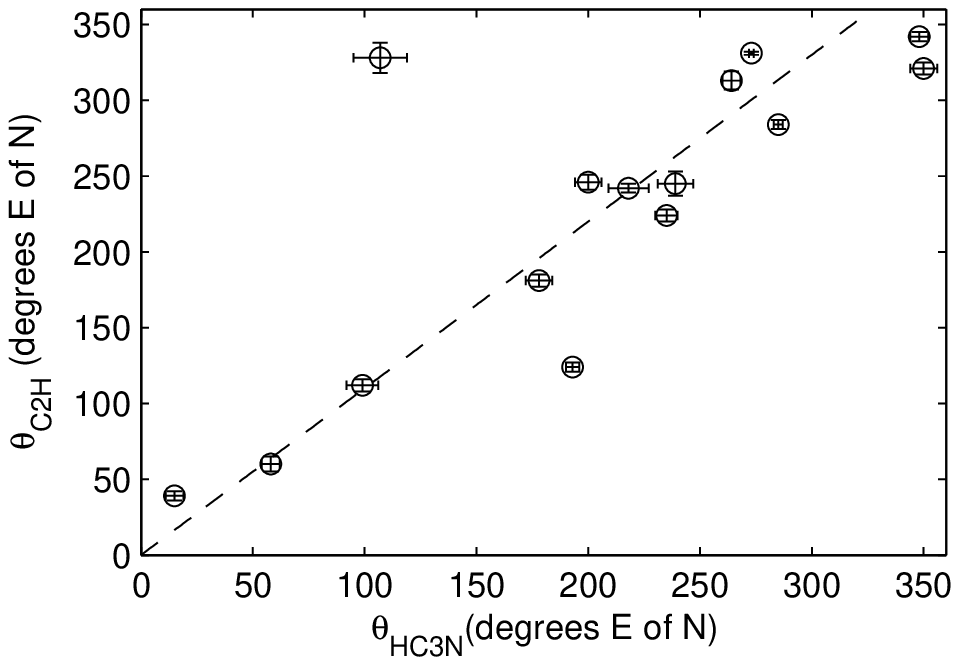}
\vspace*{-0.2 cm} \caption{\label{graddirecomp} Comparison of
gradient direction for \hcccn\ and HNC (left), \hcccn\ and \cch\
(right). The dashed lines indicate the best-fit relation, with slope
of 1.01$\pm$0.12 and 1.10$\pm$0.23, respectively.}
\end{center}
\end{figure}

\clearpage

\begin{figure}[]
\begin{center}
\includegraphics[width=3.0in]{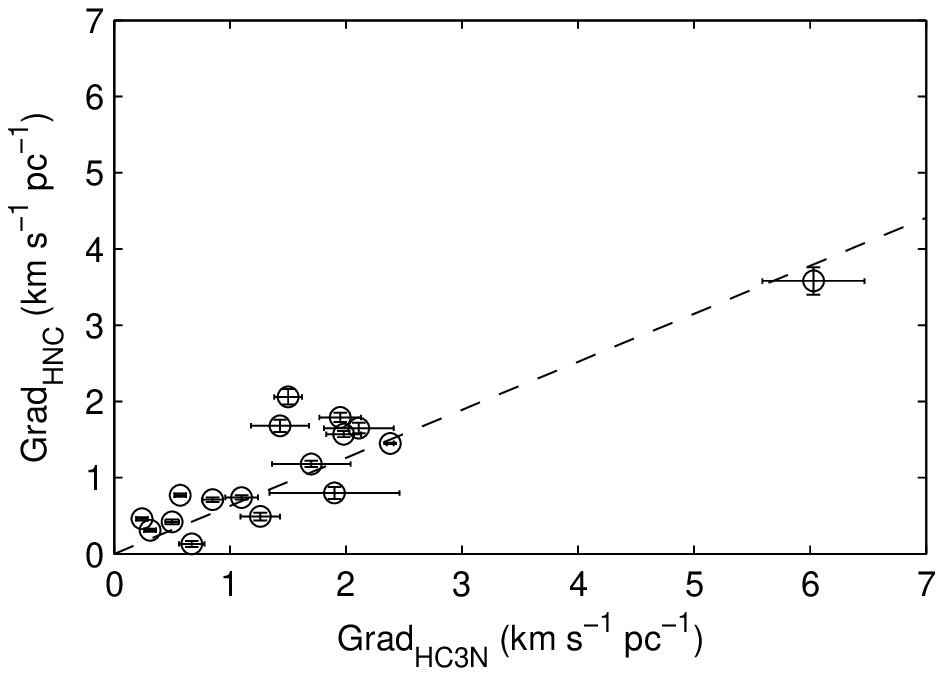}
\includegraphics[width=3.0in]{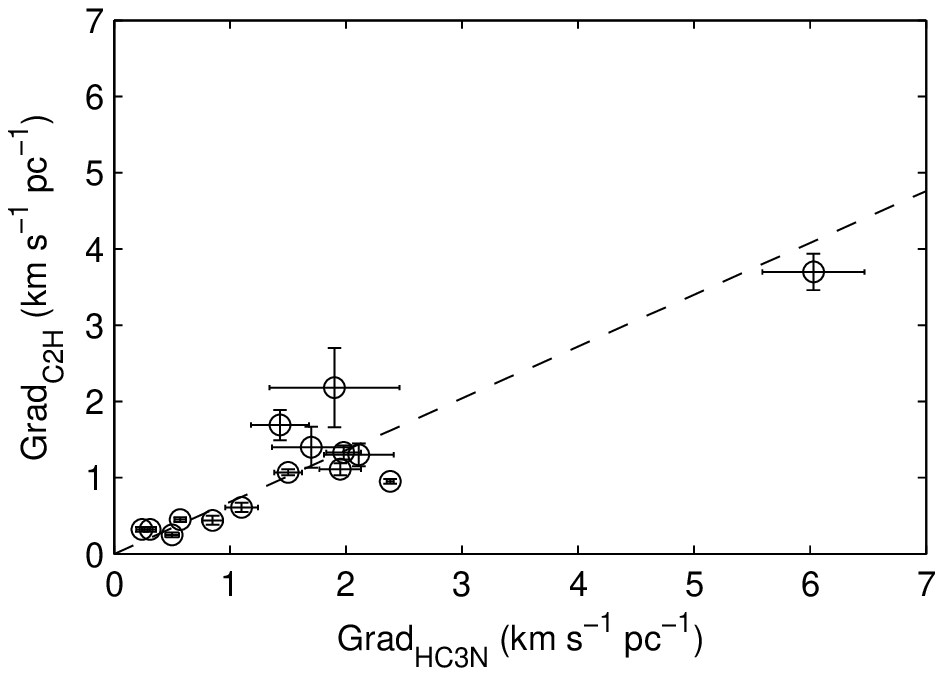}
\vspace*{-0.2 cm} \caption{\label{gradcomp} Comparison of gradient
magnitude for \hcccn\ and HNC (left), \hcccn\ and \cch\ (right). The
dashed lines indicate the best-fit relation, with slope of
0.63$\pm$0.08 and 0.68$\pm$0.10.}
\end{center}
\end{figure}

\clearpage

%%\begin{figure}[]          % not necessary
%%\begin{center}
%%\includegraphics[width=3.0in]{rotlinebroad.eps}
%%\vspace*{-0.2 cm} \caption{\label{rotlinebroad} Histogram of line
%%broadening due to rotation. The dotted vertical line represents the
%%median of $\sigma_{rot}$.}
%%\end{center}
%%\end{figure}
%%
%%\clearpage

\begin{figure}[]
\begin{center}
\includegraphics[width=3.0in]{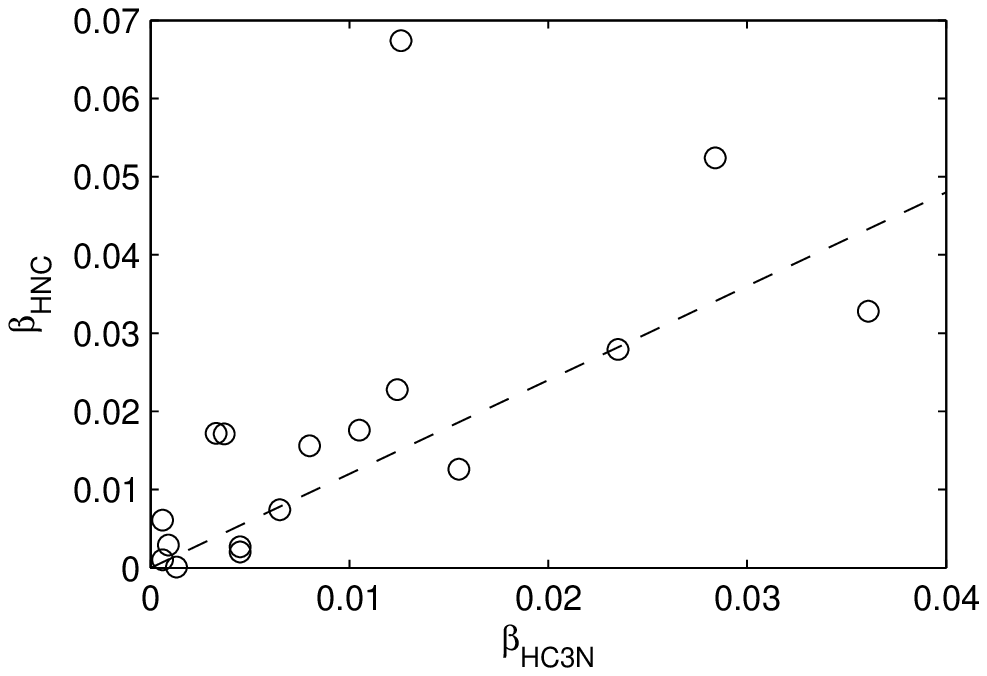}
\includegraphics[width=3.0in]{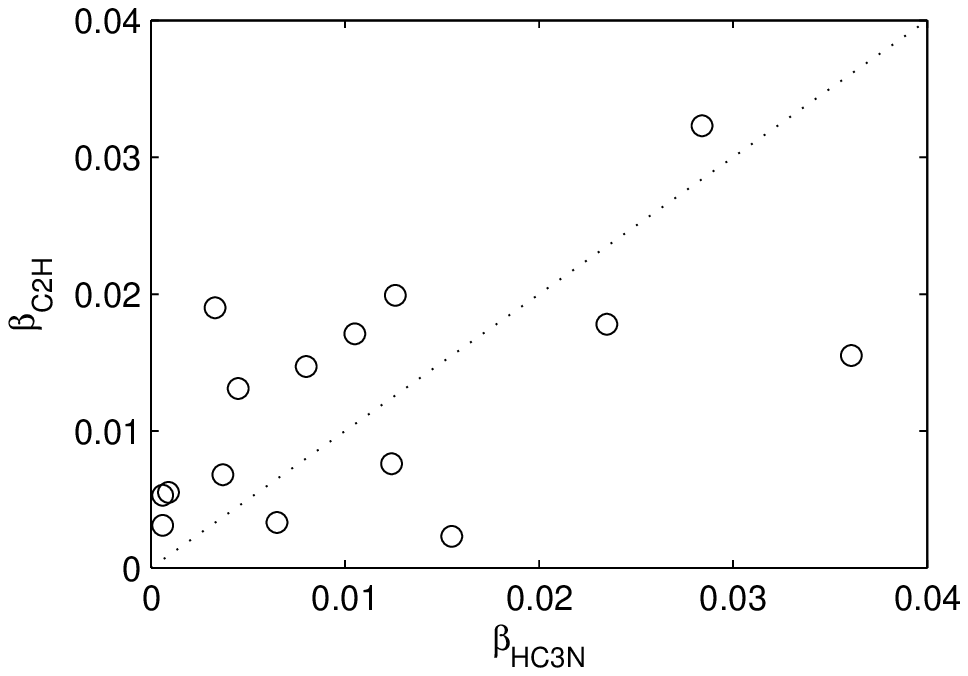}
\vspace*{-0.2 cm} \caption{\label{betacomp} Comparison of ratio of
rotational energy to gravitational energy $\beta$ for \hcccn\ and
HNC (left), \hcccn\ and \cch\ (right). The dashed line in the left
figure indicate the best-fit relation, with slopes of 1.48$\pm$0.52.
The slope of the dotted line in the right figure is 1.}
\end{center}
\end{figure}

\clearpage

\begin{figure}[]
\begin{center}
\includegraphics[width=3.0in]{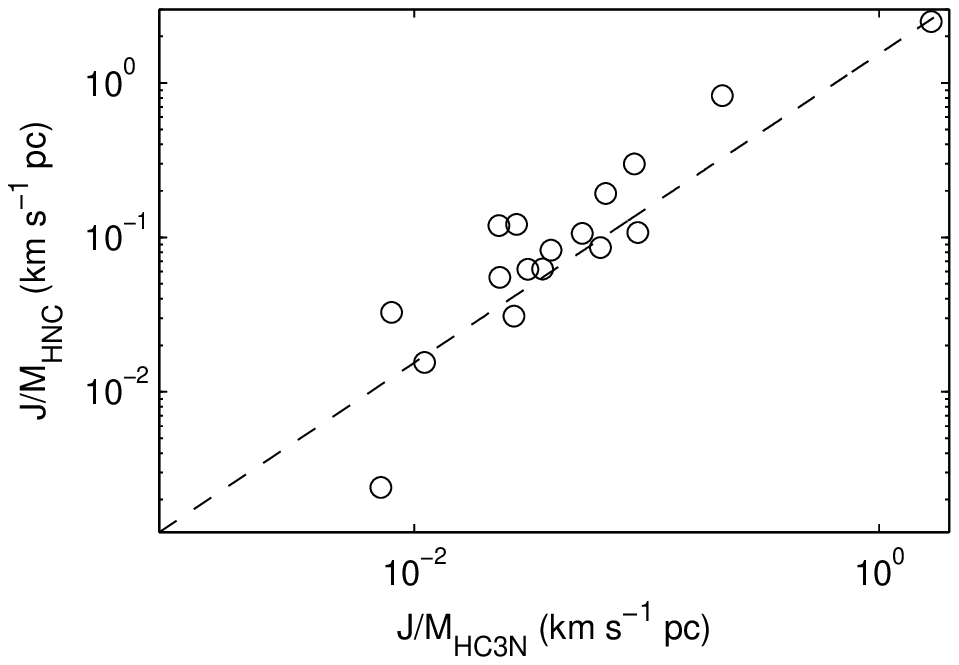}
\includegraphics[width=3.0in]{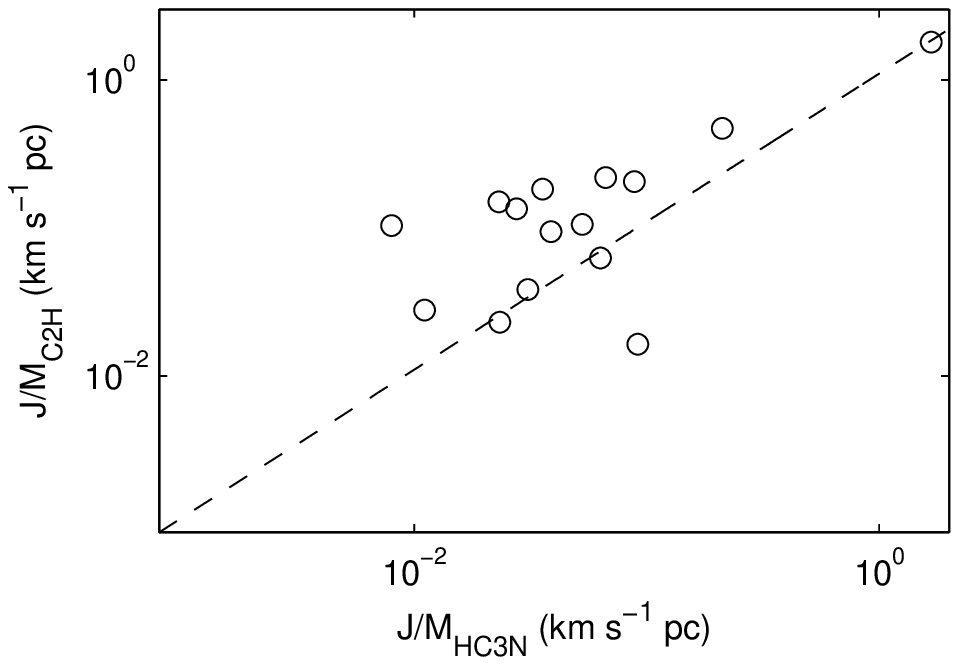}
\vspace*{-0.2 cm} \caption{\label{jmcomp} Comparison of specific
angular momentum $J/M$ for \hcccn\ and HNC (left), \hcccn\ and \cch\
(right). The dashed lines indicate the best-fit relation, with
slopes of 1.54$\pm$0.17 and 1.10$\pm$0.13, respectively. }
\end{center}
\end{figure}

\clearpage

\begin{figure}[]
\begin{center}
\includegraphics[width=3.0in]{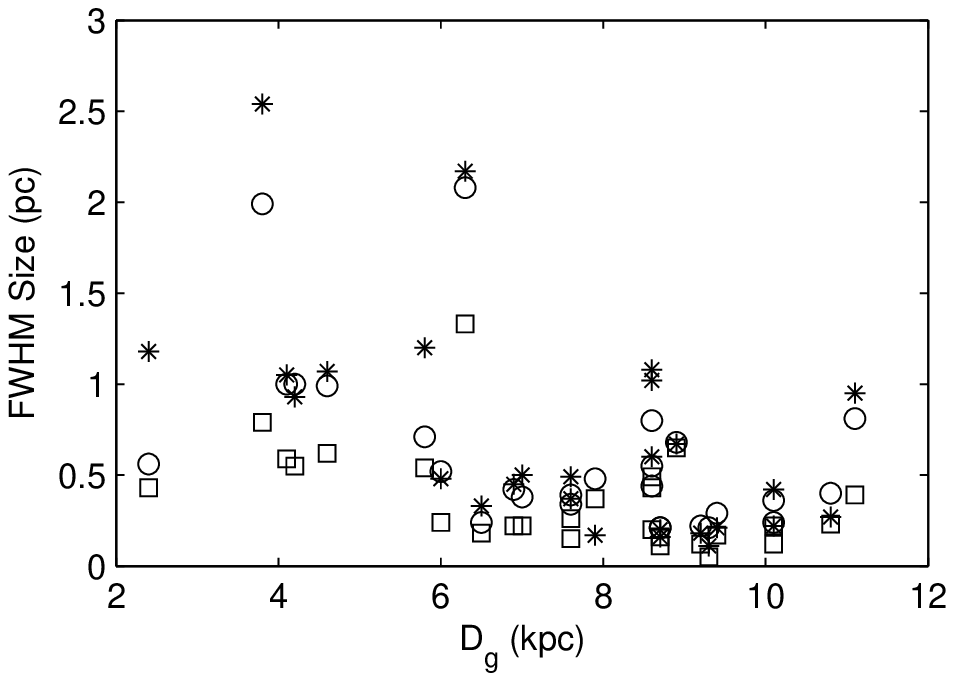}
\includegraphics[width=3.0in]{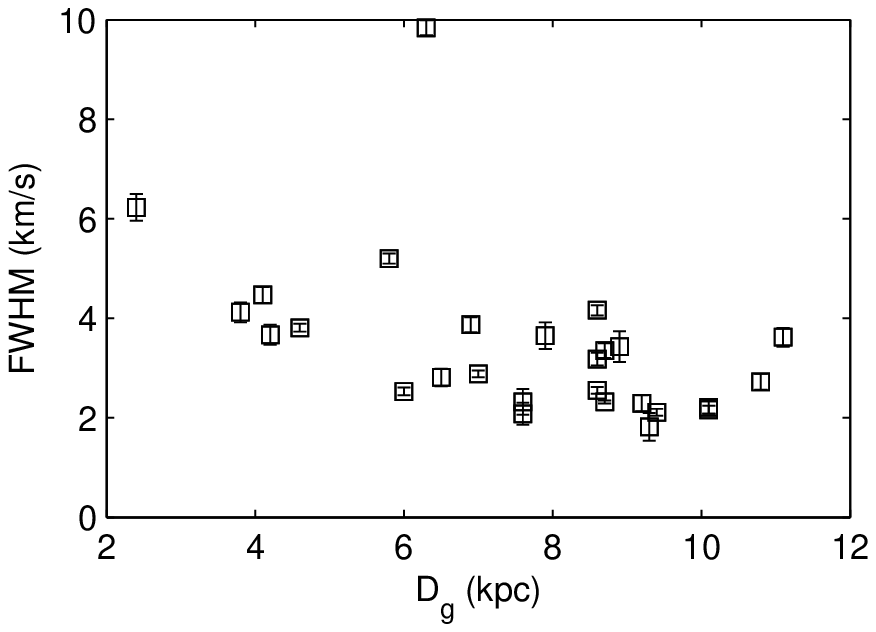}
\includegraphics[width=3.0in]{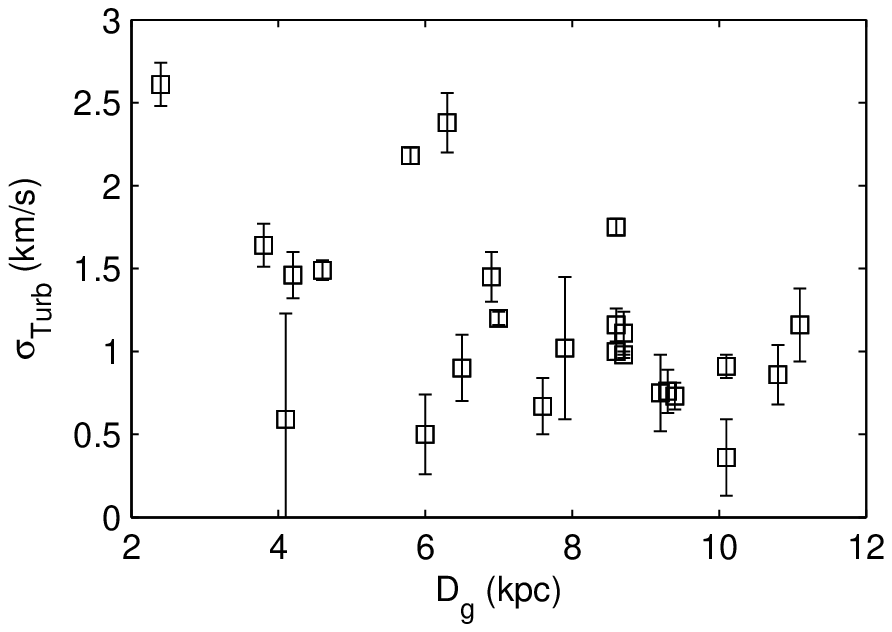}
\includegraphics[width=3.0in]{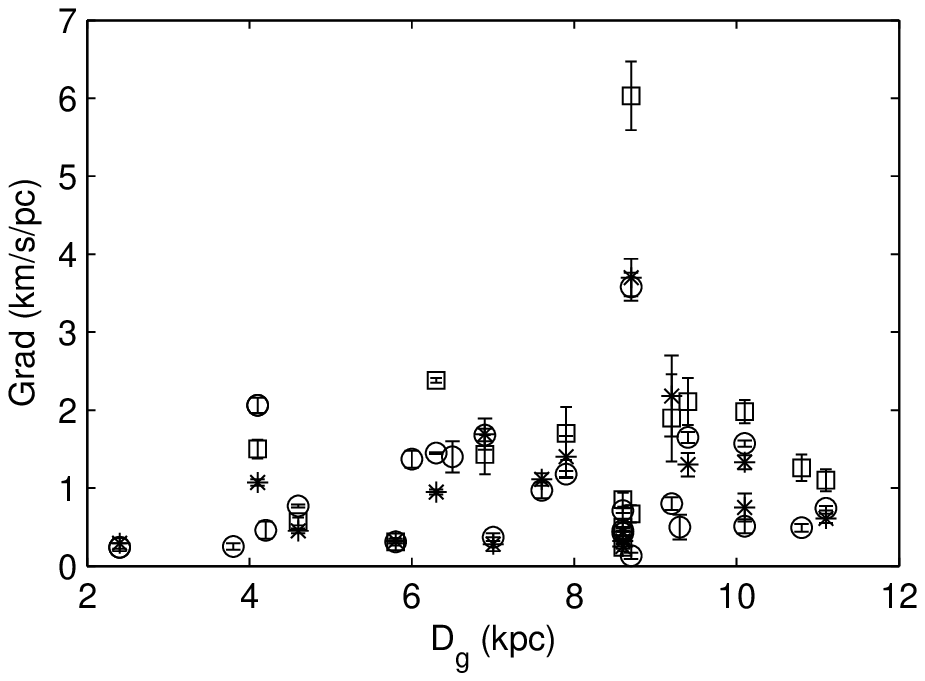}
\includegraphics[width=3.0in]{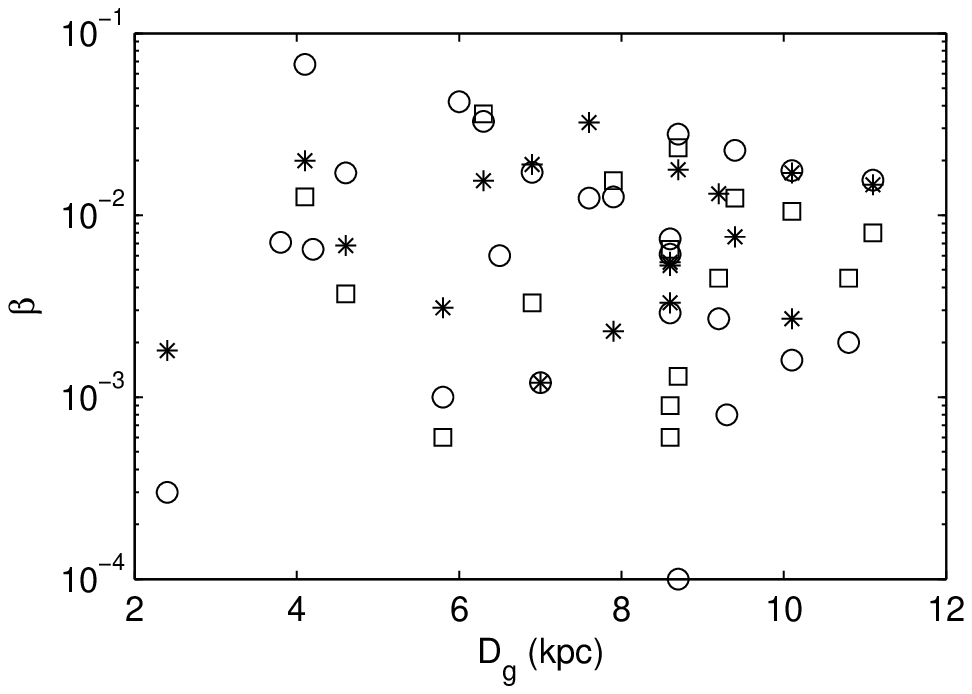}
\includegraphics[width=3.0in]{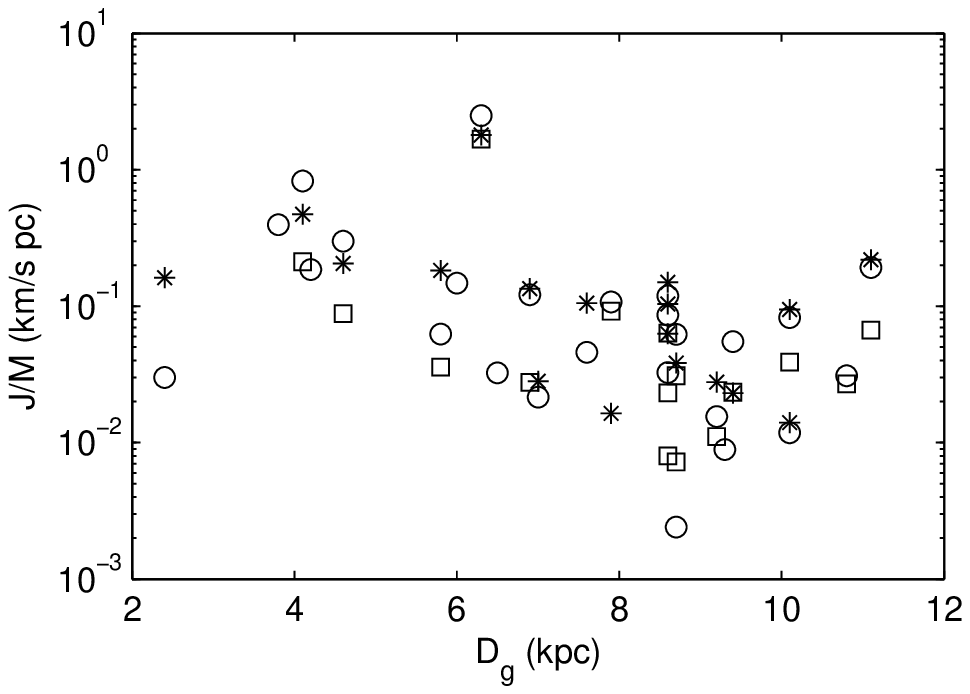}
\vspace*{-0.2 cm} \caption{\label{galactrend} Plot of FWHM sizes,
linewidths, $\sigma_{Turb}$, velocity gradients, $\beta$ and
specific angular momentum $J/M$ versus galactocentric distance for
\hcccn\ (square), HNC (circle) and \cch\ (star). }
\end{center}
\end{figure}

\end{document}